\documentclass[%
 reprint,
 superscriptaddress,
nofootinbib,
 amsmath,amssymb,
 aps,
]{revtex4-2}

\usepackage{graphicx}
\usepackage{dcolumn}
\usepackage{bm}
\usepackage{hyperref}
\usepackage{notes2bib}

\usepackage{xcolor}

\begin{document}


\title{Performance of a kinetic inductance phonon-mediated detector at the NEXUS cryogenic facility}


\author{Dylan J. Temples}
    \email{dtemples@fnal.gov}

    \affiliation{Astrophysics Department, Fermi National Accelerator Laboratory (FNAL), Batavia, IL, 60510, USA}

\author{Osmond Wen}
\author{Karthik Ramanathan}
\author{Taylor Aralis}
    \altaffiliation[Presently at ]{6}
\author{Yen-Yung Chang}
    \altaffiliation[Presently at ]{Department of Physics, University of California, Berkeley, CA, 94720, USA}
\author{Sunil Golwala}

    \affiliation{Division of Physics, Mathematics, and Astronomy, California Institute of Technology, Pasadena, CA, 91106, USA}

\author{Lauren Hsu}

    \affiliation{Astrophysics Department, Fermi National Accelerator Laboratory (FNAL), Batavia, IL, 60510, USA}


    \affiliation{Division of Physics, Mathematics, and Astronomy, California Institute of Technology, Pasadena, CA, 91106, USA}

\author{Corey Bathurst}

    \affiliation{Department of Physics, University of Florida, Gainesville, FL, 32611, USA}

\author{Daniel Baxter}

    \affiliation{Astrophysics Department, Fermi National Accelerator Laboratory (FNAL), Batavia, IL, 60510, USA}
    \affiliation{Department of Physics \& Astronomy, Northwestern University, Evanston, IL, 60208, USA}

\author{Daniel Bowring}

    \affiliation{Astrophysics Department, Fermi National Accelerator Laboratory (FNAL), Batavia, IL, 60510, USA}

\author{Ran Chen}

    \affiliation{Department of Physics \& Astronomy, Northwestern University, Evanston, IL, 60208, USA}

\author{Enectali Figueroa-Feliciano}

    \affiliation{Department of Physics \& Astronomy, Northwestern University, Evanston, IL, 60208, USA}
    \affiliation{Astrophysics Department, Fermi National Accelerator Laboratory (FNAL), Batavia, IL, 60510, USA}

\author{Matthew Hollister}
\author{Christopher James}

    \affiliation{SQMS Ultralow Temperature Cryogenics Department, Fermi National Accelerator Laboratory (FNAL), Batavia, IL, 60510, USA}

\author{Kyle Kennard}

    \affiliation{Department of Physics \& Astronomy, Northwestern University, Evanston, IL, 60208, USA}

\author{Noah Kurinsky}

    \affiliation{Kavli Institute for Particle Astrophysics and Cosmology, SLAC National Accelerator Laboratory, Menlo Park, CA, 94025, USA}
    
\author{Samantha Lewis}
    \altaffiliation[Presently at ]{Department of Physics and Astronomy, Wellesley College, Wellesley, MA, 02481, USA}
\author{Patrick Lukens}

    \affiliation{Astrophysics Department, Fermi National Accelerator Laboratory (FNAL), Batavia, IL, 60510, USA}

\author{Valentina Novati}
    \altaffiliation[Presently at ]{LPSC, National Centre for Scientific Research, Universit\'{e} Grenoble Alpes, Grenoble, France}
\author{Runze Ren}
    \altaffiliation[Presently at ]{Department of Physics, University of Toronto, Toronto, ON, Canada}
\author{Benjamin Schmidt}
    \altaffiliation[Presently at ]{IRFU, Alternative Energies and Atomic Energy Commission, Universit\'{e} Paris-Saclay, France}

    \affiliation{Department of Physics \& Astronomy, Northwestern University, Evanston, IL, 60208, USA}

\date{February 6, 2024}

\begin{abstract}
Microcalorimeters that leverage microwave kinetic inductance detectors to read out phonon signals in the particle-absorbing target, referred to as kinetic inductance phonon-mediated (KIPM) detectors, offer an attractive detector architecture to probe dark matter (DM) down to the fermionic thermal relic mass limit. %
A prototype KIPM detector featuring a single aluminum resonator patterned onto a 1-gram silicon substrate was operated in the NEXUS low-background facility at Fermilab for characterization and evaluation of this detector architecture's efficacy for a dark matter search. %
An energy calibration was performed by exposing the bare substrate to a pulsed source of 470 nm photons, resulting in a baseline resolution on the energy absorbed by the phonon sensor of $2.1\pm0.2$ eV, a factor of two better than the current state-of-the-art, enabled by quasiparticle lifetimes extending up to 6.5 ms. %
However, due to the sub-percent phonon collection efficiency, the resolution on energy deposited in the substrate is limited to $\sigma_E=318 \pm 29$ eV. %
We further model both the signal pulse shape as a function of device temperature to extract quasiparticle lifetimes, and the observed noise spectra. 

\begin{description}
\item[Report Number]
FERMILAB-PUB-23-674-LDRD-PPD
\end{description}

\end{abstract}

\keywords{Calorimeters, dark matter, kinetic inductance detectors, phonons, superconductors}
\maketitle


\section{\label{sec:intro}Introduction}

Null results from searches for particle dark matter (DM) with mass above 1 GeV$/c^2$~\cite{Cooley2022} motivate the development of detector technologies that can probe the remainder of the thermal relic parameter space down to the fermionic mass limit of a few keV. The average kinetic energy carried by DM with 1 MeV$/c^2$ mass is about 500 meV. As such, sensitivity to light DM requires the ability to detect meV-scale quanta. Detectors employing superconducting thin films to sense athermal phonons are a class of sensors that can achieve this requirement.~\cite{Essig2022}.

One detector architecture in this class is the kinetic inductance phonon-mediated (KIPM) detector, which uses microwave kinetic inductance detectors (MKIDs) to read out phonon signals in the device substrate~\cite{Golwala2008, Moore2012, Cardani2018, Golwala2022, Wen2021, Ramanathan2022}. Phonons produced by particle interactions in the substrate propagate to the superconducting films on the surface, where they break Cooper pairs into quasiparticles, provided the phonon carries an energy larger than the Cooper pair binding energy, $2\Delta$. This transient increase in quasiparticle density increases the surface impedance (AC resistance and kinetic inductance) of the resonator, shifting its resonant frequency $f_r$ and quality factor $Q$ to lower values. This shift can be measured in both magnitude and phase by interrogating the resonator with a microwave/radiofrequency (RF) probe tone at the quiescent resonant frequency and measuring the complex forward scattering parameter $S_{21}$.

KIPM detectors have a number of attractive features that motivate their use to detect low-mass dark matter. The energy carried by a phonon quantum is 2--3 orders of magnitude lower than the electronic bandgap energy of the substrate, which sets the threshold for sensing techniques that rely on ionization production~\cite{Chavarria2016}.
They are additionally non-dissipative, allowing for the use of quantum information science (QIS) techniques to reduce noise (\textit{e.g.}, vacuum squeezing~\cite{Malnou2019b}, quantum-limited amplification~\cite{Klimovich2022,Klimovich2023}). The sensors themselves are natively frequency-domain multiplexable, and their high quality factors ($> 10^5$) allow for thousands of resonators to be coupled to a common feedline enabling simple scalability to large arrays~\cite{Rantwijk2016}. 

Current state-of-the-art phonon-mediated MKID sensors have demonstrated resolution on energy deposited in the substrate ($\sigma_E$) of 26--30 eV~\cite{Cardani2021a,Cruciani2022}. For comparison, phonon-coupled transition-edge sensor (TES) detectors have achieved an order of magnitude better resolution~\cite{Ren2020}. Preliminary results have demonstrated sub-eV resolution with phonon-coupled TES detectors~\cite{Romani2023}. In order for KIPM detectors to be competitive in the low-mass DM landscape, their resolution must be improved. 

In this paper, we report a measurement of the energy resolution of a KIPM detector operated at the Northwestern EXerimental Underground Site (NEXUS), a low-background cryogenic facility at Fermilab. NEXUS is located 105 meters underground in the MINOS cavern which provides a rock overburden of 225 meters of water equivalent (m.w.e.)~\cite{Michael2008}, reducing the muon flux by a factor of 515~\cite{Mei2006} compared with the surface. The heart of the facility is a CryoConcept HEXA-DRY $^3$He/$^4$He dilution refrigerator~\cite{hexadry}, which reliably maintains a base temperature of 10 mK. The refrigerator is enshrouded by a near-hermetic radiation shield comprising five 4-inch-thick walls of Pb bricks and an internal copper-clad Pb slab (5.2" thick) below the mixing chamber. The experimental payloads are deployed on a plate thermally coupled to the mixing chamber, but below this internal lead slab. Three sides and the bottom face of the shield are mounted on a movable cart that can be mated to a vertical lead wall creating a shield closed and shield open configuration. The refrigerator also features an Amuneal A4K$^\mathrm{TM}$ magnetic shield integrated into the 4K thermal shield as well as an external Metglas$^\mathrm{TM}$ magnetic shielding blanket. 

The design of the KIPM detector under test in this work is discussed in Section~\ref{sec:kipmds}, whereas the experimental facility and setup are reviewed in Section~\ref{sec:experiment}. In Section~\ref{sec:resolution}, we present the model we use to evaluate the energy resolution of KIPM detectors. In Section~\ref{sec:calib}, we present the methodology and results for the optical photon calibration performed on our device. Measurements of quasiparticle lifetime based on the observed pulse shape, as a function of temperature, are presented in Section~\ref{sec:qplifetime}. Finally, in Section~\ref{sec:disc}, we summarize our results, discuss their importance, and outline potential near-term improvements and future work.

\section{\label{sec:kipmds}The Kinetic Inductance Phonon-Mediated Detector Design}

\begin{figure}
    \centering
    \includegraphics[width=\linewidth]{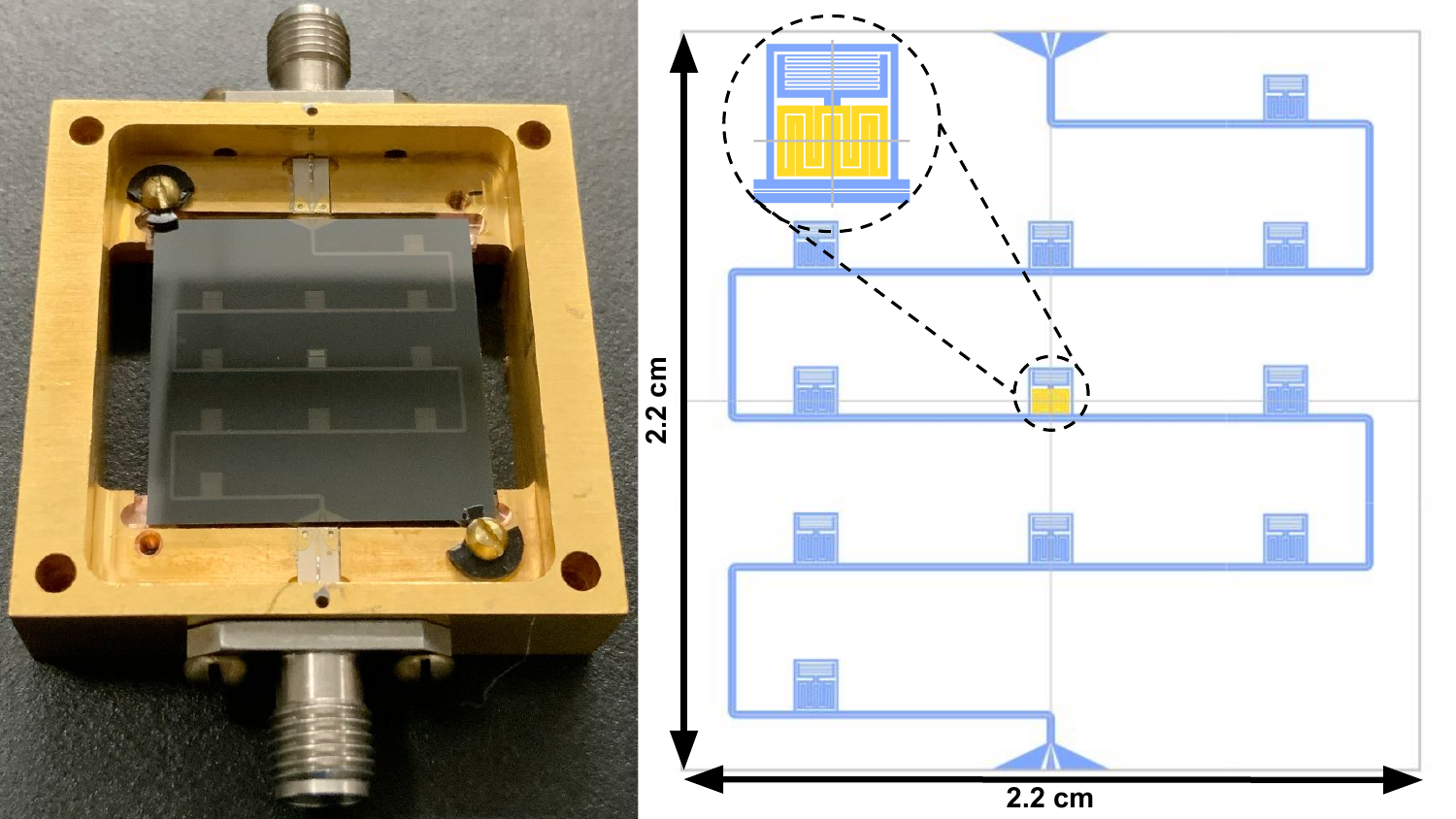}
    \caption{(Left) The KIPM detector in its housing as deployed at NEXUS for this measurement featuring a central, phonon-absorbing aluminum resonator. (Right) The design mask of the device showing the Nb (blue) and Al (yellow) features, with an inset showing the design of the primary phonon-absorbing resonator.}
    \label{fig:device}
\end{figure}

The KIPM detector under test in this work 
was fabricated at the Jet Propulsion Laboratory and features 11 resonators with resonant frequencies $f_r=1/2\pi\sqrt{LC}$ in the 3.9--4.4 GHz band deposited on a 2.2 $\times$ 2.2 cm$^2$, 1 mm thick ($\sim$1 gram) high-resistivity, double-sided polished silicon substrate. The physical device and its design mask are shown in Fig.~\ref{fig:device}. The primary phonon-absorbing resonator, shown in the inset of the right panel of Fig.~\ref{fig:device}, is formed from a 30-nm-thick Al meandered inductor and an interdigitated capacitor formed from a bilayer of 30 nm each of Al and Nb. The capacitor has a 950 $\mu$m finger length and 20 $\mu$m finger width and spacing, while the inductor's meander width is 80 $\mu$m with 10 and 20 $\mu$m spacing, creating a footprint of 0.81 mm$^2$. The inductor, which has a volume of $2.4\times10^4~\mu$m$^3$~\bibnote{The volume of the inductor is estimated in the following way.  The footprint area is known precisely by the design mask used in fabrication, while the film thickness is determined by the film deposition rate and exposure time. It was also measured via a Tencor$^\mathrm{TM}$ profilometer at the time of fabrication. These measurements confirm the thickness is as expected, with an uncertainty of 5\%. This uncertainty is subdominant to other sources of uncertainty, discussed below, but is included in the results presented.}, serves as the phonon absorber, where incident energy breaks Cooper pairs and modulates the surface impedance of the resonator. The remaining ten resonators are fabricated entirely of Nb, originally intended for a separate measurement~\cite{Wen2021}. The $T_c$ of Nb is a factor of ten larger than that of Al~\cite{Matthias1963}, though not large enough to raise its Cooper pair binding energy above the average phonon energy in silicon ($\sim$1 meV)~\cite{Martinez2019}. This introduces an avenue for phonon loss, but phonon recycling~\cite{Fyhrie2016,
Ulbricht2015} from quasiparticle recombination will inject phonons of energy $2\Delta_\mathrm{Nb} \approx 2.8$ meV~\cite{Matthias1963, Tinkham2004} back into the substrate to be absorbed by the Al ($2\Delta_\mathrm{Al} \approx 0.36$ meV~\cite{Matthias1963, Tinkham2004}) with a characteristic timescale given by the quasiparticle lifetime of the higher-gap material. This recycling was expected to maintain the efficiency for phonon collection in Al. All resonators are inductively coupled to a 20-$\mu$m-wide, 300-nm-thick, serpentine coplanar waveguide (CPW) Nb feedline with two 100$\mu$m-wide ground planes separated by 10 $\mu$m, with a 50 $\Omega$ characteristic impedance. The resonant frequency of each resonator is tuned by adjusting the length of the inductor.

The silicon chip is housed in a gold-plated copper enclosure  with a penetration for an SMA-connectorized optical fiber (see Fig.~\ref{fig:fiber-position}). Thermalization of the device is accomplished by physical contact with its enclosure on four corners. The device is held in thermal contact with the enclosure by two nylon clip washers on opposite corners fastened with two brass screws. For readout, Al wirebonds connect the CPW feedline to Duroid$^\mathrm{TM}$ launcher boards with a 50-ohm copper CPW feedline and a stainless-steel SMA connector soldered to it.

\section{\label{sec:experiment}Experimental Setup}

The KIPM detector payload was deployed on the RF payload plate of the NEXUS dilution refrigerator, which is thermally coupled to the mixing chamber plate. Infrared-absorbing inline coaxial filters~\cite{Spahn2021} were connected to the RF input and output of the KIPM device. The RF input line is thermalized at each refrigerator stage with 0 dB attenuators except at the 4 K and 10 mK stages, where 20 dB attenuators are used. RF lines with stainless steel conductors are used from 300K to 4K and NbTi from 4K to the mixing chamber. A 10 GHz low-pass filter 
(RF-Lambda RLPFTBG10) is deployed at the 1K stage. This intentional attenuation and the passive attenuation from warm and cold RF cabling sums to a total of 56.5 dB of attenuation between the RF source and the device input at the Al resonant frequency of 4.242 GHz. This attenuation is used to convert between power at the digital-to-analog converter (DAC) $P_\mathrm{DAC}$ and power on the device feedline $P_g$. The device output is routed through a Low Noise Factory LNF-ISC4\_8A single-junction isolator at the 10 mK stage and amplified with a Low Noise Factory LNF-LNC0.3\_14A high electron mobility field-effect transistor (HEMT) amplifier with noise temperature\bibnote{This value does not affect our analysis in any way. We have estimated the noise temperature of our system at the HEMT output, however there are relatively large uncertainties on the in situ gain values. Using the maximum specified attenuation of the components in our post-HEMT signal chain, we find that the PSD of our noise is compatible with 3.6 K Johnson noise to within a factor of 2.
} $T_n=3.6$ K~\cite{HEMT} mounted at the 4K stage. The output line is similarly thermalized at each refrigerator stage with 0 dB attenuators. Finally, the device output signal is amplified at room temperature using a Mini-Circuits ZX60-83LN-S$+$ amplifier. See Fig.~\ref{fig:wiring-diagram} for a wiring diagram of the system.

The RF probe tones are generated and later digitized by an Ettus Research x300 Universal Software-defined Radio Platform (USRP),  controlled using the \texttt{GPU\_SDR} package~\cite{Minutolo2019}. The USRP provides a 1 V$_\mathrm{pp}$ pulse-per-second clock output line which serves as a trigger for an arbitrary waveform generator, which then imprints a desired pulse pattern on the output of a ThorLabs EP470S04 470-nm fiber-coupled LED. The LED photons are routed via optical fiber into the cryostat and to the device, after passing through a KG3 infrared filter thermalized at the 100 mK stage. The fiber is thermalized at each temperature stage below 50 K and penetrates the KIPM device enclosure such that the optical photons impinge upon the bare silicon side of the device, ensuring no photons are directly incident on the inductor and all Cooper-pair-breaking energy comes in the form of phonons produced in the substrate (see Appendix~\ref{supp8-directhits} for a discussion of this). 

\begin{figure*}
    \centering
    \includegraphics[width=\linewidth]{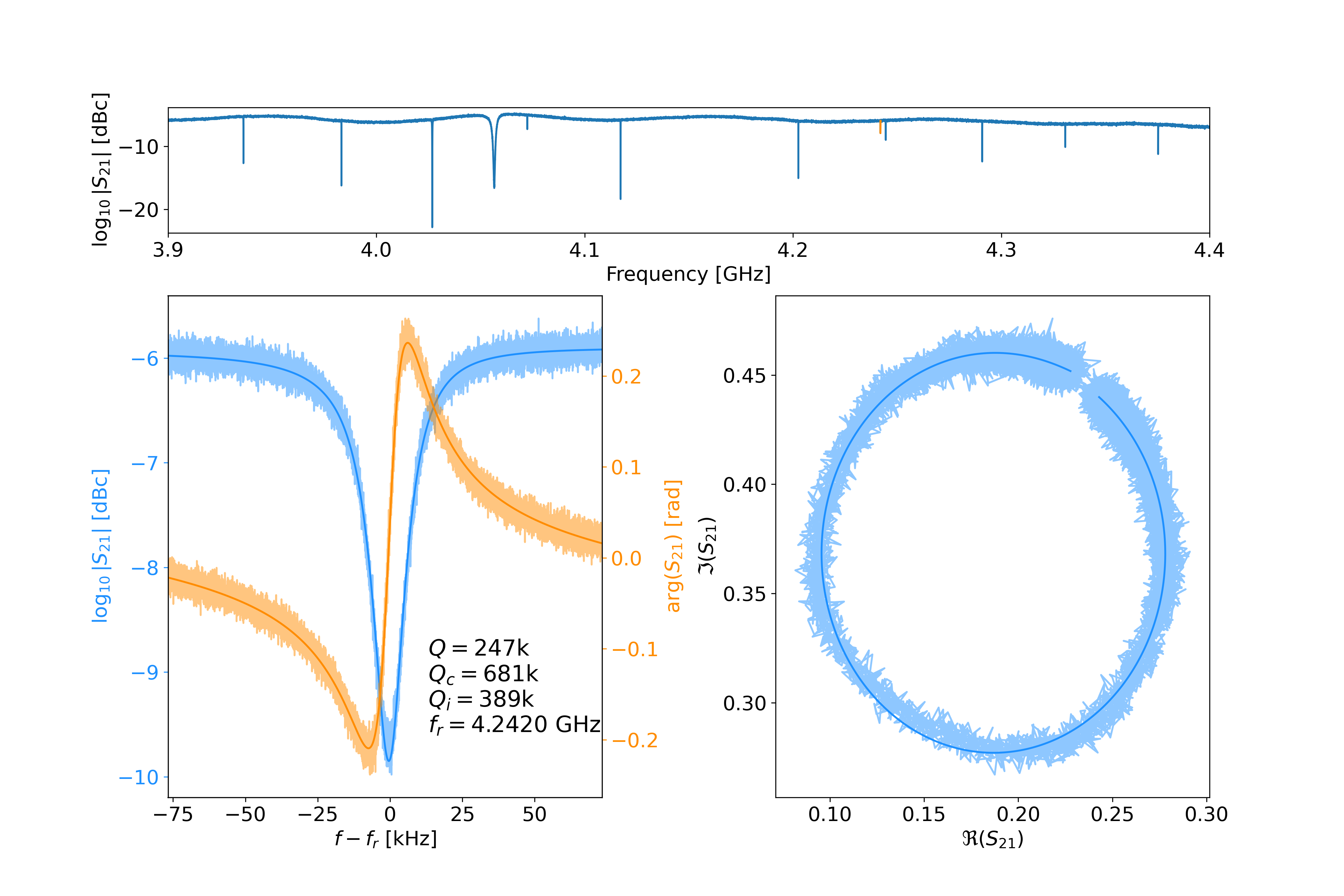}
    \caption{(Top) The logarithmic magnitude of the forward scattering parameter $S_{21}$ (transmission) of the KIPM device in the resonators' frequency band. The Al resonator is shown in orange. (Bottom left) In blue is the logarithmic magnitude of the Al resonance feature, while in orange is the phase response vs frequency relative to the resonant frequency. The resonance's parameters are listed. (Bottom right) The phonon-sensitive resonance in complex $S_{21}$ space. These data are acquired with a power on the feedline of $P_g=-106.5$ dBm at a device temperature of 10 mK. Fits to the resonance lineshape Eq.~(\ref{eq:cplx-trans-full}) are shown as solid lines.} 
    \label{fig:spectrum}
\end{figure*}

The resonance features of the device are modeled as notch filters, with the complex transmission as a function of frequency given by (\textit{e.g.}, Refs.~\cite{Mazin2002,Zmuidzinas2012}) 
\begin{align} \label{eq:s21}
S_{21}(f,T) = 1 - \frac{Q}{Q_c}\frac{1}{1+2jQ \frac{f-f_r(T)}{f_r(T)}}~,
\end{align}
where $f_r(T)$ is the resonant frequency as a function of temperature, $Q_c$ is the coupling quality factor, and $Q$ is the total quality factor, given by $Q^{-1} = Q_c^{-1} + Q_i^{-1}$ where $Q_i$ is the internal quality factor. For the Al resonator under test, we measure a total $Q$ of approximately 250,000. 
This expression represents the idealized $S_{21}$, where off-resonance transmission has been set to unity. The complex transmission as a function of stimulus frequency $S_{21}(f)$ can be measured and fit to the above equation to extract the characteristics of the resonator. The bottom-left panel of Fig~\ref{fig:spectrum} shows the shape of the Al resonance feature in both logarithmic magnitude and phase of $S_{21}$, and provides values for the $Q$ factors, while the bottom-right panel shows the resonance in the complex $S_{21}$ plane. In practice, Eq.~(\ref{eq:cplx-trans-full}) is used for fitting to account for nonideal line shapes and off-resonant transmission (see Appendix~\ref{supp-mb}). The transmission spectrum of this device in the resonators' band is shown in the top panel of Fig.~\ref{fig:spectrum}.

The shift in on-resonance transmission $\delta S_{21}$ in response to a time-dependent change in quasiparticle density $\delta n_\mathrm{qp}(t)$ is given by~\cite{Siegel2016}
\begin{align} \label{eq:s21_qpdens}
    \delta S_{21}(t) =\alpha \frac{Q_r^2}{Q_c}(\kappa_1 + i\kappa_2)\delta n_\mathrm{qp}(t)~,
\end{align}
where $t$ is time, $\alpha$ is the kinetic inductance fraction, and $\kappa_{1(2)}$ carries units of volume and is the real(imaginary) component of the fractional change in complex conductivity per unit change in quasiparticle density. The values of both $\kappa_1$ and $\kappa_2$ are a function of superconductor temperature $T$, readout frequency, and superconducting gap energy at zero temperature ($\Delta$), expressions for which are given in~\cite{Gao2008c}. A waveform is acquired by reading out the on-resonance transmission as a function of time, $S_{21}(t)$. A Cooper-pair breaking event leads to an increased $n_\mathrm{qp}$ which, in turn, leads to a fast reduction (limited by the resonator ringdown time $\tau_r = Q/2\pi f_r$) in both resonant frequency and quality factor, usually expressed as fractional shifts $\delta f/f$ and $\delta(1/Q)$. As the broken Cooper pairs recombine, the resonance returns to its quiescent point. The entire process appears as a pulse in the $S_{21}$ timestream. We adopt values for $\alpha=3.8\%$ and $\Delta=0.184$ meV from measurements of an identical device from the same fabrication wafer~\cite{Wen2021}. See Appendix~\ref{supp-mb} for a discussion on measuring the zero-temperature characteristics of this device.

\section{\label{sec:resolution}KIPM detector energy resolution} 

The energy resolution of the KIPM detector can be determined via exposure of the device to a pulsed source of optical photons. By varying the optical power delivered with a single pulse, one can exploit the broadening of the pulse amplitude distribution due to photon shot noise to infer the total energy deposited in the substrate~\cite{Isaila2012}. If an interaction in the substrate deposits an energy $E$ into the phonon channel, the resonator will absorb an energy $E_\mathrm{abs} = \eta_\mathrm{ph} E $ where $\eta_\mathrm{ph} < 1$ is the efficiency for the resonator to collect energy (in the form of phonons) from the substrate. This efficiency contains contributions from phonon downconversion to energies below the aluminum gap, phonon loss at interfaces (\textit{e.g.}, mounting hardware) and to other superconducting films, and the efficiency for phonon transmission between the substrate and the superconductor. This factor can be used to relate the device energy resolution $\sigma_E$ (resolution on energy deposited in the substrate) 
and resolution on energy absorbed by the superconductor $\sigma_E^\mathrm{abs}$(the intrinsic energy resolution of the resonator):
\begin{align} \label{eq:res_deposited}
    \sigma_E
    = \frac{\sqrt{N_r}}{\eta_\mathrm{ph}} \sigma_E^\mathrm{abs}~.
\end{align}
Here, the factor $\sqrt{N_r}$ arises from using $N_r$ resonators to measure an energy signal and adding their resolutions in quadrature. For the device under test we set $N_r=1$ since we only use the Al resonator to measure energy depositions, and fold any phonon loss due to the Nb resonators into $\eta_\mathrm{ph}$.

The baseline (zero-energy) resolution of a resonator is set by the level of on-resonance noise, contributed by sources including the thermal noise of the first-stage amplifier, noise from electrically active two-level systems (TLS) present in oxide layers of the device~\cite{Noroozian2009,Sueno2021,Wen2021}, and from Poissonian fluctuations in the density of quasiparticles in the superconductor due to thermal excitation (generation-recombination noise~\cite{Visser2011,Visser2012}). The on-resonance noise can be characterized by its power spectral density (PSD). The on-resonance noise PSDs for the two readout quadratures at base temperature, for a range of readout powers, and a brief discussion of noise sources present can be found in Appendix~\ref{supp-noise}.

The baseline resolution $\sigma_0$ can be determined with an optimal filter~\cite{Zadeh1952, Golwala2000, Pyle2012} by comparing a 
signal template in the frequency domain with complex component $\tilde{s}_n$ in the $n^\mathrm{th}$ frequency bin to the noise PSD, $J(f_n)$, in the same frequency bin. In this case, the baseline resolution is given by the square-root of 
\begin{align} \label{eq:ofres}
    \sigma_0^2 = \left[ D \sum_{n=-\frac{N}{2}}^{\frac{N}{2}-1} \frac{|\tilde{s}_n|^2}{J(f_n)} \right]^{-1}~,
\end{align}
where $D=100$~ms is the total duration of the signal timestream $s_n$, and $N=2001$ is the total number of frequency bins\bibnote{The optimal filter estimator for resolution is constructed from a $\chi^2$ fit, in frequency space, of the observed pulse to a signal template. The expression for $\chi^2$ is then minimized with respect to the pulse amplitude estimator. The expression above is simply the expected variance of the distribution of this amplitude estimator. Taken another way, this is the discrete case of the integration of a normalized signal-to-noise ratio over the full pulse shape.}. The resolution calculated in this fashion, for a unit-normalized signal template, inherits units from $J$ (\textit{e.g.}, if $J$ has units of $(\delta f/f)^2/\mathrm{Hz}$, then $\sigma_0$ carries units of $\delta f/f$). In this analysis, we consider only frequency components extending up to $10^4$ Hz.

\section{\label{sec:calib}Optical photon calibration}

\subsection{Data collection}
The KIPM detector was exposed to a pulsed source of 470 nm (2.6 eV) photons with variable optical power, up to 7 mW~\bibnote{At the input of the room-temperature optical feed-through of the dilution refrigerator.}. Note that at low temperatures, the Si bandgap is approximately 1.17 eV~\cite{Bludau1974,Ramanathan2020ab}, as such the photons used in this calibration have sufficient energy to create electron/hole pairs. For each optical power, the LED bias $V_\mathrm{LED}$ is modulated by a pulse generator enabling control of the photon pulse width and arrival time relative to the start of data acquisition. The data acquisition for the energy calibration proceeds as follows. With the LED disabled, 30-second timestreams of noise are acquired for three probe tones: one at $f_r$ and the others taken at a $\pm10$ MHz offset from $f_r$. These off-resonant tones, taken at the same RF stimulus power, are used for removal of correlated electronics noise present in the on-resonance timestream~\cite{Wen2021}. This is accomplished by subtracting from the on-resonance timestream, each of the off-resonance timestreams with a relative amplitude given by the ratio of the covariance of the on- and the particular off-resonance timestreams to the variance of the on-resonance timestream. (see Ref.~\cite{Aralis2024} for a detailed discussion of this removal). The pulse generator output is then enabled, with an output pulse voltage of $V_\mathrm{LED} = 5.0$ V, corresponding to the largest optical power used. For the calibrations presented in this work, we use a pulse width of 2 $\mu$s delivered at 5 Hz. A 5-ms delay is applied to the pulse generator such that the first optical pulse arrives 5 ms after the start of the timestream acquisition. We drive the LED driver board with a square pulse train from the AWG. 
In this calibration, the exact shape of the photon pulse is unimportant as the pulse width is short compared with all rise times one would get with a delta-function impact in the substrate. In that case, the rise time is simply the resonator ring-up time ($\tau_r$
$\sim10$ $\mu$s for the resonator of interest). We exclude frequencies above $10^4$ Hz from analysis, which thus ignores any features faster than $\sim$10$\times$ the resonator’s ring-up time. 

\begin{figure*}
    \centering
    \includegraphics[width=\linewidth]{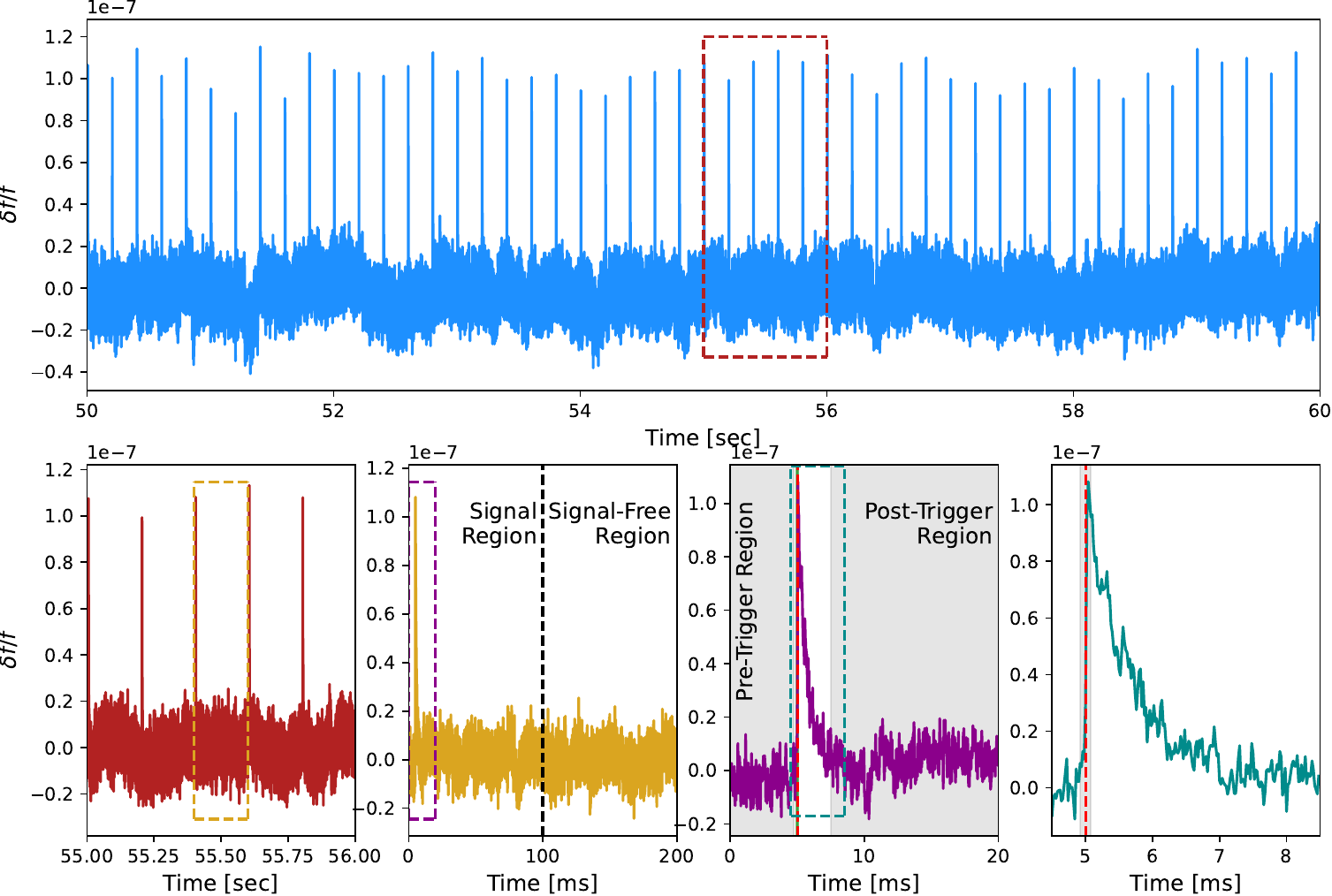}
    \caption{Example timestream of LED pulses: (top) 10 second window; (bottom left) 1 second window; (bottom middle left) a single pulse window; (bottom middle right) a zoom-in on the signal window near the pulse showing the pre-trigger and post-pulse regions; (bottom right) a closer zoom-in on the signal window to demonstrate the window in which the pulse maximum must arrive. These time regions are used to evaluate cuts as well as define the signal and noise region of each pulse window.}
    \label{fig:pulse_region_definitions}
\end{figure*}

With the LED pulsing, 200-second timestreams are acquired at the same three tones. The on-resonance timestream therefore contains 1,000 windows in which LED pulses may be found. See Fig.~\ref{fig:pulse_region_definitions} for examples of these pulse windows. The pulse voltage (LED bias) is lowered to the next setpoint (0.5 V steps) and data are acquired in the same fashion. After every three LED timestreams are acquired, the pulse generator is disabled and a noise timestream is acquired, as described above. The lowest $V_\mathrm{LED}$ is chosen such that the LED does not receive a sufficient forward voltage to emit light. This data set serves to test for electronic pickup from the LED pulsing setup. The turnoff voltage in this setup is $V_\mathrm{LED} = 2.0$ V.

This calibration was performed twice, once in February 2023 and once in July 2023, with the same device under slightly different environmental conditions. The February calibration was performed with an RF power at the output of the USRP of $P_\mathrm{DAC}=-16$ dBm at a device temperature of 15 mK, whereas the calibration in July was acquired with $P_\mathrm{DAC}=-15$ dBm and a device temperature of 10 mK.

\subsection{Energy resolution analysis}
Once the data have been acquired, they are processed offline. The first 10\% of pulse windows (a total of 20 seconds) at the start of an acquisition are discarded to avoid any transient behavior of the USRP as the continuous-wave RF stimulus begins and the RF circuitry reestablishes equilibrium. Then, each of the 900 remaining 200-ms windows is evaluated for its noise properties and pulse location in time. In the pre-trigger region, defined to be the first 4.75 ms of the pulse window (see the bottom, middle-right panel of Fig.~\ref{fig:pulse_region_definitions}), the mean and RMS of the noise are used to reject windows in which there is increased noise (RMS) or a baseline jump (mean). For each LED setting, the distribution of these parameters for all 900 windows is investigated and its 5$^\mathrm{th}$-- and 90$^\mathrm{th}$--percentile values are used to define the cut limits. Each distribution is checked manually for Gaussianity, as bimodal distributions were observed in baseline mean (indicating a discrete jump in on-resonance baseline transmission) and baseline RMS (indicating a period of increased noise). The percentile-based cut criteria were adjusted to remove events within the high-noise (RMS) periods or following a baseline shift (mean).

Events in which a point exists in the pre-trigger region with an amplitude larger than $3.5\times$ the pre-trigger RMS are removed from analysis, in order to veto events in which the baseline is noisy or otherwise undesired (\textit{e.g.}, from a pulse occurring near the end of a previous window). Similarly, events in which a point exists in the post-pulse region, defined to be the remainder of the window beyond 7.5 ms\bibnote{Note that changing the post-pulse definition to start 5 ms (or longer) after the pulse rising edge (vs the 2.5 ms used in the analysis) does not significantly change the number of events passing this cut ($<1$\% difference in number of events passing this cut).}, with an amplitude larger than $4\times$ the post-pulse RMS are removed from analysis, again to remove events with baseline aberrations, including pulses that are not associated with an LED flash\bibnote{The larger RMS factor is used due to the long length of the post-pulse region relative to the pre-trigger region.}. A further quality cut is applied to ensure that the pulse occurred at the time expected from the LED pulse. Any window in which the maximum occurs more than 4 samples away from the trigger time 
is removed from analysis if there exists a point anywhere in the window with an amplitude greater than $5\times$ the pre-trigger RMS\bibnote{This cut retains nearly 100\% of all noise-only traces where the baseline is well-behaved and no spurious pulses are present.}. The conditional on this cut is intended to retain windows in which there are no pulses (\textit{i.e.}, if the LED voltage is too low to produce a signal over noise). A graphical depiction of these various time regions is shown in Fig.~\ref{fig:pulse_region_definitions} These quality cuts are evaluated on the $\kappa_2$ readout quadrature, and retain $\sim$60\% of the pulse windows, with the bulk being removed by the percentile cuts discussed first. The RMS-based cuts are intentionally loose as they are intended solely to remove periods of high noise and pulses not associated with LED flashes (which are rare due to the low-background environment afforded by NEXUS).


\begin{figure*}
    \centering
    \includegraphics[width=\linewidth]{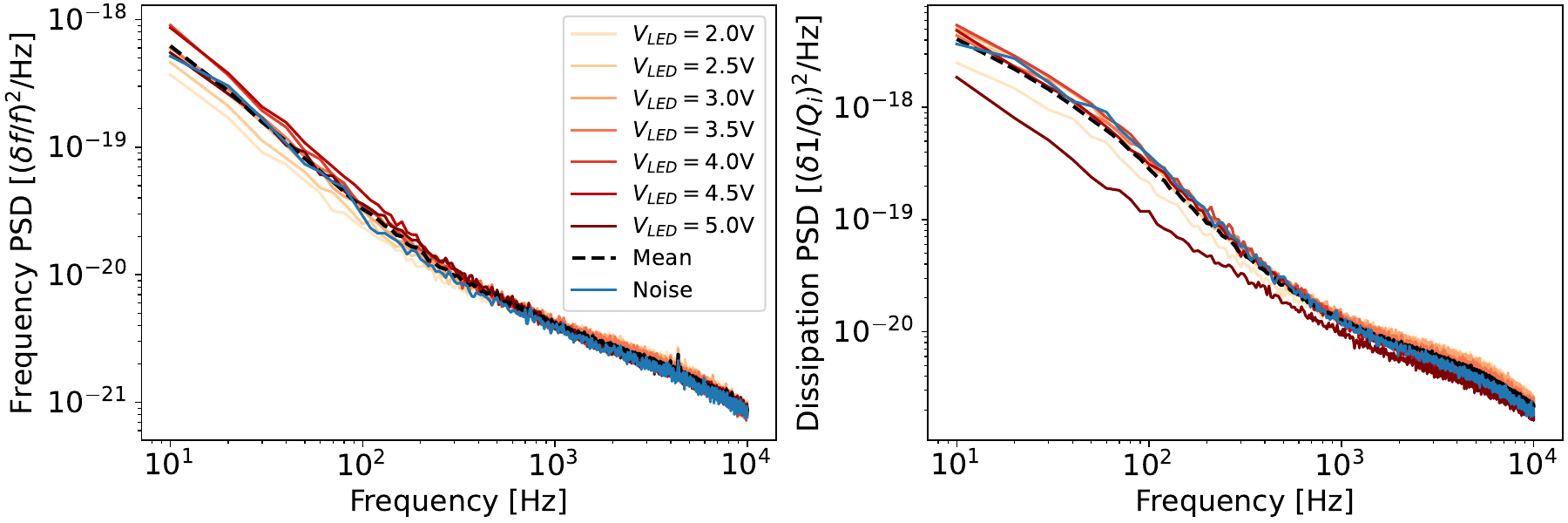}
    \caption{The average PSDs in the frequency (left) and dissipation (right) directions for the pulse-free region for each LED voltage. The mean of these PSDs is shown in the black dashed line. The PSD from a dedicated noise acquisition before the LED data acquisition is shown in blue. Only the frequency-direction readout is used for the energy resolution analysis. For a decomposition of these noise spectra, a discussion of noise sources and the voltage dependence of the PSDs, see Appendix~\ref{supp-noise}.}
    \label{fig:average_noise}
\end{figure*}

Pulse timestreams are cleaned of correlated electronic noise using measurements of the correlation in noise timestreams. For every LED voltage, each cleaned pulse window surviving the quality cuts is then split in two 100-ms halves to define a signal region (first half) and signal-free region (second half). This partitioning is shown for an example pulse window in the bottom, middle-right panel of Fig.~\ref{fig:pulse_region_definitions}. The PSD of the cleaned signal-free region, $J(f)$, is calculated for each window, then averaged over the windows for each LED setting. These average pulse-free PSDs, shown in Fig.~\ref{fig:average_noise}, are compared with the noise-only PSD to confirm no signal has leaked into the signal-free region and to each other to confirm compatibility of the data. The observed noise at this readout power is far from white and contains contributions from TLS noise and $1/f$ noise that arises at the highest readout powers (see Appendix~\ref{supp-noise}), which limits our sensor's energy resolution. Investigations into resonator design optimization for mitigating the presence of TLS sites and their impact on the overall noise are underway. A discussion on the voltage dependence of the noise PSD is presented in the same appendix.

\begin{figure*}
    \centering
    \includegraphics[width=\linewidth]{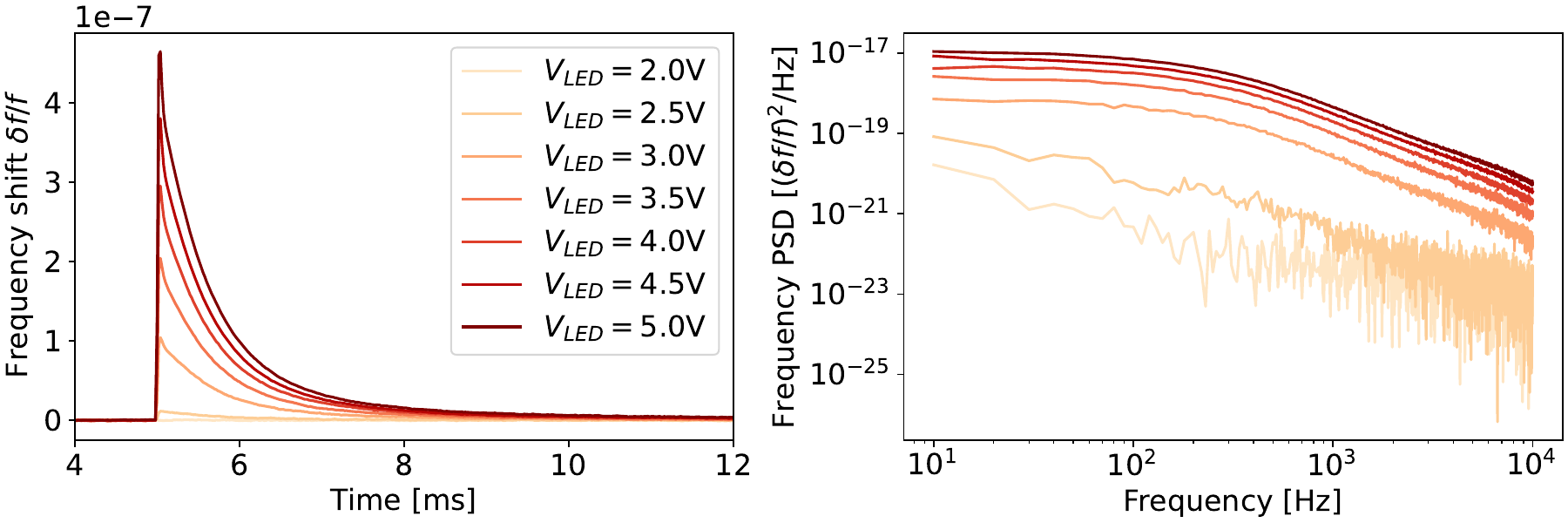}
    \caption{The average pulse shape in the time (left) and frequency (right) domains for each LED voltage. Each pulse window begins at 0 ms and extends to 100 ms.}
    \label{fig:average_pulse}
\end{figure*}

The signal region of every pulse window surviving cuts is baseline-subtracted and averaged to determine the pulse shape, shown in Fig.~\ref{fig:average_pulse}. The average pulse from the largest LED voltage defines the signal template. With the Fourier transform of this template $\tilde{s}(f)$ and the average noise PSD $J(f)$, the fractional baseline resolution $\sigma_0$ is found using Eq.~(\ref{eq:ofres}) for both the phase and magnitude readout quadratures. These are then scaled by the maximum amplitude of the unnormalized time-domain signal templates in $\delta f/f$ and $\delta(1/Q)$ to obtain the resolution in changes of the resonator's characteristics. The resolutions are converted into resolution on changes in quasiparticle density, as measured by the phase (frequency, $\kappa_2$) and magnitude (dissipation, $\kappa_1$) readout quadratures, by way of Eq.~(\ref{eq:s21_qpdens}). These results are summarized for both calibration runs in Table~\ref{tab:baselineres}. 

While the resolution calculated in this fashion carries some uncertainty due to the relationship between a particular realization of noise and its PSD (see Appendix B of~\cite{Golwala2000}), we report no baseline resolution uncertainty for two reasons: (1) this uncertainty amounts to $<1\%$ and thus the dominant uncertainty on resolution in quasiparticle density (from which the resolution on energy absorbed by the sensor $\sigma_E^\mathrm{abs}$ is calculated) arises from the calculation of $\kappa_{1,2}(T)$ due to differences between the physical device temperature and the effective temperature of the Cooper pairs, amounting to a relative uncertainty of roughly 9\% as discussed in Appendix~\ref{supp6-k2unc}; and (2) the values in Table~\ref{tab:baselineres} are not used to determine the device energy resolutions and serve only as a point of reference.

\begin{table}[b]
\caption{\label{tab:baselineres}%
The baseline resolution estimates, Eq.~(\ref{eq:ofres}), from dedicated noise acquisitions taken during each of the calibration runs. The average pulse with largest amplitude is used as the signal template. In the phase and magnitude readout quadratures, $\sigma_0$ is reported as a fraction of the amplitude of the average pulse. The resolutions are also reported in terms of fractional shifts in resonator characteristics ($\sigma_0^{df/f}$, $\sigma_0^{1/Q}$) and as fluctuations in quasiparticle density ($\sigma_0^{\kappa_2}$, $\sigma_0^{\kappa_1}$) in $\mu$m$^{-3}$. 
}
\begin{ruledtabular}
\begin{tabular}{rcc}
\textrm{Date}        & \textrm{Feb 15, 2023} & \textrm{Jul 13, 2023} \\
\colrule
$\sigma_0$ (phase)   & $7.8\times10^{-3}$    & $9.8\times10^{-3}$ \\
$\sigma_0$ (mag.)    & $3.2\times10^{-2}$    & $4.0\times10^{-2}$ \\
$\sigma_0^{df/f}$    & $3.6\times10^{-9}$    & $5.2\times10^{-9}$ \\
$\sigma_0^{1/Q}$     & $7.4\times10^{-9}$    & $1.1\times10^{-8}$ \\
$\sigma_0^{\kappa_2}$ [$\mu$m$^{-3}$] & 0.48 & 0.70 \\
$\sigma_0^{\kappa_1}$ [$\mu$m$^{-3}$] & 0.86 & 1.24 \\
\end{tabular}
\end{ruledtabular}
\end{table}

Resolution in quasiparticle density fluctuations, as measured in the $\kappa_1$ or $\kappa_2$ direction $\sigma^{\kappa_{1,2}}_0$, can be translated into a resolution on energy absorbed by the sensor: $\sigma_E^\mathrm{abs} = \Delta \sigma^{\kappa_i}_0 V$, where $\Delta$ is the superconducting gap energy and $V$ is the volume of the superconductor in which the Cooper pairs are broken. However, determination of the resolution on energy deposited in the substrate requires knowledge of $\eta_\mathrm{ph}$, provided by the LED calibration procedure described previously.

\begin{figure}
    \centering
    \includegraphics[width=\linewidth]{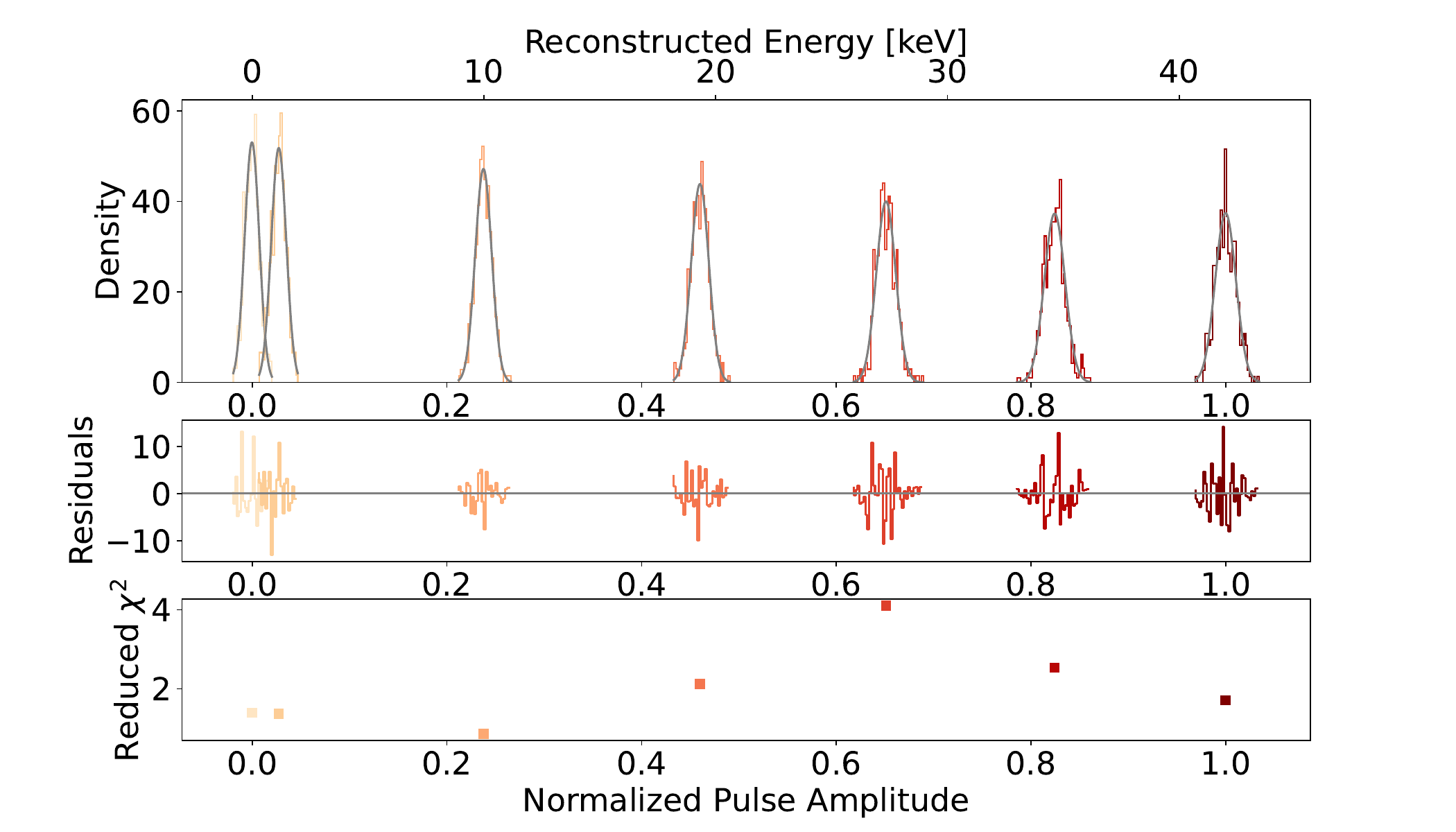}
    \caption{Results of the LED calibration: distributions of pulse amplitude estimators for each LED voltage, overlaid with Gaussian fits. The colors represent the LED bias voltage, as in Fig.~\ref{fig:average_pulse}. Fit residuals and reduced $\chi^2$ values are shown in the panels below. The amplitude values have been normalized to the amplitude of the average pulse in fractional frequency shift for the largest LED voltage. The upper horizontal axis is the reconstructed energy using the results of this calibration.}
    \label{fig:led-cal}
\end{figure}

For a single pulse window, the best estimator for the pulse amplitude $A$ is
\begin{align}
    \hat{A} = \frac{ \sum_{n=-\frac{N}{2}}^{\frac{N}{2}-1} 
    \frac{\tilde{s}_n^* v_n}{J(f_n)} }{ \sum_{n=-\frac{N}{2}}^{\frac{N}{2}-1} \frac{|\tilde{s}_n|^2}{J(f_n)} }~,
\end{align}
where $n$ is an index of frequency bins and $v_n$ is the Fourier transform of the signal-region timestream for each pulse window~\cite{Zadeh1952, Golwala2000}. In our case, we choose the frequency-domain signal template $\tilde{s}_n$ to be the average pulse from the largest LED bias voltage, the complex conjugate of which is $\tilde{s}_n^*$. The estimator $\hat{A}$ is calculated for the signal region of each pulse window, and their distribution for each LED voltage is shown in Fig.~\ref{fig:led-cal}. 

\begin{figure}
    \centering
    \includegraphics[width=\linewidth]{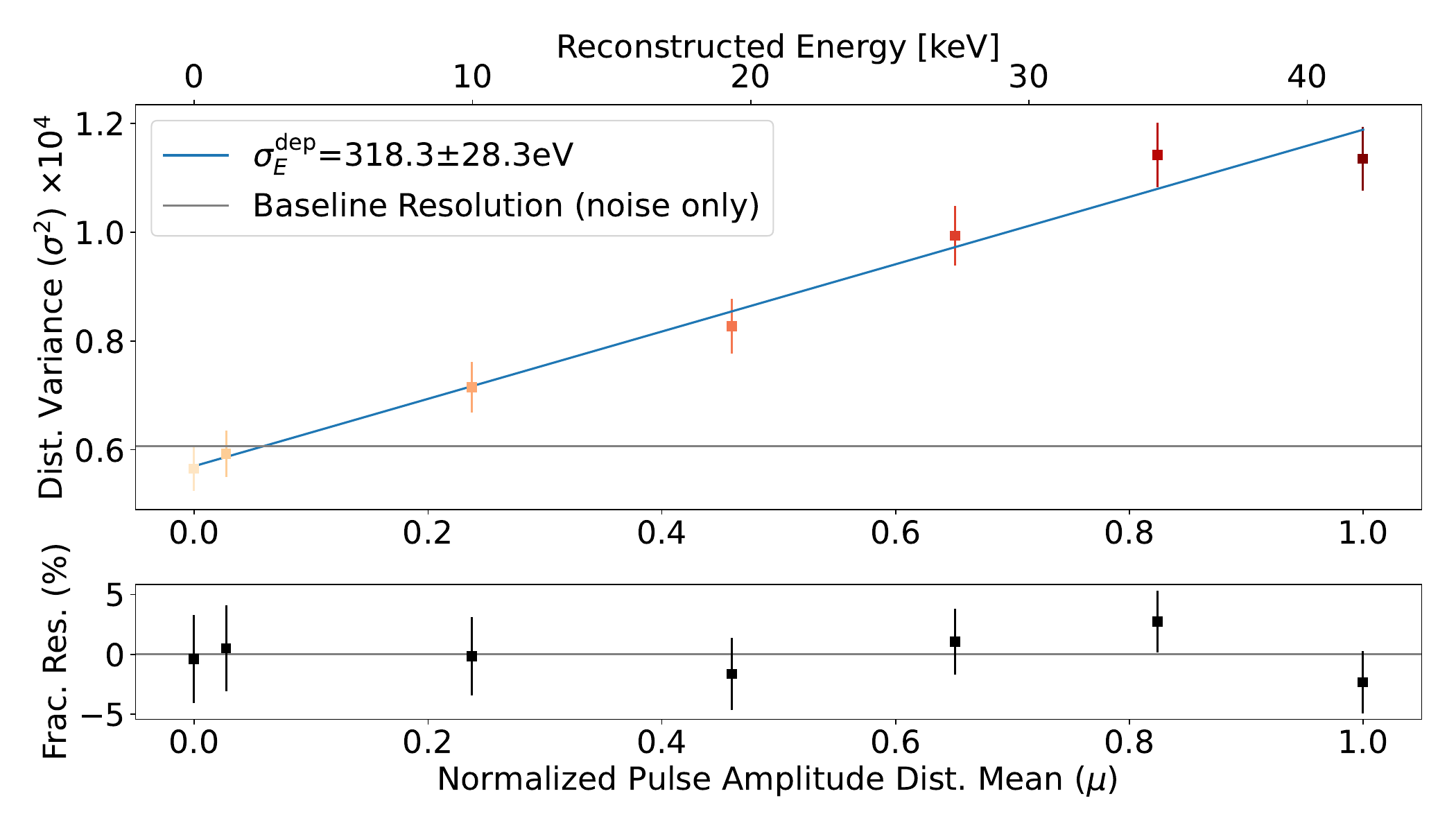}
    \caption{Results of the LED calibration: variance ($\sigma^2$) vs. mean ($\mu$) of the Gaussian fits to distribution of pulse amplitude estimators for each LED voltage (shown in Fig~\ref{fig:led-cal}), demonstrating their linear relation. The marker colors indicate LED voltage, as in Fig.~\ref{fig:average_pulse}. The fit to the two-component resolution model is shown in blue, while the baseline resolution determined from noise traces and a signal template is indicated by the grey line. Fractional residuals of the data to the fit are shown in the lower panel. The values have been normalized to the amplitude of the average pulse for the largest LED voltage. The error bars represent the total of statistical and systematic errors, the latter estimated from the variance in distribution widths from noise-only acquisitions. The upper horizontal axis is the reconstructed energy using the results of this calibration.}
    \label{fig:res-fit}
\end{figure}

The distribution of pulse amplitudes for a given voltage in our readout units, assuming a detector response linear with deposited energy, has a mean $\mu = R \bar{E}_\mathrm{abs} = R \eta_\mathrm{ph} \bar{E} = R \eta_\mathrm{ph} h \nu \bar{N}_\gamma$ where $R$ is the responsivity in readout units per unit energy absorbed by the sensor, $h\nu=2.6$ eV is the energy of a single photon, and $\bar{E}$ ($\bar{N}_\gamma$) is the average energy (number of photons) deposited in a single flash. The width $\sigma$ of each distribution is assumed to have contributions from two independent fluctuations: the intrinsic device noise $\sigma_0$ and the Poissonian fluctuation in the number of photons delivered per LED flash $\sigma_\mathrm{LED}$:
\begin{align} \label{eq:resmodel}
    \sigma^2 =  \sigma_0^2 + \sigma_\mathrm{LED}^2 ~,
\end{align}
where these various $\sigma$s all have the same units as the readout timestream. As dictated by Poisson statistics, the variance in number of photons delivered is $\sigma_{N_\gamma}^2 = \bar{N}_\gamma$, which implies (see Appendix~\ref{supp-res} for the derivation) the variance in pulse amplitudes is
\begin{align}
    \sigma_\mathrm{LED}^2 = \left(R \eta_\mathrm{ph} h \nu \right)^2 \bar{N}_\gamma~.
\end{align}
Using this expression, where $\bar{N}_\gamma$ has been substituted using the expression for the distribution mean $\mu$, Eq.~(\ref{eq:resmodel}) becomes
\begin{align}
    \sigma^2 = \sigma_0^2 + r \cdot \mu~,
\end{align}
where $r = R \eta_\mathrm{ph} h \nu$ is the responsivity per photon (\textit{i.e.}, carries the same units as readout). The $\mu$ and $\sigma$ are found for the pulse amplitude distribution at each LED bias setpoint (given in Fig.~\ref{fig:led-cal}) from unbinned Gaussian fits and are fit to the above equation to determine $r$ and $\sigma_0$, the results of which are shown in Table~\ref{tab:fitresults} and Fig.~\ref{fig:res-fit}. These carry units of fraction of largest average pulse amplitude in $\delta f/f$ (Fig.~\ref{fig:average_pulse}). The uncertainty on the width of each pulse amplitude distribution is the quadrature sum of the statistical errors returned by the fitting routine and our systematic uncertainty. The latter is estimated from the variance in pulse amplitude estimator distribution widths from the signal-free region of each pulse window. This uncertainty arises from drifts in the noise PSD, and is later combined with other sources of systematic uncertainty discussed throughout the text. Note that the baseline resolution inferred from this method and the baseline resolution measured from the signal template and noise timestreams alone are consistent to $1.3\sigma$.

The resolution on energy deposited in the substrate is given by the ratio of the baseline resolution in readout units to the responsivity per unit energy: $\sigma_E = \sigma_0 / ( r/h\nu )$. For the February calibration, we determine $\sigma_E = 318 \pm 29~\mathrm{eV}$. Following the same prescription for determining the resolution on quasiparticle fluctuations as before, we find $\sigma_\mathrm{qp}^{\kappa_2} = 0.47 \pm 0.04 ~\mu\mathrm{m}^{-3} $, from which we infer the resolution on energy absorbed by the superconductor to be $\sigma_E^\mathrm{abs} = \sigma_\mathrm{qp}^{\kappa_2} V \Delta = 2.1~\mathrm{eV} \pm 0.2$, using $V=2.4 \times 10^4 ~\mu$m$^3$ for the Al inductor. The phonon collection efficiency $\eta_\mathrm{ph}$ is determined from Eq.~(\ref{eq:res_deposited}): $ \eta_\mathrm{ph} = \sigma_E^\mathrm{abs} / \sigma_E = (0.66 \pm 0.1) \% $ for $N_r = 1$ resonators read out. These values, along with the device responsivity $R$ in fractional frequency shift per unit energy deposited in the substrate, are presented for the two calibration runs in Table~\ref{tab:fitresults}. The deposited energy scale can be reconstructed using $R$ and $\eta_\mathrm{ph}$, and is given by the upper horizontal axis in Figs.~\ref{fig:led-cal} and~\ref{fig:res-fit}. With this reconstructed energy scale, we find the average energy deposited in the substrate with each pulse at the largest $V_\mathrm{LED}$ to be 42 keV, corresponding to a mean of $1.6\times10^4$ photons absorbed by the substrate per pulse.

\begin{table}[b]
\caption{\label{tab:fitresults}%
Results of the energy resolution measurement for both calibration runs. The fit parameters $r$ and $\sigma_0$ are given as a fraction of the largest average pulse amplitude in $\delta f/f$. 
The responsivity per unit energy absorbed by the sensor $R = (r/h\nu)/\eta_\mathrm{ph}$ is given in eV$^{-1}$, and represents the fractional change in resonant frequency ($\delta f/f$) per eV of phonon energy absorbed by the sensor. 
The responsivity per unit energy deposited in the substrate is the product of $R$ and the phonon collection efficiency. The sensor and device energy resolutions and phonon collection efficiency from each calibration are also presented. The rightmost column indicates the compatibility between the two calibration results as a fraction of the combined uncertainty.
}
\begin{ruledtabular}
\begin{tabular}{rccc}
\textrm{Date}                & \textrm{Feb 15, 2023}            & \textrm{Jul 13, 2023}          & \\ 
\colrule
$r$                          & $(6.2  \pm 0.5)  \times 10^{-5}$ & $(7.3 \pm 0.6) \times 10^{-5}$ & 1.4$\sigma$\\
$\sigma_0$                   & $(7.5  \pm 0.2)  \times 10^{-3}$ & $(8.8 \pm 0.1) \times 10^{-3}$ & 2.4$\sigma$ \\
$R$ [($\delta f/f$)/eV]              & $(1.7  \pm 0.2)  \times 10^{-9}$ & $(1.7 \pm 0.3) \times 10^{-9}$ & --\\
$\sigma_E^\mathrm{abs}$ [eV] & $2.1   \pm 0.2  $                & $2.8 \pm 0.3 $                 & 1.9$\sigma$\\
$\sigma_E$ [eV]              & $318   \pm 29$                   & $315 \pm 29$                   & 0.08$\sigma$\\
$\eta_\mathrm{ph} [\%]$      & $0.66  \pm 0.1$                  & $0.89 \pm 0.11$                & 1.7$\sigma$\\
\end{tabular}
\end{ruledtabular}
\end{table}

\section{\label{sec:qplifetime}Quasiparticle lifetime measurement}

The observed pulses have ms-scale lifetimes, as shown in Fig.~\ref{fig:average_pulse}. The importance of pulse lifetime can be inferred from the construction of the optimal filter estimate for energy resolution, Eq.~(\ref{eq:ofres}): as the pulse lifetime increases, the signal template roll-off moves to lower frequency, shifting the signal weights in the sum for the resolution Eq.~(\ref{eq:ofres}) from higher frequencies to lower frequencies. This effect holds so long as the noise dependence with frequency is shallower than $f^{-1}$, though gains in resolution are highest when the noise is white. If one considers the case for white amplifier noise and a signal template given by a single falling exponential with time constant $\tau$, one determines the energy resolution scales as $\tau^{-1/2}$, as shown in~\cite{Aralis2024}. If instead, the noise is TLS-like ($f^{-1/2}$) rather than white, the energy resolution scales as $\tau^{-1/4}$. In both cases, longer pulse lifetimes lead to improved resolution, though TLS noise limits this gain. It is thus instructive for device design to understand the signal time constants at play, as well as the sources of noise, to optimize energy resolution. 

One contribution to the pulse lifetime is the quasiparticle lifetime $\tau_\mathrm{qp}$, which depends on the device temperature. The quasiparticle lifetime can be determined by the roll-off in the noise PSD in the case the noise is dominated by generation-recombination noise~\cite{Visser2012}. This is not the case for the system under test in this work\bibnote{This is due to noise from the first-stage amplifier at a power larger than that of the generation-recombination (GR) noise. See Appendix~\ref{supp-noise} for a brief discussion of GR noise.}, but $\tau_\mathrm{qp}$ can be estimated from pulse data by fitting pulses taken at varying temperature to a pulse shape model. However, this approach tends to overestimate quasiparticle lifetimes at large ($\gtrsim 250$ mK) temperatures~\cite{Visser2011}. In addition, saturation of measured quasiparticle lifetimes compared with theory are seen for $T \lesssim 170$ mK, in both pulse and noise PSD measurements.

\subsection{Data collection \& processing}

The temperature of the mixing chamber (MC) is maintained with a PID-controlled heater. The MC temperature is swept from the base temperature 25 mK to a maximum of 325 mK, in steps of 25 mK. At each temperature, noise-only and LED pulse timestreams are acquired, each with simultaneous acquisition of calibration tones at $f_r \pm$10 MHz for cleaning of correlated noise. The RF stimulus power used for this study was $P_\mathrm{DAC}=-30$ dBm. A lower readout power was used than in the energy resolution measurement, where it was desirable to minimize the baseline resolution contribution in order to resolve the photon shot noise. Once the temperature PID stabilized, the LED bias is set to $V_\mathrm{LED}=4.0$ V, pulse output is enabled, and timestreams are acquired. These data were acquired directly following the February energy calibration data acquisition. Once these data were collected, the same quality cuts, cleaning process, and average pulse identification were performed as in the energy calibration. The average pulses were then fit to a pulse shape model, defined below, to extract physical parameters of interest, including $\tau_\mathrm{qp}$.

Due to slow thermalization of the RF payloads to the MC, where the temperature control thermometer and heater are situated, the device temperature is not the same as the MC temperature $T_\mathrm{MC}$. To account for this, a map from MC temperature to device temperature was developed using data from an auxiliary thermometer installed on the RF payload plate during later runs of the NEXUS dilution refrigerator. The discrepancy in temperature is worse at higher MC temperatures, leading to an increased uncertainty in device temperature. The uncertainties on temperature presented in this section represent the standard deviation of the temperature as read by the auxiliary thermometer for a given MC temperature setpoint. 

\subsection{Empirical pulse shape model}\label{sec:empiricalpulse}
The average pulse shapes shown in Fig.~\ref{fig:average_pulse} show structure not described by a single exponential fall time, but rather a shape with two fall time constants: one prompt ($\tau_p$) and one delayed ($\tau_d$)\bibnote{In fact, one can see in Fig~\ref{fig:average_pulse} that there is a small contribution of a third, very fast, time constant which gives rise to the ``kink" shortly after the maximal pulse amplitude. The time span for this kink is of order 10 $\mu$s which corresponds to a frequency of $10^5$ Hz. The upper limit for the frequency PSDs that enter the optimal filter formalism is $10^4$ Hz, and as such any effects on timescales smaller than 100 $\mu$s do not have an effect on our result.
}. As in~ \cite{Moore2012a}, we choose to model these pulses as the sum of independent prompt and delayed components
\begin{align}
    s_{p}(t) &= (1 - e^{-(t-t_0)/\kappa_p}) e^{-(t-t_0)/\tau_p} \label{eq:promptpulse} \\
    s_{d}(t) &= (1 - e^{-(t-t_0)/\kappa_d}) e^{-(t-t_0)/\tau_d} ~,
\end{align}
which produce an average pulse with shape
\begin{align}
    s(t) = A \left( s_{p}(t) + w_d s_{d}(t) \right)~.
\end{align}
This model has 6 free parameters: the two fall time constants, the two rise time constants $\kappa_p$ and $\kappa_d$, an overall amplitude $A$, and the weight $w_d$ of the delayed component relative to the prompt component. The trigger time $t_0$ is known and fixed to 5 ms. 

\begin{figure*}
    \centering
    \includegraphics[width=\linewidth]{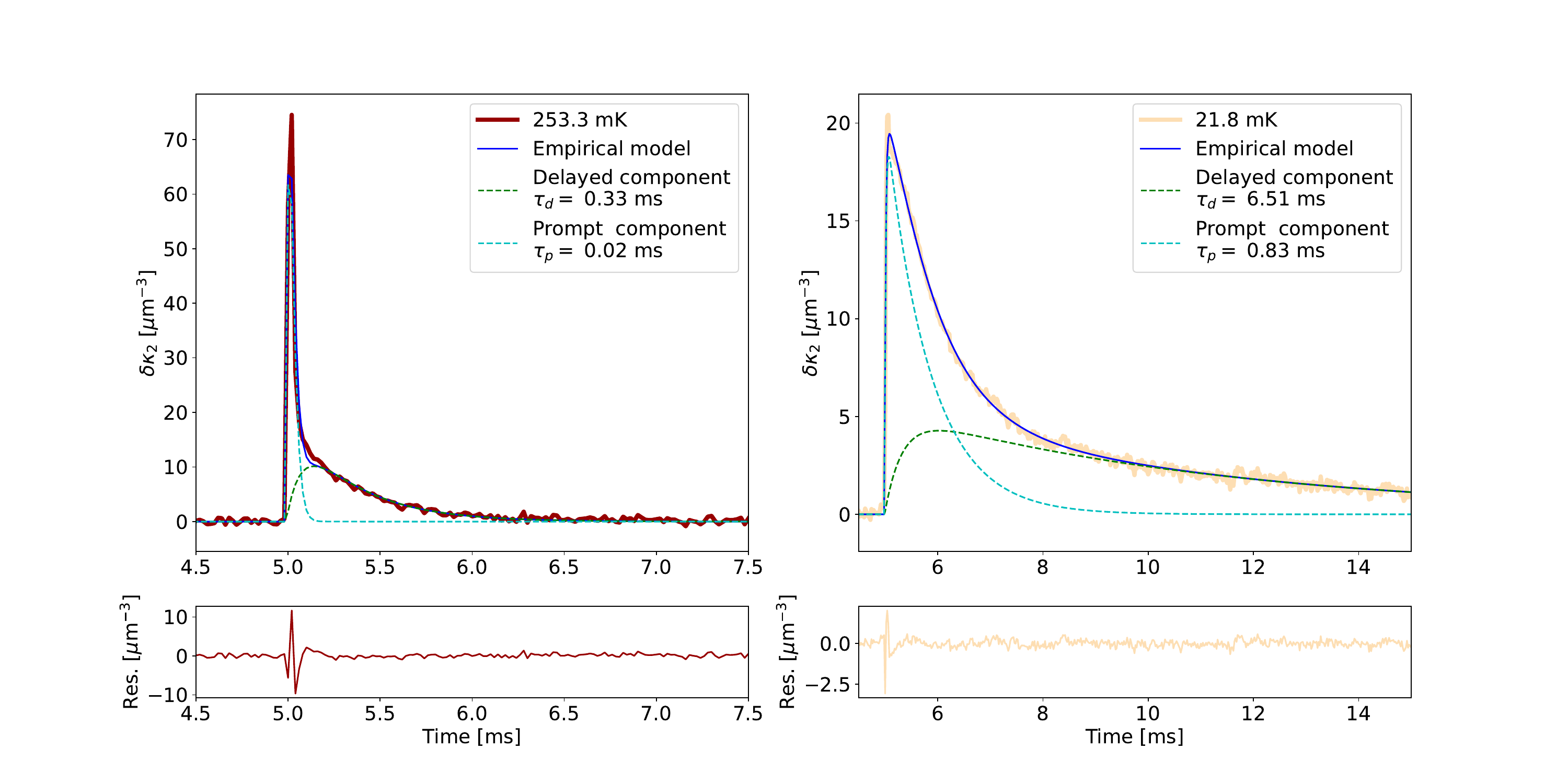}
    \caption{The average quasiparticle density response in the $\kappa_2$ direction for the same LED voltage (optical power) at $T=253$ mK (left) and $T_\mathrm{MC}=22$ mK (right). At both temperatures, a fit to the two-component empirical model is shown in blue, with the prompt and delayed components in dashed cyan and dashed green, respectively. The data are fitted out to 15.0 ms. The lower panels show the residuals of the fit. This model fits pulse data well across all temperatures investigated.}
    \label{fig:pulsefit_emp}
\end{figure*}

\begin{figure}
    \centering
    \includegraphics[width=\linewidth]{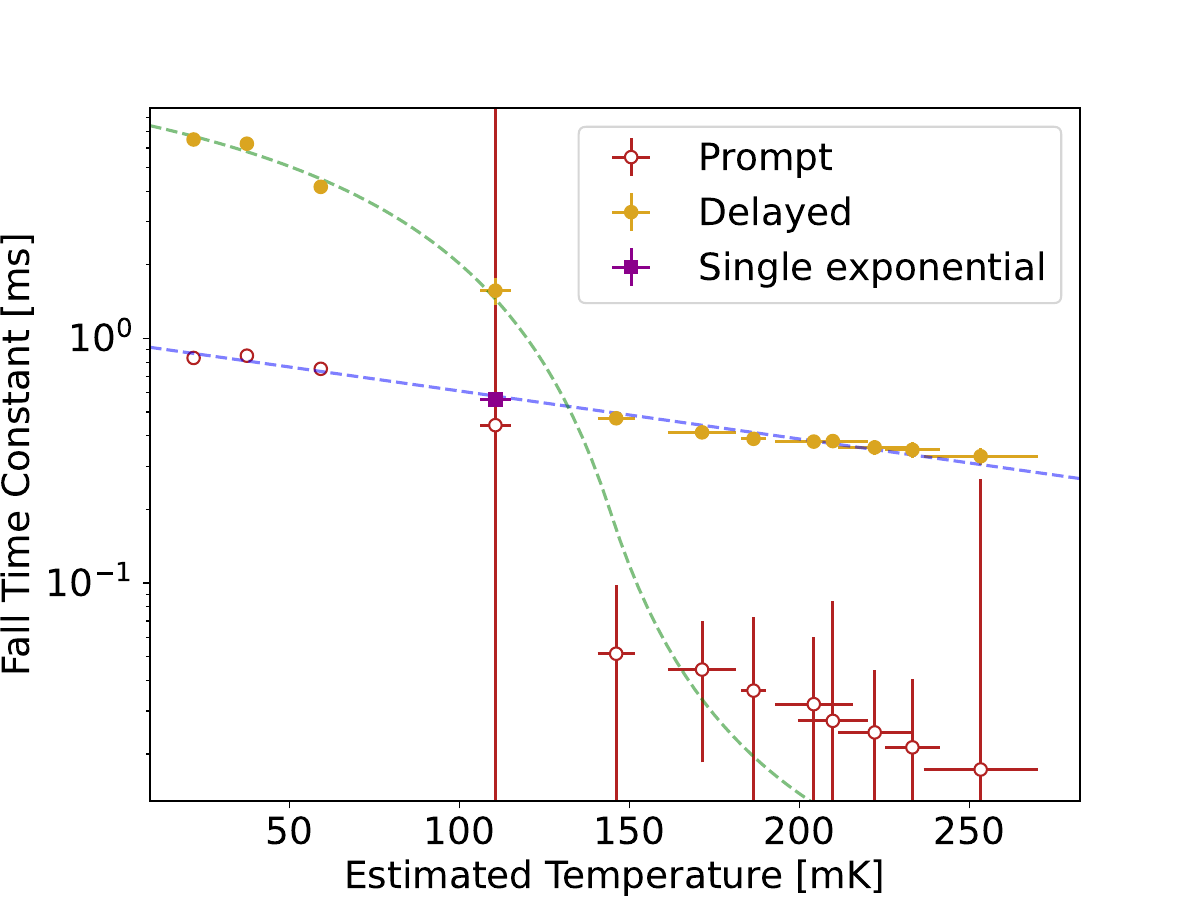}
    \caption{The prompt (open red circles) and delayed (filled yellow circles) fall-time constants for the empirical model as a function of device temperature. Time constant uncertainties are statistical and are returned by the fitting routine. Temperature uncertainties are derived from our temperature map. Analytic curves (blue and green dashed) have been fit to the data under the hypothesis that the role of $\tau_\mathrm{qp}$ changes from $\tau_d$ to $\tau_p$ around 125 mK. The green curve, which tracks the quasiparticle lifetime, is qualitatively consistent with the shape of the dependence reported in~\cite{Visser2011}. The blue curve tracks the effective phonon lifetime. The 110 mK pulse was also fit to a single component model, Eq.~(\ref{eq:promptpulse}), whose fall time constant is indicated by the purple square. Note that this point was not used in the determination of the blue curve, but nonetheless falls right on top.} 
    \label{fig:temp-times_emp}
\end{figure}

Examples of average pulses fit to this model for $T_\mathrm{MC}=325$ mK and $T_\mathrm{MC}=25$ mK are shown in Fig.~\ref{fig:pulsefit_emp} and demonstrate good agreement with the data, which is observed across all temperatures. The time constants extracted from the above model are shown as a function of estimated device temperature in Fig.~\ref{fig:temp-times_emp}. Both the prompt and delayed fall-time constants decrease monotonically as the temperature increases. The peak $\delta n_\mathrm{qp}$ (maximal pulse amplitude) was observed to change with temperature, as seen in Fig.~\ref{fig:pulsefit_emp}. While we account for the temperature dependence of $\kappa_2$, there is possible discrepancy between the quasiparticle temperature and the measured device temperature. Any variation in $\kappa_2$ due to this is not large enough to explain this, and any difference in $Q_i$ due to temperature is incorporated already in the conversion to the quasiparticle basis. The process of constructing a quantitative model is ongoing.

The data acquired at 110 mK, while fit well by the two-component model, returned with large uncertainty on the prompt component fall time, due to the degeneracy of delayed amplitude and prompt lifetime (i.e. increasing $w_d$ can compensate for shorter $\tau_p$). For this reason, we also fit this pulse to the single-component model in Eq.~(\ref{eq:promptpulse}). For this temperature only, the fit to this model accurately captures the shape, except at the latest times before returning to baseline, where the fit underestimates the pulse. This allows us to identify the single fall time constant as the prompt component, which is indicated by the purple triangle in Fig.~\ref{fig:temp-times_emp}. The fall time extracted from this falls upon the same line as the low-temperature prompt fall time constants, and was not included in the fit to determine the blue line in Fig.~\ref{fig:temp-times_emp}.

\textit{A priori} (and in a manner consistent with \cite{Moore2012}), we had ascribed respectively, $\tau_p$ and $\tau_d$ to the quasiparticle and phonon lifetimes. We depart from the interpretation in~\cite{Moore2012} and present the following novel interpretation for the two lifetime model: (1) there is one time constant that shows a relatively weak dependence on temperature, as would be expected of an athermal phonon population; and (2) there is a second time constant that shows a much stronger dependence on temperature, precipitously rising at low temperature. The first time constant is drawn suggestively as following the blue line in Fig.~\ref{fig:temp-times_emp}. The trend of the second time constant is drawn suggestively as the green line. In such a model, the lifetime that varies slowly with temperature is the phonon lifetime while the one that varies quickly is the quasiparticle lifetime, $\tau_\mathrm{qp}$. The phonon lifetime represents the characteristic time for the phonon population in the substrate above the Al pair-breaking energy to decay, presumably due to anharmonic and surface-promoted downconversion in the substrate, absorption in the resonator, loss of energy absorbed in Nb films to sub-$2\Delta_\mathrm{Al}$ phonons, and absorption in the mounting hardware. The quasiparticle lifetime becomes short at high temperature because of the elevated thermal quasiparticle density: $\tau_\mathrm{qp} \sim (2 \Gamma n_\mathrm{qp})^{-1}$ where $\Gamma$ is the quasiparticle recombination constant~\cite{Gao2008c, Visser2011} and $n_\mathrm{qp}$ increases with temperature. It begins to rise precipitously due to the exponential suppression of thermal quasiparticles by the superconducting gap. It saturates at low temperature due to various mechanisms that may stop this exponential suppression: readout power generation of quasiparticles, generation of quasiparticles by ambient ionizing radiation, or some other mechanism that maintains a non-thermal quasiparticle density. Under this hypothesis, the $\tau_\mathrm{qp}$ measurements, while discrepant with theoretical calculation from~\cite{Kaplan1976}
, show qualitative agreement in magnitude and shape with~\cite{Visser2011}. Under this assumption, quasiparticle lifetimes extend up to the several-millisecond scale at the lowest temperatures (6.5 ms), also consistent with~\cite{Visser2011,Visser2012}, corresponding to a quiescent thermal quasiparticle population of $n_\mathrm{qp,0} = 18.5~\mu$m$^{-3}$~\cite{Kaplan1976}, the lowest that has ever been measured, to the best of the authors' knowledge.

This pulse shape model, while adequately describing the observed data, is not without shortcomings. Primarily, as the sensor only measures the quasiparticle response in the superconductor, the phonon and quasiparticle components should not be separable. Rather, one should expect the pulse shape to be a convolution of the phonon time dependence and the quasiparticle lifetime. Evaluation of a class of such models is left for a future investigation.

\section{\label{sec:disc}Discussion}

The device under test demonstrated a best baseline resolution on energy absorbed by the sensor of $\sigma_E^\mathrm{abs}=2.1 \pm 0.2$ eV, and a resolution on energy deposited in the substrate of $\sigma_E=318 \pm 29$ eV. From these values, we measure the phonon collection efficiency $\eta_\mathrm{ph}$ to be at the sub-percent level. While the overall energy resolution is poor, the measured resolution on energy absorbed by the sensor ($\sigma_E^\mathrm{abs}$) is a factor 1.5--3 better than demonstrated previously in the literature for KIPM detectors~\cite{Cruciani2022, Cardani2018}. The model used to evaluate the energy resolution provides a clear two-pronged approach toward achieving the sub-eV resolutions required to probe MeV-scale DM models: (1) improving $\sigma_0$ by reducing the noise temperature of the first-stage amplifier and mitigating the presence of TLS sites to lower the intrinsic device noise; and (2) improving the phonon collection efficiency $\eta_\mathrm{ph}$. The latter can be accomplished by reducing high-gap metal on the substrate and improving device mounting to reduce phonon loss. Collection efficiencies as high as 30\% have been demonstrated with other superconducting sensors~\cite{Ren2020}. Given the comparable collection efficiency achieved by other KIPM detectors, this should be achievable with our device architecture, and alone would improve our $\sigma_E$ to roughly 10 eV. 

While a reduction of the noise temperature of the first-stage amplifier can be achieved by use of a parametric amplifier (\textit{e.g.}, a Kinetic-Inductance Traveling Wave parametric amplifier)~\cite{Ramanathan2023}, this alone does not translate directly to an improvement in resolution, due to the presence of TLS noise at a power above that of the amplifier white noise. However, one may evade the impact of TLS noise by using the dissipation quadrature for readout and offsetting the elevated white noise level with the improved noise temperature of the parametric amplifier. See Appendix~\ref{supp-noise} for a detailed discussion of the noise sources present in this experiment.

All parameters measured in the energy calibration are statistically compatible to $2\sigma$, with the exceptions of the baseline resolution, which is marginally worse for the later calibration. However, despite this difference, the resolution on energy deposited in the substrate is in good agreement. While $\sigma_E^\mathrm{abs}$ depends on specifics of the superconductor's properties that can change with time, the measured $\sigma_E$ is only dependent on the increased broadening in pulse amplitude distributions from the photon shot noise. This consistency confirms we have correctly reconstructed the absolute energy scale in both calibrations, as the device is detecting the same number of photons for the same optical calibration platform settings despite small differences in readout power and device temperature. Discrepancies in $\sigma_E$ would indicate either a problem in our analysis or an instability in our optical calibration system. The reason for the change in device performance between the two LED calibrations is not known with certainty, but we note that the device underwent six thermal cycles and exposure to atmosphere between calibrations, potentially leading to changes in film structure or oxide growth.

The resolution achieved by our sensor ($\sigma_E^\mathrm{abs}$) is primarily due to the ms-scale pulse lifetimes, which enable resolution of photon shot noise despite the presence of TLS noise. For comparison, in a nominally identical device operated in a dilution refrigerator at Caltech pulse lifetimes were limited to a few hundred $\mu$s between 50 and 100 mK. Potential reasons for the longer lifetimes in NEXUS are the lower base temperature, lower ionizing particle background, and better blackbody radiation environment. The latter will be tested by improving the blackbody radiation shield in the Caltech facility in the future.

We have introduced a signal model comprising two fall time constants: one prompt and one delayed. This model allows us to extract $\tau_\mathrm{qp}$ from our pulse data. Quasiparticle lifetimes as high as 6.5 ms were observed. 

We are developing a simulation using the G4CMP~\cite{Kelsey2023} package to understand the phonon collection efficiency by modeling phonon downconversion in high-gap materials, phonon loss at interfaces, and the dependence of phonon collection on fill fraction (ratio of phonon-absorbing inductor surface area to substrate surface area). The results of this simulation will both inform design choices for new KIPM detector architecture and allow us to investigate signal pulse models in depth, including phonon recycling effects. Furthermore, the same simulation can be used to investigate the pulse model introduced in section~\ref{sec:qplifetime}, as we can directly tune $\tau_\mathrm{qp}$ and the phonon lifetime and compare them with the results of fitting to this model. A new batch of KIPM detectors was recently fabricated to understand the dependence of $\eta_\mathrm{ph}$ on the fill fraction.

\begin{acknowledgments}
The authors would like to thank Bruce Bumble of the Jet Propulsion laboratory for his guidance and expertise in fabricating the KIPM device studied in this work. Additional thanks are due to Ryan Linehan for useful discussions of quasiparticle and phonon dynamics. This manuscript has been authored by Fermi Research Alliance, LLC under Contract No. DE-AC02-07CH11359 with the U.S. Department of Energy, Office of Science, Office of High Energy Physics. Primary funding for this project was furnished by the FNAL Laboratory Directed Research \& Development (LDRD) program award number LDRD2020-040. This work was supported by the U.S. Department of Energy, Office of Science, National Quantum Information Science Research Centers, Quantum Science Center and the U.S. Department of Energy, Office of Science, High-Energy Physics Program Office. OW was also supported by NASA fellowship NSTGRO80NSSC20K1223.

DJT commissioned the KIPMD data acquisition system, collected the data, performed the analysis, and drafted the manuscript. OW fabricated the device under test in this work, using funds provided by SG. OW and KR motivated the physical interpretation of the two-component signal model, and in conjunction with SG, contributed expertise in phonon and quasiparticle dynamics and MKID operation. LH led the management and supervision of this project. NK conceived of this project on the original LDRD award and participated in the RF upgrade of NEXUS. SL and DBo participated in the RF upgrade of NEXUS. DBa led the QSC project which supported this work following conclusion of the initial project funding (LDRD). EF-F led the management of the NEXUS facility. MH, CJ, and PL contributed to the management and operation of the NEXUS facility. RC, KK, and RR contributed to the operation of the NEXUS facility. VN and BS contributed to the management, upgrade and operation of the NEXUS facility, and led the work to characterize the radiogenic background present at this facility. CB contributed to the design and simulation of the magnetic shield at NEXUS.
\end{acknowledgments}

\appendix

\section{Fits to Mattis-Bardeen Theory} \label{supp-mb}

The zero-temperature parameters of the resonator, $f_r(T=0)$, $Q_i(T=0)$, $\alpha$, and $\Delta$, can be extracted by measuring the resonance shape, in frequency-space, as a function of temperature. For a given temperature, the resonance line shape is fit to a model given~\cite{Gao2008c} by
\begin{align} \label{eq:cplx-trans-full}
    S_{21}(f) = a e^{-2 \pi j f \tau} \left[ 1 - \frac{(Q_r/Q_c \cos \phi) e^{j\phi}}{2 j Q_r x} \right]~,
\end{align}
where $x \equiv (f-f_r)/f_r$, the quality factors are as in the main text, $a$ is a complex feedline attenuation, and $\tau$ is the feedline delay. This model takes into account any impedance mismatch that gives rise to an asymmetric transmission line shape by allowing $Q_c$ to be complex, with a phase $\phi$ which represents an angle of rotation between the resonance circle in the complex plane and the off-resonance baseline transmission.

\begin{figure*}
    \centering
    \includegraphics[width=\linewidth]{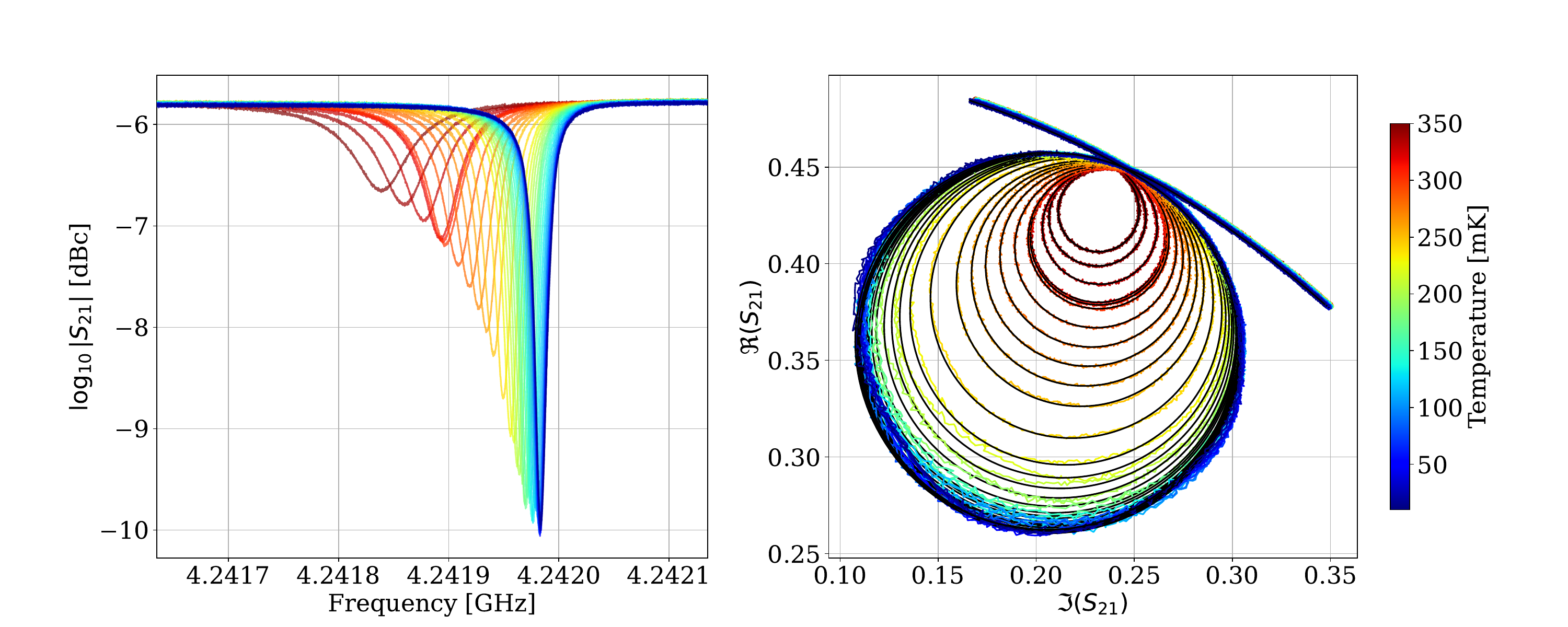}
    \caption{The forward transmission $S_{21}(f)$ of the phonon-absorbing resonator on FNAL-I at $P_\mathrm{DAC}=-30$ dBm, shown as $\log_{10}|S_{21}(f)|$ vs $f$ (left) and in the complex plane (right) as a function of mixing chamber temperature. Black lines indicate fits of Eq.~(\ref{eq:cplx-trans-full}) to the resonance circle.}
    \label{fig:S21-vs-T}
\end{figure*}

The temperature of the mixing chamber was ramped from the 10 mK base temperature to 350 mK in steps of 10 mK. At each temperature, the complex transmission $S_{21}(f)$ of the Al resonance is measured using a Copper Mountain M5090 vector network analyzer for RF stimulus powers ranging from $P_\mathrm{DAC}=-5$ dBm to -75 dBm. The remainder of the RF readout chain is identical to that described in the main text. The acquired frequency spectra are fit to Eq.~(\ref{eq:cplx-trans-full}) and the optimal parameters recorded for each temperature--power pair. The resonance feature in $S_{21}(f)$ as a function of temperature is shown for an RF stimulus power of -30 dBm in Fig.~\ref{fig:S21-vs-T}. These data were collected in a temperature sweep of the refrigerator performed after the February calibration run.

The collected data are fit as a function of temperature to the following models for the dependence of resonator line shape on temperatures:

\begin{align} 
    f_r(T) &= f_r(0) \left(1 - \frac12 \alpha \kappa_2(T, f_r(0), \Delta) n_\mathrm{qp}(T,\Delta) \right) \label{eq:MBeqs1} \\
    Q_i(T) &= \left( \alpha \kappa_1(T, f_r(0), \Delta) n_\mathrm{qp}(T,\Delta) + \frac{1}{Q_i(0)} \right)^{-1}~, \label{eq:MBeqs2}
\end{align}
with $\kappa_{1,2}$ as in~\cite{Siegel2016}, and the quiescent density of quasiparticles at temperature $T$ taken to be 
\begin{align}
    n_\mathrm{qp} = 2 N_0 \sqrt{2 \pi k_B T \Delta} e^{-\Delta / k_B T}~,
\end{align}
where $N_0$ is the single-spin density of states~\cite{Gao2008c}. The fit results are summarized in Table~\ref{tab:MBfitresults}. The values of $\alpha$ and $\Delta$ used in the main text are adopted from~\cite{Wen2021}. We adopt canonical values: $f_t(0) = 4.24198$ GHz for all RF readout powers, and $Q_i(0) = 3.7\times10^5 (4.3\times10^7)$ for $P_\mathrm{DAC}=-15(-30)$ dBm, for the analyses in the main text, as measured for FNAL-I at NEXUS. This choice was made due to poor temperature instrumentation on the RF payload plate which led to an overestimation of device temperature for these data. This deficit was corrected in future runs of the NEXUS dilution refrigerator. No such issue was present in~\cite{Wen2021}, and their measured values of $\alpha$ and $\Delta$ match more closely with literature~\cite{Matthias1963, Tinkham2004}. 


\begin{table*}
\caption{\label{tab:MBfitresults}%
Results of the resonator characteristic fits to Mattis-Bardeen theory for $P_\mathrm{DAC}=-15$ and $-30$ dBm from this work compared to the measurement in~\cite{Wen2021} of an identical device from the same fabrication batch at -30 dBm. Note that for the analysis presented herein, we adopt the values from~\cite{Wen2021}, due to the poor temperature instrumentation during data acquisition leading to poor fits to Mattis-Bardeen theory. Given that the NEXUS data optimized for values of $\alpha$ consistent with 0, the optimized value of $\Delta$ is not trustworthy despite the (statistical-only) uncertainties quoted.}
\begin{ruledtabular}
\begin{tabular}{rcccc}
& \textrm{NEXUS (-15 dBm)}             & \textrm{NEXUS (-30 dBm)}             & \textrm{Caltech (-30 dBm)~\cite{Wen2021}} \\ 
\colrule
$f_r(0)$ [GHz] & $4.241968 \pm (7\times10^{-6})$ & $4.241974 \pm (8\times10^{-6})$ & 4.2420                   \\
$Q_i(0)$       & $(3.7 \pm 0.3) \times10^5$      & $(4.2 \pm 0.3) \times10^5$      & $4.1\times10^5$          \\
$\alpha$ [\%]  & $ 0.2 \pm 0.3$                  & $ 0.2 \pm 0.2$                  & 3.80                     \\
$\Delta$ [meV] & $0.14 \pm 0.03$                 & $0.13 \pm 0.03$                 & 0.184                    \\   
\end{tabular}
\end{ruledtabular}
\end{table*}

\section{\label{supp-noise}Noise Characterization} 

\begin{figure*}
    \centering
    \includegraphics[width=\linewidth]{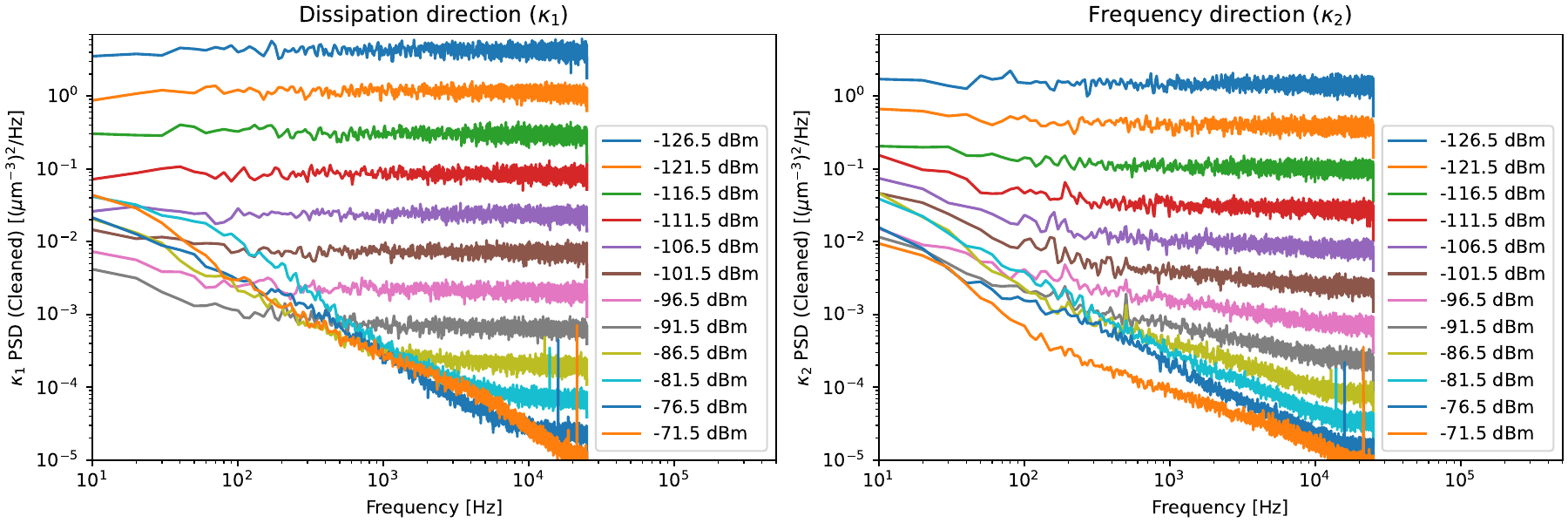}
    \caption{The noise PSDs, in units of quasiparticle density fluctuations, after correlated noise removal as measured by the (left) dissipation and (right) frequency readout quadratures. Note that the $\kappa_1$($\kappa_2$) PSDs are multiplicative rescalings of the $\delta(1/Q)$($\delta f/f$) PSDs. The color indicates the RF power on the feedline of the device for each noise acquisition.}
    \label{fig:noise-spectra}
\end{figure*}

The noise present in an on-resonance timestream directly impacts the baseline energy resolution of a KIPM detector. As such, understanding the noise sources present and their impact is important for optimizing detectors for energy resolution. The on-resonance timestream is acquired simultaneously with two calibration tones at $f_r \pm 10$ MHz for removal of correlated noise. This is repeated at RF stimulus powers ranging from $P_\mathrm{DAC}=-15$ dBm to -75 dBm. The PSD $J(f)$ of the noise in the phase and magnitude readout quadratures are then generated for each RF power. Using the resonance line shape, these are converted to PSDs of fluctuations in fractional resonant frequency shift $\delta f_r / f_r$ and fluctuations on inverse quality factor $\delta(1/Q)$. They are further converted to fluctuations in quasiparticle density as measured by the frequency and dissipation shifts, as shown in Fig.~\ref{fig:noise-spectra}. The pulse data for the energy resolution measurement in this work is taken at the highest power shown. 

The observed noise spectra are white at the lowest readout power, where amplifier noise dominates. As readout power increases ($P_g=$-111.5 dBm to $P_g=$-96.5 dBm on the device feedline), the presence of noise from electrically active two-level systems (TLS) appears in the frequency quadrature, while the dissipation quadrature remains white. At the highest readout powers, extraneous correlated noise arises in both the $\kappa_1$(dissipation) and $\kappa_2$(frequency) readout quadratures. The source of this noise is currently unknown, and is not the topic of the following discussion.

The noise power from TLS, $J_\mathrm{TLS}(f)$, at a given frequency $f_0$ should scale as a power law:
\begin{align}
    J^\mathrm{TLS}(f=f_0|P_g) \propto P_g^{-1/2}~,
\end{align}
for frequency quadrature readout~\cite{Zmuidzinas2012}. The power-law fit to the dissipation quadrature data returns an exponent of $-1.01 \pm 0.03$, consistent with the scaling of white amplifier noise with readout power. The best power-law fit to the noise power at 1 kHz for the aforementioned range of RF stimulus powers where TLS is present 
returns an exponent of $-0.81 \pm 0.03$. 
The uncertainties are statistical only, and are returned by the fitting routine. The PSDs contain contributions from both the TLS noise of interest and the white noise, and thus to extract the power dependence of the TLS contribution, these PSDs must be decomposed into their constituent contributions, which is discussed below.

\begin{figure}
    \centering
    \includegraphics[width=\linewidth]{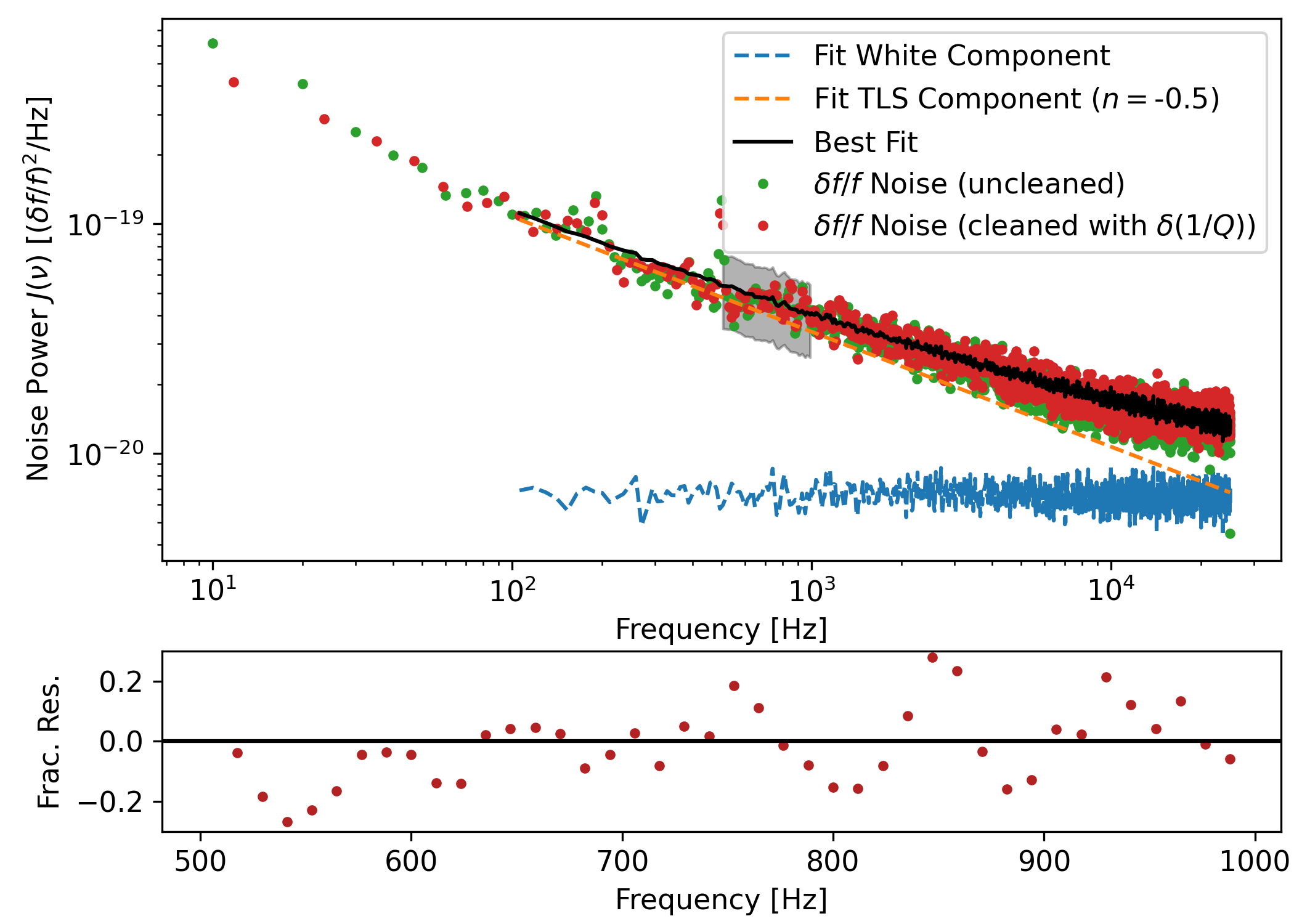}
    \caption{The $\delta f/f$ Noise PSD for $P_g=$-91.5 dBm, the highest readout power before the correlated $\kappa_1-\kappa_2$ noise arises, is shown before cleaning with $\delta (1/Q)$ (green points) and after cleaning (red points). These data are fit (black line) to a model comprising a TLS-noise term (orange dashed line) and a white-noise term (blue dashed line), for $f>100$ Hz. Below this, a slightly steeper frequency dependence in the noise power is observed. The lower panel shows the fractional residual of the cleaned data to our model in the 0.5 -- 1 kHz region, indicated by the grey shaded region in the upper panel.}
    \label{fig:noise-model}
\end{figure}

As shown in~\cite{Wen2021}, the $\delta f/f$ timestream can be cleaned of correlated noise with the $\delta (1/Q)$ timestream where TLS noise is not present, using the same formalism for cleaning on-resonance tones with the off-resonant calibration tones. For our data, in the power range of interest ($-111.5 ~\mathrm{dBm} \le P_g \le -96.5 ~\mathrm{dBm} $), this noise reduction is insignificant, indicating this noise 
is largely uncorrelated. The cleaned $\delta f/f$ PSD can then be fit to a noise model containing contributions from white amplifier noise and TLS noise, given by
\begin{align} 
    J(f) &= J_\mathrm{TLS}(f) + A_{w} J_{w}(f) \\ &= A_\mathrm{TLS} f^{n} \left| \frac{1}{1+2 j Q_r (f/f_r)} \right|^2 + A_{w} J_{w}(f)~, \label{eq:tlsmodel}
\end{align}
where $A_\mathrm{TLS}$ is the magnitude of the noise power from TLS, $n$ is the TLS frequency-dependence exponent, $f_r$ is the resonant frequency, and $Q_r$ is the total quality factor. The white noise component $J_{w}(f)$ in the model is taken to be the $\delta f/f$ PSD for the lowest readout power and its contribution to the overall noise power $A_w$ is fixed by the $P_g^{-1}$ scaling for white noise. The TLS contribution is modified by the resonator's response function (equivalently, a low-pass filter) which rolls off at a frequency $f_r/(2Q_r)$~\cite{Zmuidzinas2012}. In the fit routine, we fix the value of $2 Q_r/f_r$ for our resonator, leaving $A_\mathrm{TLS}$ and $n$ the only free parameters. This cleaning and fit to the noise model is shown in Fig.~\ref{fig:noise-model} for the largest readout power before the correlated $\kappa_1-\kappa_2$ noise arises. At this readout power, we find a frequency dependence for the TLS contribution of $n=-0.501 \pm 0.007$, in good agreement with $n=-0.5$ as is typically expected for TLS noise~\cite{Zmuidzinas2012}. 

\begin{figure}
    \centering
    \includegraphics[width=\linewidth]{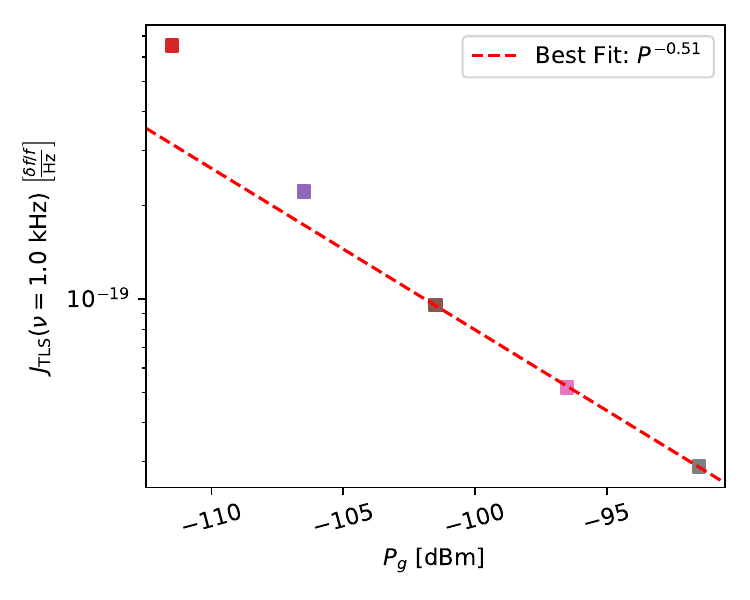}
    \caption{Dependence of the magnitude of TLS noise at $f_0=1$ kHz as a function of readout power present on the device feedline $P_g$. The circles and squares are the fluctuation powers of the frequency ($\delta f/f$) and dissipation ($\delta (1/Q$) readout quadratures, respectively. The dashed and dotted lines indicate the best fit of the these data to power laws. Note that only the three highest readout powers are used in the fit shown, as those exhibit the expected $f^{-1/2}$ scaling of the TLS noise power when fitting the PSD to the model in Eq.~\ref{eq:tlsmodel}.}
    \label{fig:TLS-noise-highpower}
\end{figure}

This process is repeated for all readout powers in the region of interest. We find that for the two lowest readout powers, the best fit indicates TLS noise is not present, opting for $n=0$. For the highest three readout powers, the optimizer selects frequency dependence in the range $n=0.51-0.55$ with $\pm0.1$ statistical uncertainty. The noise power of the TLS component at 1 kHz is extracted as a function of feedline power and fit to a power law. When all PSDs in the readout power range specified above are considered, the power law exponent is found to be $0.66 \pm 0.05$. However, when only the three highest readout powers are considered (where $f^{-1/2}$ dependence is observed), the exponent becomes $n=-0.519 \pm 0.005$, consistent with the power scaling for TLS noise. This is shown in Fig.~\ref{fig:TLS-noise-highpower}

The absence of this noise in the dissipation direction, its frequency dependence, and dependence with readout power ($P_g^{-1/2}$) indicate this is TLS noise. For comparison to TLS noise as observed in different device architectures and materials, see~\cite{Zmuidzinas2012}. We conclude the noise in our device is TLS-dominated at readout powers below those at which the unknown correlated $\kappa_1-\kappa_2$ noise arises. 

\subsection{Generation-Recombination Noise}

Determining the quasiparticle lifetime by way of the noise PSD is not possible with our device. This is due to noise from the first-stage amplifier at a power larger than that of the generation-recombination (GR) noise. To infer the quasiparticle lifetime from the PSD requires a resolvable shoulder in the PSD, after which the GR noise dies out. 


The power spectrum of generation-recombination noise, for the case of thermal generation of quasiparticles only (i.e., no optical or phonon loading) can be expressed in units of quasiparticle density fluctuations~\cite{Gao2008c} as 
\begin{align}
    S(\omega) = \frac{1/(\Gamma V)}{1 + (\omega \tau_\mathrm{qp})^2}~,
\end{align}
where $V$ is the resonator volume, $\Gamma$ is the quasiparticle recombination constant, and $\tau_\mathrm{qp}$  is the quasiparticle lifetime. In these units, the GR noise power is inversely proportional to resonator volume. The device under test has a relatively large inductor volume, which makes the GR noise power very small. For low-frequency ($< 10$ Hz), the GR noise in our device is white with a power of approximately $10^{-5}$ ($\mu$m$^{-3}$)$^2$/Hz, falling quickly off above a few 10s of Hz. The observed noise power in units of quasiparticle density fluctuations is at the $10^{-3}$ ($\mu$m$^{-3}$)$^2$/Hz level for frequencies below 100 Hz (see Fig~\ref{fig:noise-spectra} for the observed noise PSDs in units of quasiparticle density for comparison). However, this is due to the observed $1/f$ noise that appears in both readout quadratures for the highest readout powers. If not for this noise, and we were limited by amplifier white noise, we would approach $10^{-5}$ ($\mu$m$^{-3}$)$^2$/Hz. Further reduction in amplifier noise (for instance, by using a parametric amplifier) would enable resolving GR noise in our device, provided the $\kappa_{1}$--$\kappa_{2}$ noise that rises steeply with decreasing frequency for high readout powers can be mitigated. GR noise has been observed in smaller volume devices fabricated and studied by this group for millimeter-wave astronomy~\cite{Shu2022}.

\subsection{Voltage Dependence of Noise PSDs}
The noise PSDs shown in Fig.~\ref{fig:average_noise} (left) display some scatter with respect to LED voltage setpoint. However, they do not exhibit a clear trend vs voltage. Their scatter was investigated and it was found that fitting the lowest frequency bin value of each PSD vs LED voltage to a line resulted in a slope that was consistent with zero. While we do not have a clear understanding of why this scatter is present, it is unlikely to be due to long tails of the pulse since the ``pulse-free” region begins 95 ms after the LED flash (and consequently, the rising edge of the pulse). These pulses return to baseline well before 20 ms after the LED flash. Furthermore, the scatter in the noise PSDs are inconsequential to our result. For this analysis we chose our noise PSD that enters the optimal filter formalism ($J(f_n)$), for all LED voltage settings, to be that from the 2.0 V LED setting. If instead, we picked any other noise PSD for the optimal filter, this would shift our resulting resolution on energy deposited in the substrate by $<$2\%, about an order of magnitude smaller than other sources of uncertainty ($\sim$10\%), discussed in the main text. Due to the dependence on the baseline resolution from the choice of noise realization, the resolution on quasiparticle density and energy absorbed by the resonator both would shift by $<0.5$\%. As such, this would shift the phonon collection efficiency also by $<2$\%.

\section{\label{supp-res}Resolution Model}
For a set of LED flashes with the same settings, the substrate will absorb an average number of photons $\bar{N}_\gamma$ in each flash, corresponding to an average energy deposition of $\bar{E}=\bar{N}_\gamma h\nu$, where $h\nu$ is the energy of an individual photon. Due to the Poisson statistics of this process, the variance in the number of photons deposited in the substrate is $\sigma_{N_\gamma}^2 = \bar{N}_\gamma$, giving a fluctuation in number of photons deposited of
\begin{align}
    \sigma_{N_\gamma} = \sqrt{\bar{N}_\gamma}~,
\end{align}
which corresponds to fluctuations in deposited energy of
\begin{align}
    \sigma_E
    = h \nu \sqrt{\bar{N}_\gamma} ~.
\end{align}
Only a fraction $\eta_\mathrm{ph}<1$ of the energy deposited is absorbed by the superconductor: $\bar{E}_\mathrm{abs} = \eta_\mathrm{ph} \bar{E}$. The resolution in deposited energy is scaled by the phonon collection efficiency $\eta_\mathrm{ph}$ to find the resolution on energy absorbed by the superconductor:
\begin{align}
    \sigma_E^\mathrm{abs} = (h \nu) \eta_\mathrm{ph} \sqrt{\bar{N}_\gamma} ~.
\end{align}
A further scaling factor is applied to determine the fluctuations in device readout units due to the photon shot noise. Assuming linearity in device response to absorbed energy, we can define $R$ to be the responsivity per unit energy absorbed by the superconductor, and thus the mean device response to photon bursts carrying a mean number of photons $\bar{N}_\gamma$ is
\begin{align} \label{eq:led-mean}
    \mu = R \bar{E}_\mathrm{abs} &= R \eta_\mathrm{ph} \bar{E} = R \eta_\mathrm{ph} \bar{N}_\gamma h\nu~.
\end{align}
Similarly, the fluctuation in device readout is
\begin{align} \label{eq:led-flucts}
    \sigma_\mathrm{LED} = R \sigma_E^\mathrm{abs} = (h \nu) \eta_\mathrm{ph} R \sqrt{\bar{N}_\gamma}~.
\end{align}
Inverting Eq.~(\ref{eq:led-mean}) and inserting into Eq.~(\ref{eq:led-flucts}) for $\bar{N}_\gamma$ yields
\begin{align}
    \sigma_\mathrm{LED}^2 = \mu \cdot (h\nu) \eta_\mathrm{ph} R~.
\end{align}
The distribution of pulse amplitudes from a set of LED pulses has a two contributions to its width $\sigma$: a term from the intrinsic, zero-energy, noise of the device ($\sigma_0$) and the broadening due to photon shot noise:
\begin{align}
    \sigma &= \sqrt{ \sigma_0^2 + \sigma_\mathrm{LED}^2 } = \sqrt{ \sigma_0^2 + \mu \cdot (h\nu) \eta_\mathrm{ph} R } \\
    &= \sqrt{ \sigma_0^2 + \mu \cdot r } ~,
\end{align}
where we have defined 
\begin{align}
    r = (h\nu) \eta_\mathrm{ph} R~,
\end{align}
to be the device responsivity per photon deposited in the substrate (as $\mu = r \bar{N}_\gamma$).

If we then take the resonator equation for fractional frequency shift due to a change in quasiparticle density $\delta n_\mathrm{qp}$ (Eq.~\ref{eq:res_qpdens001}), and differentiate it with respect to energy, we find\bibnote{From~\cite{Siegel2016} Eq. 2.59, we see $\delta f/f = -\frac{\alpha}{2}\kappa_2 n_\mathrm{qp}$. We are interested in determining how this changes with respect to energy absorbed by the superconductor. Since $\alpha,\kappa_2$ are independent of deposited energy, we must find $\mathrm{d}/\mathrm{d}E_\mathrm{abs} (n_\mathrm{qp})$. The density of quasiparticles increases by $2/V$ for every $2\Delta$ deposited, thus $\mathrm{d}/\mathrm{d}E_\mathrm{abs} (n_\mathrm{qp}) = \frac{1}{V\Delta}$.}
\begin{align} \label{eq:res_eq_dff}
    R = \left| \frac{\mathrm{d}(\delta f/f)}{\mathrm{d}E_\mathrm{abs}} \right| = \frac{\alpha}{2} \kappa_2 \frac{d (\delta n_\mathrm{qp})}{dE_\mathrm{abs}} = \frac{\alpha}{2} \kappa_2 \frac{1}{V \Delta}~,
\end{align}
where $\alpha$ is the kinetic inductance fraction, $V$ is the volume of the phonon absorber, and $\kappa_2$ is a function of readout frequency, device temperature, and the Cooper pair binding energy $\Delta$, as described in the main text. The responsivity of the device per unit energy deposited in the substrate is then
\begin{align}
    r = \alpha \frac{\kappa_2}{V} \frac{h\nu}{2 \Delta} \eta_\mathrm{ph}~.
\end{align}

\section{\label{supp6-k2unc}Uncertainty in Resolution of Quasiparticle Density}

The fractional shift in resonant frequency and quality factor induced by an increase in  quasiparticle density by an amount $\delta n_\mathrm{qp}$ are given~\cite{Siegel2016} by 
\begin{align} 
    \frac{\delta f_r}{f_r} 
    &= - \frac12 \alpha \kappa_2(T,\omega,\Delta) \delta n_\mathrm{qp} \label{eq:res_qpdens001}\\
    \delta\frac{1}{Q_i}    
    &= -\alpha \kappa_1(T,\omega,\Delta) \delta n_\mathrm{qp}~, 
\end{align}
where $\alpha$ is the kinetic inductance fraction, and $\kappa_{1(2)}(T,\omega,\Delta)$ carries units of volume and is the real(imaginary) component of the fractional change in complex conductivity per unit change in quasiparticle density as a function of temperature, readout frequency ($\omega$), and superconducting gap energy at zero temperature ($\Delta$), expressions for which are given in~\cite{Siegel2016}. As such, inferring $\delta n_\mathrm{qp}$ (and subsequently the resolution on this value) from shifts in resonator characteristics requires calculation of $\kappa_1$ or $\kappa_2$:
\begin{align}
    \sigma_\mathrm{qp}^{\kappa_2} &= \frac{2}{\alpha \kappa_2(T)} \sigma^{\delta f_r/f_r} \\
    \sigma_\mathrm{qp}^{\kappa_1} &= \frac{1}{\alpha \kappa_1(T)} \sigma^{1/Q_i}~,
\end{align}
for a fixed $\Delta$ and readout frequency. 

The question is, at what temperature does one evaluate $\kappa_{1,2}$? Given that this resolution is on excess quasiparticle density over the quiescent, thermal density of quasiparticles $n_\mathrm{qp,0}$, one should insert the temperature of the Cooper pairs in the superconductor. It is obvious that the Cooper pair temperature cannot be lower than that of its environment, but what is the upper bound? The quasiparticle lifetime $\tau_\mathrm{qp}$ is set by $n_\mathrm{qp,0}$, as their product is a constant~\cite{Kaplan1976}. Thus, by measuring $\tau_\mathrm{qp}$, one knows $n_\mathrm{qp,0}$, from which an effective Cooper pair temperature can be determined using theoretical predictions~\cite{Visser2012}:
\begin{align}
    n_\mathrm{qp}(T) = 2 N_0 \sqrt{2 \pi k_B T} e^{-\Delta / k_B T}~.
\end{align}
As discussed in section~\ref{sec:qplifetime}, at the lowest temperatures, $\tau_\mathrm{qp}$ plateaued at 6.5 ms, the corresponding thermal quasiparticle density is $n_\mathrm{qp,0} = 18.5~\mu$m$^{-3}$, from which we extract an effective temperature of the Cooper pairs of $T_\mathrm{eff}=172$ mK. The value of $\kappa_2$ over the temperature range between the environment temperature (10 mK) up to $T_\mathrm{eff}$ changes by 9.1\%. This is the fractional uncertainty we adopt in the text when converting between resolution on changes in resonator properties and resolution on quasiparticle density fluctuations.

\section{\label{supp7-diagrams}Electrical and Optical Wiring Diagrams}
\begin{figure}
    \centering
    \includegraphics[width=\linewidth]{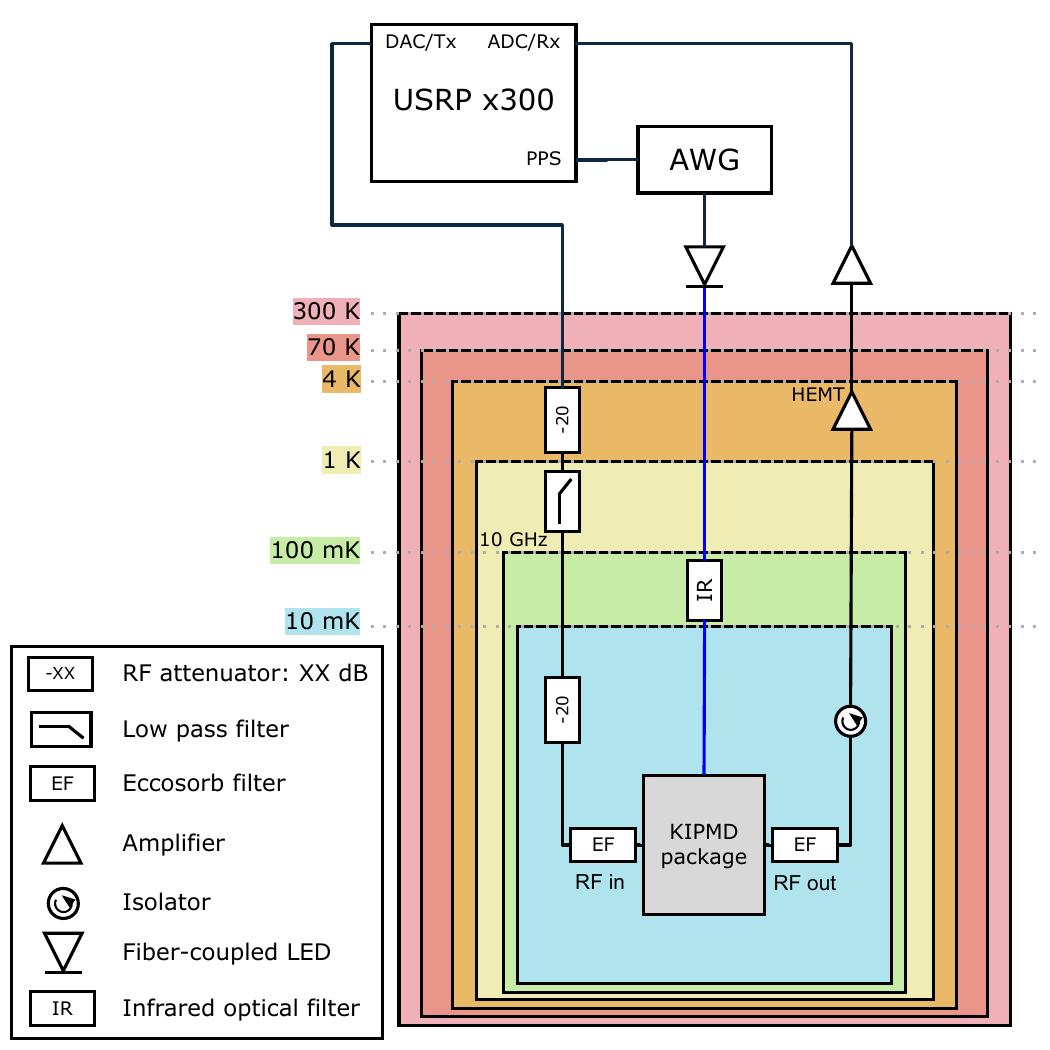}
    \caption{The simplified wiring diagram for the device under test in this work. RF connections are indicated by solid black lines while optical fiber connections are indicated by solid blue lines. The in-refrigerator components are thermally coupled to the temperature stage inside which they appear. Note that only the nonzero attenuators at each temperature stage are shown. Both the input and output lines are thermalized at each temperature stage with 0 dB attenuators. The schematic of the warm electronics has been vastly simplified. Components that are not shown include the VNA, RF switch and the DC power supply used to bias the fiber-coupled LED that is modulated by the arbitrary waveform generator (AWG).}
    \label{fig:wiring-diagram}
\end{figure}

\begin{figure}
    \centering
    \includegraphics[width=\linewidth]{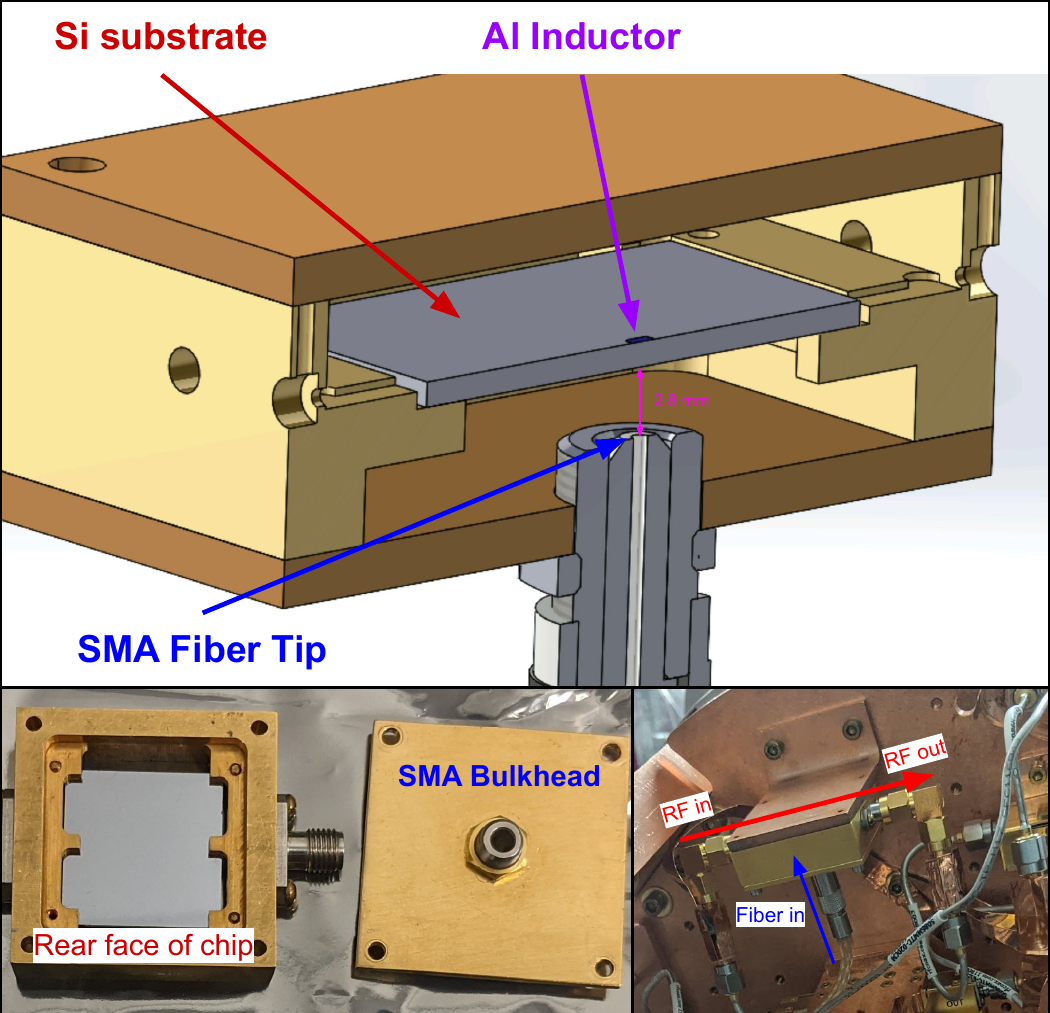}
    \caption{The relative positioning of the optical fiber tip and the inductor of the aluminum resonator. (Top) A cross-sectional 3D rendering of the KIPM detector chip, enclosure, and fiber connection. Note the chip geometry has been vastly simplified to only show the aluminum resonator's inductor as a monolithic layer. The thickness of this layer has been exaggerated for visibility. (Bottom left) A picture of the KIPM detector chip in its housing (rear side is facing up) and the lid with the optical fiber penetration. (Bottom right) A picture of the KIPM detector installed in the NEXUS dilution refrigerator showing both the RF input and output lines as well as the optical fiber connection.}
    \label{fig:fiber-position}
\end{figure}

Fig.~\ref{fig:wiring-diagram} shows the in-refrigerator wiring diagram for electrical and optical lines as well as a simplified schematic of the warm electronics used for RF readout and LED control. Fig.~\ref{fig:fiber-position} shows a 3D rendering of the KIPM detector in its enclosure and the relative positioning of the optical fiber. In this setup, the optical photons impinge upon the surface of the bare substrate opposite the inductor of the Al resonator at the center of the chip, leaving only two narrow slots (visible in Fig.~\ref{fig:device}) for reflected photons to enter the side of the enclosure where the superconductor is exposed.

\section{\label{supp8-directhits}Constraining Direct Impacts on Inductor}
One may be concerned about the effect of photons directly incident on the inductor rather than on the substrate, as energy directly deposited in the inductor is much more efficient at breaking Cooper Pairs than energy deposited in the substrate. Given our $\eta_\mathrm{ph}$ of $\sim$1\%, direct photon hits on the inductor produce a $\sim$100$\times$ larger signal than hits in the substrate, for the same total energy. For direct photon impacts to be negligible in this calibration, the efficiency for direct hits must be small compared to $\eta_\mathrm{ph}$, or $\ll 1\%$. 
Due to the geometry of the device, its enclosure, and fiber position, this is a very inefficient process (see Figs~\ref{fig:device} and~\ref{fig:fiber-position}), as motivated below. 

For context, in our largest LED voltage setting (most light produced), approximately 42 keV is deposited in the chip through photons, of which the resonator collects roughly 420 eV in phonons. A single direct photon impact would impart 2.6 keV into the superconductor. Two direct photon hits would correspond to an energy of roughly 1\% the amount deposited via phonons. 42 keV corresponds to roughly $1.6\times10^4$ photons, so to have an order unity number of photons impinge directly on the superconductor requires an efficiency at the $10^{-4}$ scale.

We constrain the contribution to our signal from photons directly incident on the inductor with three methods: (1) investigating the rise time of our observed pulse shape; and (2) a back-of-the envelope estimate on intensity of reflected light. 
We discuss these methods below.

\subsection{Pulse Rise Times}

\begin{figure*}
    \centering
    \includegraphics[width=0.49\textwidth]{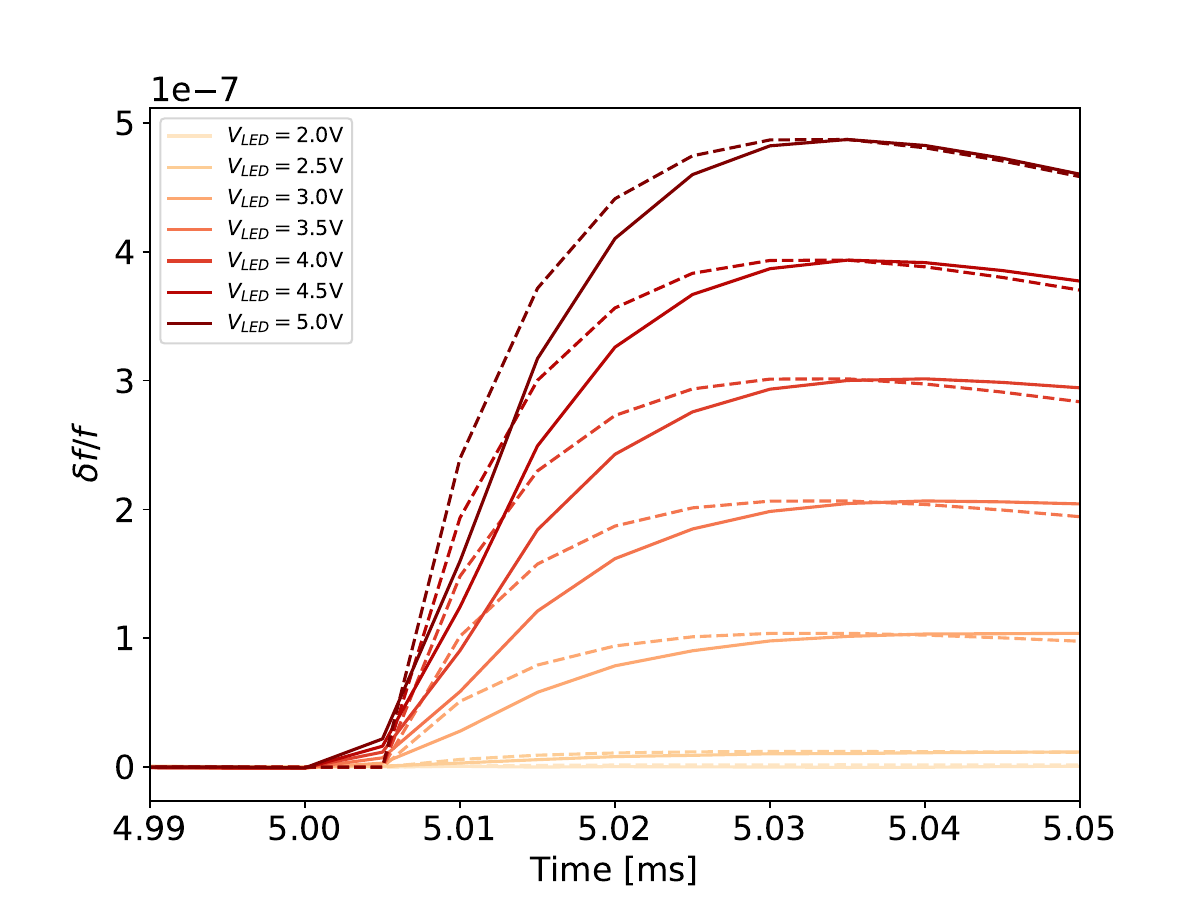}
    ~
    \includegraphics[width=0.49\textwidth]{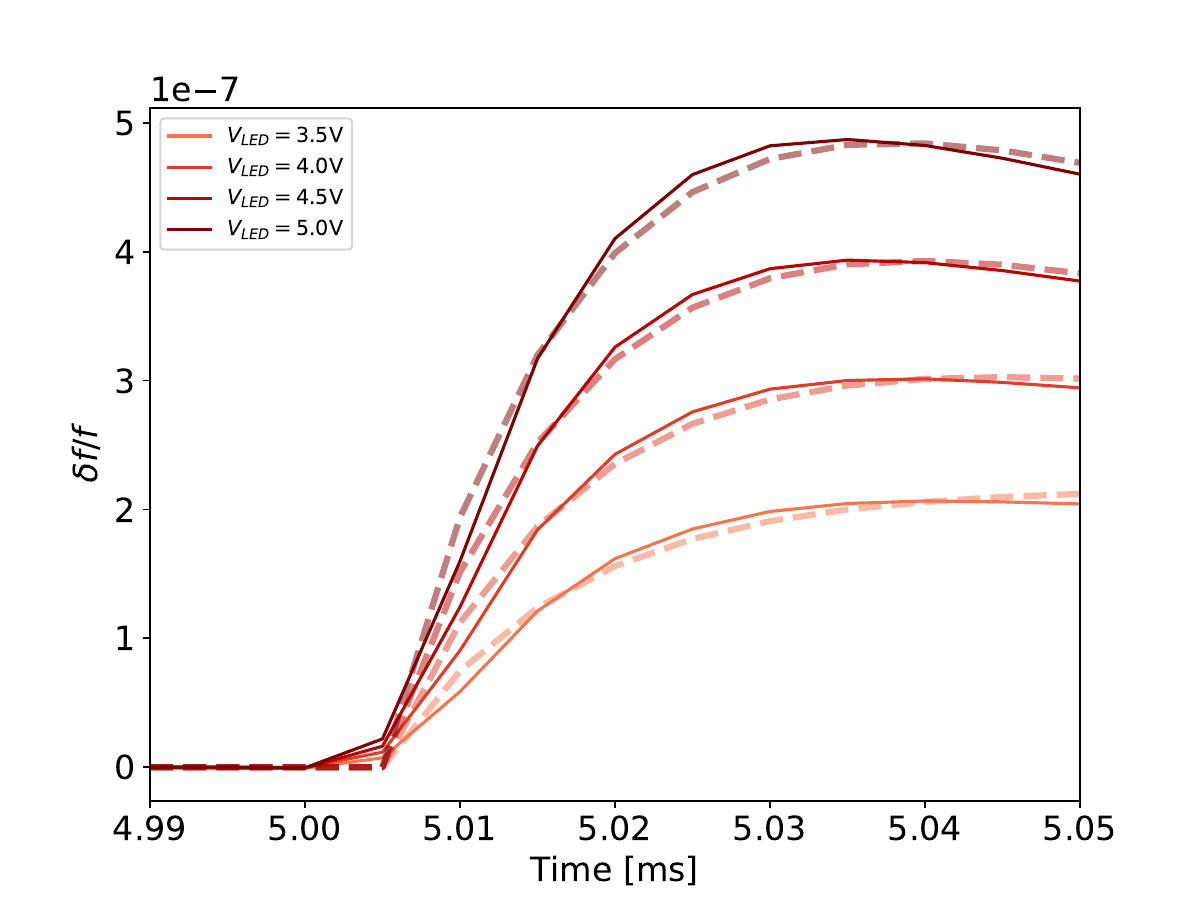}
    \caption{The rising edges of the average pulses observed for various LED powers (solid lines). (Left) Comparing the average pulses to the pulse shape expected for a delta-function deposition of energy in the inductor (dashed lines), where the rise time is set completely by the ring time of the resonator of interest: $\tau_\mathrm{ring}=9.3~\mu$s. (Right) Fits of each average pulse to the delta-function impact model, from which we extract a resonator ring time that is $>30\%$ larger than that for the resonator of interest.}
    \label{fig:pulse-rise}
\end{figure*}

Given the finite rise time constant of the observed pulses (time-to-peak $>35~\mu$s), we can be confident that the signal is dominated by the phonon channel. In Fig~\ref{fig:pulse-rise} (left), we compare the rise time of our observed pulses (solid lines) with that of an analytic pulse shape derived assuming a delta-function impact of energy in the inductor (as from a photon burst), smeared by the ring time of the resonator, $\tau_\mathrm{ring} = Q/\omega$ (dashed lines). In this plot, for all voltages, we have set the resonator ring time to 9.3 $\mu$s, as is expected for our resonator’s parameters, and the fall time constant to be 150 $\mu$s. We have fixed the amplitude of the analytic pulse shape to match that of the observed data. In all cases, the analytic pulse shape rises faster than the data, indicating that the rise time is due to phonon arrival, rather than an effect from direct photon impacts convolved with the resonator ring time. 

It is important to note that as phonons or photons are absorbed, the $Q$ value decreases, in turn shortening the ring time of the resonator, leading to a faster rising edge than is seen in the plot for this model.

Next, in Fig.~\ref{fig:pulse-rise} (right), we fit the average pulses for each LED voltage to the delta-function impact pulse model and extract the resonator ring time (assuming all energy comes in the form of direct photon hits). For the four largest LED voltages (the fit routine would not converge for this model on the lower LED powers), this model selects resonator ring times in the 12--15 $\mu$s range. However, notice that the fit to this model wants the peak of the pulse to occur at later times than we see. Under the assumption that the pulses in the data come from direct photon impacts on the resonator, the ring time would need to be even longer than these values to correctly reconstruct the time of the pulse maximum. Note that this model does not capture the falling-edge behavior of the pulse shape well at all, hence the truncation of the fits shortly after the peak.

Both of these checks motivate resonator ring times that are over 30\% longer than expected for our resonator (9.3 $\mu$s).


\subsection{Reflected Light Intensity Estimates}

In the following, we perform a back-of-the-envelope estimate on the intensity of un-absorbed light after some number of reflections. We make the following assumptions.
\begin{enumerate}
    \item All photons emitted from the fiber first are incident on the bare Si. Due to the numerical aperture of the fiber and the distance from the fiber to the substrate, the spot size is significantly smaller than the substrate dimension, which makes this a robust assumption.
    \item All photons reflected from the Si are next incident on the gold-plated copper. Given the shape of the Si chip, there is no physical way a photon reflected from the surface could later impinge back upon the Si without first reflecting off the gold-plated copper.
    \item All photons reflected from gold-plated copper are next incident on Si (and so on). This is a conservative assumption in this estimate, as the reflectance of Si is higher than that of Au (see below), so an Au-Au reflection pair has a lower final intensity than an Au-Si reflection pair.
    \item For 475 nm light, the reflectance of Si is $r_\mathrm{Si}=40\%$~\cite{Green2008}, and the reflectance of Au is $r_\mathrm{Au}=30\%$~\cite{Loebich1972}.
\end{enumerate}

Assuming the intensity of light first incident on the Si is $I_0$, after the first reflection, the intensity of un-absorbed light is $r_\mathrm{Si} I_0$. This intensity of light is now incident on gold-plated copper, and after being reflected has an intensity $r_\mathrm{Si} r_\mathrm{Au} I_0$. Each pair of reflections adds a factor ($r_\mathrm{Si} r_\mathrm{Au}$), thus after $n$ pairs of reflections, the ratio of the un-absorbed light intensity to the initial intensity is $(r_\mathrm{Si} r_\mathrm{Au})^n$. After three(ten) pairs of reflections, the un-absorbed intensity is less than 0.2\%(1 ppb) the initial intensity incident on the Si. Thus,  the condition that direct photon impact efficiency is small compared to 1\% is met after 3--4 pairs of reflections.

The geometry of the enclosure, seen in Figs~\ref{fig:device} and~\ref{fig:fiber-position}, makes it inefficient for photons to impinge upon the inductor directly without numerous reflections. Small grooves on either side of the chip ($\sim$3.5 mm $\times$ $\sim$1.6 cm) are the only paths for light to make it from the rear face of the chip to the front. From a solid-angle consideration, the amount of  light reflected from the Si that is later incident on the inductor can be estimated to be $r_\mathrm{Si} \Omega_\mathrm{open} \Omega_\mathrm{inductor} I_0$, where $\Omega_\mathrm{open}$ is the fraction of the solid angle subtended by the open slots on either side of the chips and $\Omega_\mathrm{inductor}$ is the ratio of areas of the inductor to the Si face. While $\Omega_\mathrm{open}$ is slightly more tedious to estimate, $\Omega_\mathrm{inductor}$ is known well from the geometry of our device (0.81 mm$^2$ / 4.84 cm$^2$). However, the product $r_\mathrm{Si} \Omega_\mathrm{inductor}$ evaluates to $\sim7 \times 10^{-4}$, (and factoring in $\Omega_\mathrm{open}$ will only diminish it further), highlighting the inefficiency for this process.


Light transmitted through the chip to be directly incident on the superconductor is also negligible. The absorption length for 475 nm light in silicon is about 1 $\mu$m~\cite{Green2008}. Thus, at a depth of 1 $\mu$m into the substrate only 36\% of the initial light intensity remains unabsorbed. After 10 $\mu$m, or 1\% of the total substrate thickness, the intensity of unabsorbed light is reduced to 45 ppm. 



\begin{thebibliography}{60}%
\makeatletter
\providecommand \@ifxundefined [1]{%
 \@ifx{#1\undefined}
}%
\providecommand \@ifnum [1]{%
 \ifnum #1\expandafter \@firstoftwo
 \else \expandafter \@secondoftwo
 \fi
}%
\providecommand \@ifx [1]{%
 \ifx #1\expandafter \@firstoftwo
 \else \expandafter \@secondoftwo
 \fi
}%
\providecommand \natexlab [1]{#1}%
\providecommand \enquote  [1]{``#1''}%
\providecommand \bibnamefont  [1]{#1}%
\providecommand \bibfnamefont [1]{#1}%
\providecommand \citenamefont [1]{#1}%
\providecommand \href@noop [0]{\@secondoftwo}%
\providecommand \href [0]{\begingroup \@sanitize@url \@href}%
\providecommand \@href[1]{\@@startlink{#1}\@@href}%
\providecommand \@@href[1]{\endgroup#1\@@endlink}%
\providecommand \@sanitize@url [0]{\catcode `\\12\catcode `\$12\catcode `\&12\catcode `\#12\catcode `\^12\catcode `\_12\catcode `\%12\relax}%
\providecommand \@@startlink[1]{}%
\providecommand \@@endlink[0]{}%
\providecommand \url  [0]{\begingroup\@sanitize@url \@url }%
\providecommand \@url [1]{\endgroup\@href {#1}{\urlprefix }}%
\providecommand \urlprefix  [0]{URL }%
\providecommand \Eprint [0]{\href }%
\providecommand \doibase [0]{https://doi.org/}%
\providecommand \selectlanguage [0]{\@gobble}%
\providecommand \bibinfo  [0]{\@secondoftwo}%
\providecommand \bibfield  [0]{\@secondoftwo}%
\providecommand \translation [1]{[#1]}%
\providecommand \BibitemOpen [0]{}%
\providecommand \bibitemStop [0]{}%
\providecommand \bibitemNoStop [0]{.\EOS\space}%
\providecommand \EOS [0]{\spacefactor3000\relax}%
\providecommand \BibitemShut  [1]{\csname bibitem#1\endcsname}%
\let\auto@bib@innerbib\@empty
\bibitem [{\citenamefont {Cooley}\ \emph {et~al.}(2022)\citenamefont {Cooley}, \citenamefont {Lin}, \citenamefont {Lippincott}, \citenamefont {Slatyer}, \citenamefont {Yu}, \citenamefont {Akerib}, \citenamefont {Aramaki}, \citenamefont {Baxter}, \citenamefont {Bringmann}, \citenamefont {Bunker}, \citenamefont {Carney}, \citenamefont {Cebrián}, \citenamefont {Chen}, \citenamefont {Cushman}, \citenamefont {Dahl}, \citenamefont {Essig}, \citenamefont {Fan}, \citenamefont {Gaitskell}, \citenamefont {Galbiati}, \citenamefont {Gelmini}, \citenamefont {Giovanetti}, \citenamefont {Giroux}, \citenamefont {Grandi}, \citenamefont {Harding}, \citenamefont {Haselschwardt}, \citenamefont {Hsu}, \citenamefont {Horiuchi}, \citenamefont {Kahn}, \citenamefont {Kim}, \citenamefont {Kim}, \citenamefont {Kravitz}, \citenamefont {Kudryavtsev}, \citenamefont {Kurinsky}, \citenamefont {Lang}, \citenamefont {Leane}, \citenamefont {Lehmann}, \citenamefont {Levy}, \citenamefont {Li}, \citenamefont {Loer}, \citenamefont {Manalaysay},
  \citenamefont {Martoff}, \citenamefont {Mohlabeng}, \citenamefont {Monzani}, \citenamefont {Murphy}, \citenamefont {Neilson}, \citenamefont {Nelson}, \citenamefont {O'Hare}, \citenamefont {Palladino}, \citenamefont {Parikh}, \citenamefont {Park}, \citenamefont {Perez}, \citenamefont {Profumo}, \citenamefont {Raj}, \citenamefont {Roach}, \citenamefont {Saab}, \citenamefont {Sarsa}, \citenamefont {Schnee}, \citenamefont {Shaw}, \citenamefont {Shin}, \citenamefont {Sinha}, \citenamefont {Stifter}, \citenamefont {Suzuki}, \citenamefont {Szydagis}, \citenamefont {Tait}, \citenamefont {Takhistov}, \citenamefont {Tsai}, \citenamefont {Vahsen}, \citenamefont {Vitagliano}, \citenamefont {von Doetinchem}, \citenamefont {Wang}, \citenamefont {Westerdale}, \citenamefont {Williams}, \citenamefont {Xiang},\ and\ \citenamefont {Yang}}]{Cooley2022}%
  \BibitemOpen
  \bibfield  {author} {\bibinfo {author} {\bibfnamefont {J.}~\bibnamefont {Cooley}}, \bibinfo {author} {\bibfnamefont {T.}~\bibnamefont {Lin}}, \bibinfo {author} {\bibfnamefont {W.~H.}\ \bibnamefont {Lippincott}}, \bibinfo {author} {\bibfnamefont {T.~R.}\ \bibnamefont {Slatyer}}, \bibinfo {author} {\bibfnamefont {T.-T.}\ \bibnamefont {Yu}}, \bibinfo {author} {\bibfnamefont {D.~S.}\ \bibnamefont {Akerib}}, \bibinfo {author} {\bibfnamefont {T.}~\bibnamefont {Aramaki}}, \bibinfo {author} {\bibfnamefont {D.}~\bibnamefont {Baxter}}, \bibinfo {author} {\bibfnamefont {T.}~\bibnamefont {Bringmann}}, \bibinfo {author} {\bibfnamefont {R.}~\bibnamefont {Bunker}}, \bibinfo {author} {\bibfnamefont {D.}~\bibnamefont {Carney}}, \bibinfo {author} {\bibfnamefont {S.}~\bibnamefont {Cebrián}}, \bibinfo {author} {\bibfnamefont {T.~Y.}\ \bibnamefont {Chen}}, \bibinfo {author} {\bibfnamefont {P.}~\bibnamefont {Cushman}}, \bibinfo {author} {\bibfnamefont {C.~E.}\ \bibnamefont {Dahl}}, \bibinfo {author} {\bibfnamefont
  {R.}~\bibnamefont {Essig}}, \bibinfo {author} {\bibfnamefont {A.}~\bibnamefont {Fan}}, \bibinfo {author} {\bibfnamefont {R.}~\bibnamefont {Gaitskell}}, \bibinfo {author} {\bibfnamefont {C.}~\bibnamefont {Galbiati}}, \bibinfo {author} {\bibfnamefont {G.~B.}\ \bibnamefont {Gelmini}}, \bibinfo {author} {\bibfnamefont {G.~K.}\ \bibnamefont {Giovanetti}}, \bibinfo {author} {\bibfnamefont {G.}~\bibnamefont {Giroux}}, \bibinfo {author} {\bibfnamefont {L.}~\bibnamefont {Grandi}}, \bibinfo {author} {\bibfnamefont {J.~P.}\ \bibnamefont {Harding}}, \bibinfo {author} {\bibfnamefont {S.}~\bibnamefont {Haselschwardt}}, \bibinfo {author} {\bibfnamefont {L.}~\bibnamefont {Hsu}}, \bibinfo {author} {\bibfnamefont {S.}~\bibnamefont {Horiuchi}}, \bibinfo {author} {\bibfnamefont {Y.}~\bibnamefont {Kahn}}, \bibinfo {author} {\bibfnamefont {D.}~\bibnamefont {Kim}}, \bibinfo {author} {\bibfnamefont {G.-B.}\ \bibnamefont {Kim}}, \bibinfo {author} {\bibfnamefont {S.}~\bibnamefont {Kravitz}}, \bibinfo {author} {\bibfnamefont {V.~A.}\
  \bibnamefont {Kudryavtsev}}, \bibinfo {author} {\bibfnamefont {N.}~\bibnamefont {Kurinsky}}, \bibinfo {author} {\bibfnamefont {R.~F.}\ \bibnamefont {Lang}}, \bibinfo {author} {\bibfnamefont {R.~K.}\ \bibnamefont {Leane}}, \bibinfo {author} {\bibfnamefont {B.~V.}\ \bibnamefont {Lehmann}}, \bibinfo {author} {\bibfnamefont {C.}~\bibnamefont {Levy}}, \bibinfo {author} {\bibfnamefont {S.}~\bibnamefont {Li}}, \bibinfo {author} {\bibfnamefont {B.}~\bibnamefont {Loer}}, \bibinfo {author} {\bibfnamefont {A.}~\bibnamefont {Manalaysay}}, \bibinfo {author} {\bibfnamefont {C.~J.}\ \bibnamefont {Martoff}}, \bibinfo {author} {\bibfnamefont {G.}~\bibnamefont {Mohlabeng}}, \bibinfo {author} {\bibfnamefont {M.~E.}\ \bibnamefont {Monzani}}, \bibinfo {author} {\bibfnamefont {A.~S.~J.}\ \bibnamefont {Murphy}}, \bibinfo {author} {\bibfnamefont {R.}~\bibnamefont {Neilson}}, \bibinfo {author} {\bibfnamefont {H.~N.}\ \bibnamefont {Nelson}}, \bibinfo {author} {\bibfnamefont {C.~A.~J.}\ \bibnamefont {O'Hare}}, \bibinfo {author}
  {\bibfnamefont {K.~J.}\ \bibnamefont {Palladino}}, \bibinfo {author} {\bibfnamefont {A.}~\bibnamefont {Parikh}}, \bibinfo {author} {\bibfnamefont {J.-C.}\ \bibnamefont {Park}}, \bibinfo {author} {\bibfnamefont {K.}~\bibnamefont {Perez}}, \bibinfo {author} {\bibfnamefont {S.}~\bibnamefont {Profumo}}, \bibinfo {author} {\bibfnamefont {N.}~\bibnamefont {Raj}}, \bibinfo {author} {\bibfnamefont {B.~M.}\ \bibnamefont {Roach}}, \bibinfo {author} {\bibfnamefont {T.}~\bibnamefont {Saab}}, \bibinfo {author} {\bibfnamefont {M.~L.}\ \bibnamefont {Sarsa}}, \bibinfo {author} {\bibfnamefont {R.}~\bibnamefont {Schnee}}, \bibinfo {author} {\bibfnamefont {S.}~\bibnamefont {Shaw}}, \bibinfo {author} {\bibfnamefont {S.}~\bibnamefont {Shin}}, \bibinfo {author} {\bibfnamefont {K.}~\bibnamefont {Sinha}}, \bibinfo {author} {\bibfnamefont {K.}~\bibnamefont {Stifter}}, \bibinfo {author} {\bibfnamefont {A.}~\bibnamefont {Suzuki}}, \bibinfo {author} {\bibfnamefont {M.}~\bibnamefont {Szydagis}}, \bibinfo {author} {\bibfnamefont
  {T.~M.~P.}\ \bibnamefont {Tait}}, \bibinfo {author} {\bibfnamefont {V.}~\bibnamefont {Takhistov}}, \bibinfo {author} {\bibfnamefont {Y.-D.}\ \bibnamefont {Tsai}}, \bibinfo {author} {\bibfnamefont {S.~E.}\ \bibnamefont {Vahsen}}, \bibinfo {author} {\bibfnamefont {E.}~\bibnamefont {Vitagliano}}, \bibinfo {author} {\bibfnamefont {P.}~\bibnamefont {von Doetinchem}}, \bibinfo {author} {\bibfnamefont {G.}~\bibnamefont {Wang}}, \bibinfo {author} {\bibfnamefont {S.}~\bibnamefont {Westerdale}}, \bibinfo {author} {\bibfnamefont {D.~A.}\ \bibnamefont {Williams}}, \bibinfo {author} {\bibfnamefont {X.}~\bibnamefont {Xiang}},\ and\ \bibinfo {author} {\bibfnamefont {L.}~\bibnamefont {Yang}},\ }\bibfield  {title} {\bibinfo {title} {Report of the topical group on particle dark matter for snowmass 2021},\ }\href@noop {} {\  (\bibinfo {year} {2022})},\ \Eprint {https://arxiv.org/abs/2209.07426} {arXiv:2209.07426 [hep-ph]} \BibitemShut {NoStop}%
\bibitem [{\citenamefont {Essig}\ \emph {et~al.}(2022)\citenamefont {Essig}, \citenamefont {Giovanetti}, \citenamefont {Kurinsky}, \citenamefont {McKinsey}, \citenamefont {Ramanathan}, \citenamefont {Stifter}, \citenamefont {Yu}, \citenamefont {Aboubrahim}, \citenamefont {Adams}, \citenamefont {Alves}, \citenamefont {Aralis}, \citenamefont {Araújo}, \citenamefont {Baxter}, \citenamefont {Berghaus}, \citenamefont {Berlin}, \citenamefont {Blanco}, \citenamefont {Bloch}, \citenamefont {Bonivento}, \citenamefont {Bunker}, \citenamefont {Burdin}, \citenamefont {Caminata}, \citenamefont {Carmona-Benitez}, \citenamefont {Chaplinsky}, \citenamefont {Chen}, \citenamefont {Derenzo}, \citenamefont {de~Viveiros}, \citenamefont {Dick}, \citenamefont {Marco}, \citenamefont {Du}, \citenamefont {Dutta}, \citenamefont {Ebadi}, \citenamefont {Emken}, \citenamefont {Esposito}, \citenamefont {Etzion}, \citenamefont {Feng}, \citenamefont {Fernandez}, \citenamefont {Ge}, \citenamefont {Ghosh}, \citenamefont {Giroux},
  \citenamefont {Hamaide}, \citenamefont {Hertel}, \citenamefont {Herrera}, \citenamefont {Hochberg}, \citenamefont {Kahn}, \citenamefont {Kavanagh}, \citenamefont {Khan}, \citenamefont {Kluck}, \citenamefont {Kravitz}, \citenamefont {von Krosigk}, \citenamefont {Kumar}, \citenamefont {Lawson}, \citenamefont {Lehmann}, \citenamefont {Lin}, \citenamefont {Liao}, \citenamefont {Lyon}, \citenamefont {Majewski}, \citenamefont {Manzari}, \citenamefont {Monroe}, \citenamefont {Monzani}, \citenamefont {Morrissey}, \citenamefont {Norcini}, \citenamefont {Orly}, \citenamefont {Parikh}, \citenamefont {Park}, \citenamefont {Patel}, \citenamefont {Paul}, \citenamefont {Pèrez-Ríos}, \citenamefont {Phipps}, \citenamefont {Pocar}, \citenamefont {Ritz}, \citenamefont {Sarkis}, \citenamefont {Schuster}, \citenamefont {Schwetz}, \citenamefont {Shaw}, \citenamefont {Shin}, \citenamefont {Singal}, \citenamefont {Singh}, \citenamefont {Slone}, \citenamefont {Sorensen}, \citenamefont {Sun}, \citenamefont {Szydagis},
  \citenamefont {Temples}, \citenamefont {Testera}, \citenamefont {Thieme}, \citenamefont {Toro}, \citenamefont {Trickle}, \citenamefont {Uemura}, \citenamefont {Velan}, \citenamefont {Vitagliano}, \citenamefont {Wagner}, \citenamefont {Wang}, \citenamefont {Westerdale},\ and\ \citenamefont {Zurek}}]{Essig2022}%
  \BibitemOpen
  \bibfield  {author} {\bibinfo {author} {\bibfnamefont {R.}~\bibnamefont {Essig}}, \bibinfo {author} {\bibfnamefont {G.~K.}\ \bibnamefont {Giovanetti}}, \bibinfo {author} {\bibfnamefont {N.}~\bibnamefont {Kurinsky}}, \bibinfo {author} {\bibfnamefont {D.}~\bibnamefont {McKinsey}}, \bibinfo {author} {\bibfnamefont {K.}~\bibnamefont {Ramanathan}}, \bibinfo {author} {\bibfnamefont {K.}~\bibnamefont {Stifter}}, \bibinfo {author} {\bibfnamefont {T.-T.}\ \bibnamefont {Yu}}, \bibinfo {author} {\bibfnamefont {A.}~\bibnamefont {Aboubrahim}}, \bibinfo {author} {\bibfnamefont {D.}~\bibnamefont {Adams}}, \bibinfo {author} {\bibfnamefont {D.~S.~M.}\ \bibnamefont {Alves}}, \bibinfo {author} {\bibfnamefont {T.}~\bibnamefont {Aralis}}, \bibinfo {author} {\bibfnamefont {H.~M.}\ \bibnamefont {Araújo}}, \bibinfo {author} {\bibfnamefont {D.}~\bibnamefont {Baxter}}, \bibinfo {author} {\bibfnamefont {K.~V.}\ \bibnamefont {Berghaus}}, \bibinfo {author} {\bibfnamefont {A.}~\bibnamefont {Berlin}}, \bibinfo {author} {\bibfnamefont
  {C.}~\bibnamefont {Blanco}}, \bibinfo {author} {\bibfnamefont {I.~M.}\ \bibnamefont {Bloch}}, \bibinfo {author} {\bibfnamefont {W.~M.}\ \bibnamefont {Bonivento}}, \bibinfo {author} {\bibfnamefont {R.}~\bibnamefont {Bunker}}, \bibinfo {author} {\bibfnamefont {S.}~\bibnamefont {Burdin}}, \bibinfo {author} {\bibfnamefont {A.}~\bibnamefont {Caminata}}, \bibinfo {author} {\bibfnamefont {M.~C.}\ \bibnamefont {Carmona-Benitez}}, \bibinfo {author} {\bibfnamefont {L.}~\bibnamefont {Chaplinsky}}, \bibinfo {author} {\bibfnamefont {T.~Y.}\ \bibnamefont {Chen}}, \bibinfo {author} {\bibfnamefont {S.~E.}\ \bibnamefont {Derenzo}}, \bibinfo {author} {\bibfnamefont {L.}~\bibnamefont {de~Viveiros}}, \bibinfo {author} {\bibfnamefont {R.}~\bibnamefont {Dick}}, \bibinfo {author} {\bibfnamefont {N.~D.}\ \bibnamefont {Marco}}, \bibinfo {author} {\bibfnamefont {P.}~\bibnamefont {Du}}, \bibinfo {author} {\bibfnamefont {B.}~\bibnamefont {Dutta}}, \bibinfo {author} {\bibfnamefont {R.}~\bibnamefont {Ebadi}}, \bibinfo {author}
  {\bibfnamefont {T.}~\bibnamefont {Emken}}, \bibinfo {author} {\bibfnamefont {A.}~\bibnamefont {Esposito}}, \bibinfo {author} {\bibfnamefont {E.}~\bibnamefont {Etzion}}, \bibinfo {author} {\bibfnamefont {J.~L.}\ \bibnamefont {Feng}}, \bibinfo {author} {\bibfnamefont {N.}~\bibnamefont {Fernandez}}, \bibinfo {author} {\bibfnamefont {S.~F.}\ \bibnamefont {Ge}}, \bibinfo {author} {\bibfnamefont {S.}~\bibnamefont {Ghosh}}, \bibinfo {author} {\bibfnamefont {G.}~\bibnamefont {Giroux}}, \bibinfo {author} {\bibfnamefont {L.}~\bibnamefont {Hamaide}}, \bibinfo {author} {\bibfnamefont {S.~A.}\ \bibnamefont {Hertel}}, \bibinfo {author} {\bibfnamefont {G.}~\bibnamefont {Herrera}}, \bibinfo {author} {\bibfnamefont {Y.}~\bibnamefont {Hochberg}}, \bibinfo {author} {\bibfnamefont {Y.}~\bibnamefont {Kahn}}, \bibinfo {author} {\bibfnamefont {B.~J.}\ \bibnamefont {Kavanagh}}, \bibinfo {author} {\bibfnamefont {A.~N.}\ \bibnamefont {Khan}}, \bibinfo {author} {\bibfnamefont {H.}~\bibnamefont {Kluck}}, \bibinfo {author}
  {\bibfnamefont {S.}~\bibnamefont {Kravitz}}, \bibinfo {author} {\bibfnamefont {B.}~\bibnamefont {von Krosigk}}, \bibinfo {author} {\bibfnamefont {J.}~\bibnamefont {Kumar}}, \bibinfo {author} {\bibfnamefont {I.~T.}\ \bibnamefont {Lawson}}, \bibinfo {author} {\bibfnamefont {B.~V.}\ \bibnamefont {Lehmann}}, \bibinfo {author} {\bibfnamefont {T.}~\bibnamefont {Lin}}, \bibinfo {author} {\bibfnamefont {J.}~\bibnamefont {Liao}}, \bibinfo {author} {\bibfnamefont {S.~A.}\ \bibnamefont {Lyon}}, \bibinfo {author} {\bibfnamefont {P.~M.}\ \bibnamefont {Majewski}}, \bibinfo {author} {\bibfnamefont {C.~A.}\ \bibnamefont {Manzari}}, \bibinfo {author} {\bibfnamefont {J.}~\bibnamefont {Monroe}}, \bibinfo {author} {\bibfnamefont {M.~E.}\ \bibnamefont {Monzani}}, \bibinfo {author} {\bibfnamefont {D.~E.}\ \bibnamefont {Morrissey}}, \bibinfo {author} {\bibfnamefont {D.}~\bibnamefont {Norcini}}, \bibinfo {author} {\bibfnamefont {A.}~\bibnamefont {Orly}}, \bibinfo {author} {\bibfnamefont {A.}~\bibnamefont {Parikh}}, \bibinfo
  {author} {\bibfnamefont {J.~C.}\ \bibnamefont {Park}}, \bibinfo {author} {\bibfnamefont {P.~K.}\ \bibnamefont {Patel}}, \bibinfo {author} {\bibfnamefont {S.}~\bibnamefont {Paul}}, \bibinfo {author} {\bibfnamefont {J.}~\bibnamefont {Pèrez-Ríos}}, \bibinfo {author} {\bibfnamefont {A.}~\bibnamefont {Phipps}}, \bibinfo {author} {\bibfnamefont {A.}~\bibnamefont {Pocar}}, \bibinfo {author} {\bibfnamefont {A.}~\bibnamefont {Ritz}}, \bibinfo {author} {\bibfnamefont {Y.}~\bibnamefont {Sarkis}}, \bibinfo {author} {\bibfnamefont {P.}~\bibnamefont {Schuster}}, \bibinfo {author} {\bibfnamefont {T.}~\bibnamefont {Schwetz}}, \bibinfo {author} {\bibfnamefont {S.}~\bibnamefont {Shaw}}, \bibinfo {author} {\bibfnamefont {S.}~\bibnamefont {Shin}}, \bibinfo {author} {\bibfnamefont {A.}~\bibnamefont {Singal}}, \bibinfo {author} {\bibfnamefont {R.}~\bibnamefont {Singh}}, \bibinfo {author} {\bibfnamefont {O.}~\bibnamefont {Slone}}, \bibinfo {author} {\bibfnamefont {P.}~\bibnamefont {Sorensen}}, \bibinfo {author} {\bibfnamefont
  {C.}~\bibnamefont {Sun}}, \bibinfo {author} {\bibfnamefont {M.}~\bibnamefont {Szydagis}}, \bibinfo {author} {\bibfnamefont {D.~J.}\ \bibnamefont {Temples}}, \bibinfo {author} {\bibfnamefont {G.}~\bibnamefont {Testera}}, \bibinfo {author} {\bibfnamefont {K.}~\bibnamefont {Thieme}}, \bibinfo {author} {\bibfnamefont {N.}~\bibnamefont {Toro}}, \bibinfo {author} {\bibfnamefont {T.}~\bibnamefont {Trickle}}, \bibinfo {author} {\bibfnamefont {S.}~\bibnamefont {Uemura}}, \bibinfo {author} {\bibfnamefont {V.}~\bibnamefont {Velan}}, \bibinfo {author} {\bibfnamefont {E.}~\bibnamefont {Vitagliano}}, \bibinfo {author} {\bibfnamefont {F.}~\bibnamefont {Wagner}}, \bibinfo {author} {\bibfnamefont {G.}~\bibnamefont {Wang}}, \bibinfo {author} {\bibfnamefont {S.}~\bibnamefont {Westerdale}},\ and\ \bibinfo {author} {\bibfnamefont {K.~M.}\ \bibnamefont {Zurek}},\ }\bibfield  {title} {\bibinfo {title} {Snowmass2021 cosmic frontier: The landscape of low-threshold dark matter direct detection in the next decade},\ }\href@noop {} {\
   (\bibinfo {year} {2022})},\ \Eprint {https://arxiv.org/abs/2203.08297} {arXiv:2203.08297 [hep-ph]} \BibitemShut {NoStop}%
\bibitem [{\citenamefont {Golwala}\ \emph {et~al.}(2008)\citenamefont {Golwala}, \citenamefont {Gao}, \citenamefont {Moore}, \citenamefont {Mazin}, \citenamefont {Eckart}, \citenamefont {Bumble}, \citenamefont {Day}, \citenamefont {LeDuc},\ and\ \citenamefont {Zmuidzinas}}]{Golwala2008}%
  \BibitemOpen
  \bibfield  {author} {\bibinfo {author} {\bibfnamefont {S.}~\bibnamefont {Golwala}}, \bibinfo {author} {\bibfnamefont {J.}~\bibnamefont {Gao}}, \bibinfo {author} {\bibfnamefont {D.}~\bibnamefont {Moore}}, \bibinfo {author} {\bibfnamefont {B.}~\bibnamefont {Mazin}}, \bibinfo {author} {\bibfnamefont {M.}~\bibnamefont {Eckart}}, \bibinfo {author} {\bibfnamefont {B.}~\bibnamefont {Bumble}}, \bibinfo {author} {\bibfnamefont {P.}~\bibnamefont {Day}}, \bibinfo {author} {\bibfnamefont {H.~G.}\ \bibnamefont {LeDuc}},\ and\ \bibinfo {author} {\bibfnamefont {J.}~\bibnamefont {Zmuidzinas}},\ }\bibfield  {title} {\bibinfo {title} {A {WIMP} dark matter detector using {MKIDs}},\ }\href {https://doi.org/10.1007/s10909-007-9687-0} {\bibfield  {journal} {\bibinfo  {journal} {Journal of Low Temperature Physics}\ }\textbf {\bibinfo {volume} {151}},\ \bibinfo {pages} {550} (\bibinfo {year} {2008})}\BibitemShut {NoStop}%
\bibitem [{\citenamefont {Moore}\ \emph {et~al.}(2012)\citenamefont {Moore}, \citenamefont {Golwala}, \citenamefont {Bumble}, \citenamefont {Cornell}, \citenamefont {Day}, \citenamefont {LeDuc},\ and\ \citenamefont {Zmuidzinas}}]{Moore2012}%
  \BibitemOpen
  \bibfield  {author} {\bibinfo {author} {\bibfnamefont {D.~C.}\ \bibnamefont {Moore}}, \bibinfo {author} {\bibfnamefont {S.~R.}\ \bibnamefont {Golwala}}, \bibinfo {author} {\bibfnamefont {B.}~\bibnamefont {Bumble}}, \bibinfo {author} {\bibfnamefont {B.}~\bibnamefont {Cornell}}, \bibinfo {author} {\bibfnamefont {P.~K.}\ \bibnamefont {Day}}, \bibinfo {author} {\bibfnamefont {H.~G.}\ \bibnamefont {LeDuc}},\ and\ \bibinfo {author} {\bibfnamefont {J.}~\bibnamefont {Zmuidzinas}},\ }\bibfield  {title} {\bibinfo {title} {Position and energy-resolved particle detection using phonon-mediated microwave kinetic inductance detectors},\ }\href {https://doi.org/10.1063/1.4726279} {\bibfield  {journal} {\bibinfo  {journal} {Applied Physics Letters}\ }\textbf {\bibinfo {volume} {100}},\ \bibinfo {pages} {232601} (\bibinfo {year} {2012})},\ \Eprint {https://arxiv.org/abs/1203.4549} {arXiv:1203.4549 [astro-ph.IM]} \BibitemShut {NoStop}%
\bibitem [{\citenamefont {Cardani}\ \emph {et~al.}(2018)\citenamefont {Cardani}, \citenamefont {Casali}, \citenamefont {Cruciani}, \citenamefont {le~Sueur}, \citenamefont {Martinez}, \citenamefont {Bellini}, \citenamefont {Calvo}, \citenamefont {Castellano}, \citenamefont {Colantoni}, \citenamefont {Cosmelli}, \citenamefont {D'Addabbo}, \citenamefont {Domizio}, \citenamefont {Goupy}, \citenamefont {Minutolo}, \citenamefont {Monfardini},\ and\ \citenamefont {Vignati}}]{Cardani2018}%
  \BibitemOpen
  \bibfield  {author} {\bibinfo {author} {\bibfnamefont {L.}~\bibnamefont {Cardani}}, \bibinfo {author} {\bibfnamefont {N.}~\bibnamefont {Casali}}, \bibinfo {author} {\bibfnamefont {A.}~\bibnamefont {Cruciani}}, \bibinfo {author} {\bibfnamefont {H.}~\bibnamefont {le~Sueur}}, \bibinfo {author} {\bibfnamefont {M.}~\bibnamefont {Martinez}}, \bibinfo {author} {\bibfnamefont {F.}~\bibnamefont {Bellini}}, \bibinfo {author} {\bibfnamefont {M.}~\bibnamefont {Calvo}}, \bibinfo {author} {\bibfnamefont {M.~G.}\ \bibnamefont {Castellano}}, \bibinfo {author} {\bibfnamefont {I.}~\bibnamefont {Colantoni}}, \bibinfo {author} {\bibfnamefont {C.}~\bibnamefont {Cosmelli}}, \bibinfo {author} {\bibfnamefont {A.}~\bibnamefont {D'Addabbo}}, \bibinfo {author} {\bibfnamefont {S.~D.}\ \bibnamefont {Domizio}}, \bibinfo {author} {\bibfnamefont {J.}~\bibnamefont {Goupy}}, \bibinfo {author} {\bibfnamefont {L.}~\bibnamefont {Minutolo}}, \bibinfo {author} {\bibfnamefont {A.}~\bibnamefont {Monfardini}},\ and\ \bibinfo {author} {\bibfnamefont
  {M.}~\bibnamefont {Vignati}},\ }\bibfield  {title} {\bibinfo {title} {{Al/Ti/Al} phonon-mediated {KIDs} for {UV}{\textendash}vis light detection over large areas},\ }\href {https://doi.org/10.1088/1361-6668/aac1d4} {\bibfield  {journal} {\bibinfo  {journal} {Superconductor Science and Technology}\ }\textbf {\bibinfo {volume} {31}},\ \bibinfo {pages} {075002} (\bibinfo {year} {2018})},\ \Eprint {https://arxiv.org/abs/https://arxiv.org/abs/1801.08403} {https://arxiv.org/abs/1801.08403} \BibitemShut {NoStop}%
\bibitem [{\citenamefont {Golwala}\ and\ \citenamefont {Figueroa-Feliciano}(2022)}]{Golwala2022}%
  \BibitemOpen
  \bibfield  {author} {\bibinfo {author} {\bibfnamefont {S.}~\bibnamefont {Golwala}}\ and\ \bibinfo {author} {\bibfnamefont {E.}~\bibnamefont {Figueroa-Feliciano}},\ }\bibfield  {title} {\bibinfo {title} {Novel quantum sensors for light dark matter and neutrino detection},\ }\href {https://doi.org/10.1146/annurev-nucl-102020-112133} {\bibfield  {journal} {\bibinfo  {journal} {Annual Review of Nuclear and Particle Science}\ }\textbf {\bibinfo {volume} {72}},\ \bibinfo {pages} {419} (\bibinfo {year} {2022})}\BibitemShut {NoStop}%
\bibitem [{\citenamefont {Wen}\ \emph {et~al.}(2021)\citenamefont {Wen}, \citenamefont {Aralis}, \citenamefont {Thakur}, \citenamefont {Bumble}, \citenamefont {Chang}, \citenamefont {Ramanathan},\ and\ \citenamefont {Golwala}}]{Wen2021}%
  \BibitemOpen
  \bibfield  {author} {\bibinfo {author} {\bibfnamefont {O.}~\bibnamefont {Wen}}, \bibinfo {author} {\bibfnamefont {T.}~\bibnamefont {Aralis}}, \bibinfo {author} {\bibfnamefont {R.~B.}\ \bibnamefont {Thakur}}, \bibinfo {author} {\bibfnamefont {B.}~\bibnamefont {Bumble}}, \bibinfo {author} {\bibfnamefont {Y.-Y.}\ \bibnamefont {Chang}}, \bibinfo {author} {\bibfnamefont {K.}~\bibnamefont {Ramanathan}},\ and\ \bibinfo {author} {\bibfnamefont {S.}~\bibnamefont {Golwala}},\ }\bibfield  {title} {\bibinfo {title} {Performance of a phonon-mediated detector using {KIDs} optimized for sub-{GeV} dark matter},\ }\href {https://doi.org/10.1007/s10909-022-02764-2} {\bibfield  {journal} {\bibinfo  {journal} {Journal of Low Temperature Physics}\ }\textbf {\bibinfo {volume} {209}},\ \bibinfo {pages} {510} (\bibinfo {year} {2021})},\ \Eprint {https://arxiv.org/abs/2111.08064} {arXiv:2111.08064 [physics.ins-det]} \BibitemShut {NoStop}%
\bibitem [{\citenamefont {Ramanathan}\ \emph {et~al.}(2022)\citenamefont {Ramanathan}, \citenamefont {Aralis}, \citenamefont {Thakur}, \citenamefont {Bumble}, \citenamefont {Chang}, \citenamefont {Wen},\ and\ \citenamefont {Golwala}}]{Ramanathan2022}%
  \BibitemOpen
  \bibfield  {author} {\bibinfo {author} {\bibfnamefont {K.}~\bibnamefont {Ramanathan}}, \bibinfo {author} {\bibfnamefont {T.}~\bibnamefont {Aralis}}, \bibinfo {author} {\bibfnamefont {R.~B.}\ \bibnamefont {Thakur}}, \bibinfo {author} {\bibfnamefont {B.}~\bibnamefont {Bumble}}, \bibinfo {author} {\bibfnamefont {Y.-Y.}\ \bibnamefont {Chang}}, \bibinfo {author} {\bibfnamefont {O.}~\bibnamefont {Wen}},\ and\ \bibinfo {author} {\bibfnamefont {S.~R.}\ \bibnamefont {Golwala}},\ }\bibfield  {title} {\bibinfo {title} {Identifying drivers of energy resolution variation in a multi-{KID} phonon-mediated detector},\ }\bibfield  {journal} {\bibinfo  {journal} {Journal of Low Temperature Physics}\ }\href {https://doi.org/10.1007/s10909-022-02753-5} {10.1007/s10909-022-02753-5} (\bibinfo {year} {2022})\BibitemShut {NoStop}%
\bibitem [{\citenamefont {Chavarria}\ \emph {et~al.}(2016)\citenamefont {Chavarria}, \citenamefont {Collar}, \citenamefont {Peña}, \citenamefont {Privitera}, \citenamefont {Robinson}, \citenamefont {Scholz}, \citenamefont {Sengul}, \citenamefont {Zhou}, \citenamefont {Estrada}, \citenamefont {Izraelevitch}, \citenamefont {Tiffenberg}, \citenamefont {de~Mello~Neto},\ and\ \citenamefont {Machado}}]{Chavarria2016}%
  \BibitemOpen
  \bibfield  {author} {\bibinfo {author} {\bibfnamefont {A.~E.}\ \bibnamefont {Chavarria}}, \bibinfo {author} {\bibfnamefont {J.~I.}\ \bibnamefont {Collar}}, \bibinfo {author} {\bibfnamefont {J.~R.}\ \bibnamefont {Peña}}, \bibinfo {author} {\bibfnamefont {P.}~\bibnamefont {Privitera}}, \bibinfo {author} {\bibfnamefont {A.~E.}\ \bibnamefont {Robinson}}, \bibinfo {author} {\bibfnamefont {B.}~\bibnamefont {Scholz}}, \bibinfo {author} {\bibfnamefont {C.}~\bibnamefont {Sengul}}, \bibinfo {author} {\bibfnamefont {J.}~\bibnamefont {Zhou}}, \bibinfo {author} {\bibfnamefont {J.}~\bibnamefont {Estrada}}, \bibinfo {author} {\bibfnamefont {F.}~\bibnamefont {Izraelevitch}}, \bibinfo {author} {\bibfnamefont {J.}~\bibnamefont {Tiffenberg}}, \bibinfo {author} {\bibfnamefont {J.~R.~T.}\ \bibnamefont {de~Mello~Neto}},\ and\ \bibinfo {author} {\bibfnamefont {D.~T.}\ \bibnamefont {Machado}},\ }\bibfield  {title} {\bibinfo {title} {Measurement of the ionization produced by sub-kev silicon nuclear recoils in a ccd dark matter
  detector},\ }\href {https://doi.org/10.1103/PhysRevD.94.082007} {\bibfield  {journal} {\bibinfo  {journal} {Phys. Rev. D}\ }\textbf {\bibinfo {volume} {94}},\ \bibinfo {pages} {082007} (\bibinfo {year} {2016})},\ \Eprint {https://arxiv.org/abs/1608.00957v2} {1608.00957v2 [astro-ph.IM]} \BibitemShut {NoStop}%
\bibitem [{\citenamefont {Malnou}\ \emph {et~al.}(2019)\citenamefont {Malnou}, \citenamefont {Palken}, \citenamefont {Brubaker}, \citenamefont {Vale}, \citenamefont {Hilton},\ and\ \citenamefont {Lehnert}}]{Malnou2019b}%
  \BibitemOpen
  \bibfield  {author} {\bibinfo {author} {\bibfnamefont {M.}~\bibnamefont {Malnou}}, \bibinfo {author} {\bibfnamefont {D.~A.}\ \bibnamefont {Palken}}, \bibinfo {author} {\bibfnamefont {B.~M.}\ \bibnamefont {Brubaker}}, \bibinfo {author} {\bibfnamefont {L.~R.}\ \bibnamefont {Vale}}, \bibinfo {author} {\bibfnamefont {G.~C.}\ \bibnamefont {Hilton}},\ and\ \bibinfo {author} {\bibfnamefont {K.~W.}\ \bibnamefont {Lehnert}},\ }\bibfield  {title} {\bibinfo {title} {Squeezed vacuum used to accelerate the search for a weak classical signal},\ }\href {https://doi.org/10.1103/PhysRevX.9.021023} {\bibfield  {journal} {\bibinfo  {journal} {Phys. Rev. X}\ }\textbf {\bibinfo {volume} {9}},\ \bibinfo {pages} {021023} (\bibinfo {year} {2019})}\BibitemShut {NoStop}%
\bibitem [{\citenamefont {Klimovich}(2022)}]{Klimovich2022}%
  \BibitemOpen
  \bibfield  {author} {\bibinfo {author} {\bibfnamefont {N.~S.}\ \bibnamefont {Klimovich}},\ }\emph {\bibinfo {title} {Traveling Wave Parametric Amplifiers and Other Nonlinear Kinetic Inductance Devices}},\ \href {https://doi.org/10.7907/W980-RS97} {Ph.D. thesis},\ \bibinfo  {school} {California Institute of Technology} (\bibinfo {year} {2022})\BibitemShut {NoStop}%
\bibitem [{\citenamefont {Klimovich}\ \emph {et~al.}(2023)\citenamefont {Klimovich}, \citenamefont {Day}, \citenamefont {Shu}, \citenamefont {Eom}, \citenamefont {Leduc},\ and\ \citenamefont {Beyer}}]{Klimovich2023}%
  \BibitemOpen
  \bibfield  {author} {\bibinfo {author} {\bibfnamefont {N.}~\bibnamefont {Klimovich}}, \bibinfo {author} {\bibfnamefont {P.}~\bibnamefont {Day}}, \bibinfo {author} {\bibfnamefont {S.}~\bibnamefont {Shu}}, \bibinfo {author} {\bibfnamefont {B.~H.}\ \bibnamefont {Eom}}, \bibinfo {author} {\bibfnamefont {H.}~\bibnamefont {Leduc}},\ and\ \bibinfo {author} {\bibfnamefont {A.}~\bibnamefont {Beyer}},\ }\bibfield  {title} {\bibinfo {title} {Demonstration of a quantum noise limited traveling-wave parametric amplifier},\ }\href@noop {} {\  (\bibinfo {year} {2023})},\ \Eprint {https://arxiv.org/abs/2306.11028} {arXiv:2306.11028 [quant-ph]} \BibitemShut {NoStop}%
\bibitem [{\citenamefont {van Rantwijk}\ \emph {et~al.}(2016)\citenamefont {van Rantwijk}, \citenamefont {Grim}, \citenamefont {van Loon}, \citenamefont {Yates}, \citenamefont {Baryshev},\ and\ \citenamefont {Baselmans}}]{Rantwijk2016}%
  \BibitemOpen
  \bibfield  {author} {\bibinfo {author} {\bibfnamefont {J.}~\bibnamefont {van Rantwijk}}, \bibinfo {author} {\bibfnamefont {M.}~\bibnamefont {Grim}}, \bibinfo {author} {\bibfnamefont {D.}~\bibnamefont {van Loon}}, \bibinfo {author} {\bibfnamefont {S.}~\bibnamefont {Yates}}, \bibinfo {author} {\bibfnamefont {A.}~\bibnamefont {Baryshev}},\ and\ \bibinfo {author} {\bibfnamefont {J.}~\bibnamefont {Baselmans}},\ }\bibfield  {title} {\bibinfo {title} {Multiplexed readout for 1000-pixel arrays of microwave kinetic inductance detectors},\ }\href {https://doi.org/10.1109/tmtt.2016.2544303} {\bibfield  {journal} {\bibinfo  {journal} {IEEE Transactions on Microwave Theory and Techniques}\ }\textbf {\bibinfo {volume} {64}},\ \bibinfo {pages} {1876} (\bibinfo {year} {2016})}\BibitemShut {NoStop}%
\bibitem [{\citenamefont {Cardani}\ \emph {et~al.}(2021)\citenamefont {Cardani}, \citenamefont {Casali}, \citenamefont {Colantoni}, \citenamefont {Cruciani}, \citenamefont {Domizio}, \citenamefont {Martinez}, \citenamefont {Pettinacci}, \citenamefont {Pettinari},\ and\ \citenamefont {Vignati}}]{Cardani2021a}%
  \BibitemOpen
  \bibfield  {author} {\bibinfo {author} {\bibfnamefont {L.}~\bibnamefont {Cardani}}, \bibinfo {author} {\bibfnamefont {N.}~\bibnamefont {Casali}}, \bibinfo {author} {\bibfnamefont {I.}~\bibnamefont {Colantoni}}, \bibinfo {author} {\bibfnamefont {A.}~\bibnamefont {Cruciani}}, \bibinfo {author} {\bibfnamefont {S.~D.}\ \bibnamefont {Domizio}}, \bibinfo {author} {\bibfnamefont {M.}~\bibnamefont {Martinez}}, \bibinfo {author} {\bibfnamefont {V.}~\bibnamefont {Pettinacci}}, \bibinfo {author} {\bibfnamefont {G.}~\bibnamefont {Pettinari}},\ and\ \bibinfo {author} {\bibfnamefont {M.}~\bibnamefont {Vignati}},\ }\bibfield  {title} {\bibinfo {title} {Final results of {CALDER}: Kinetic inductance light detectors to search for rare events},\ }\bibfield  {journal} {\bibinfo  {journal} {The European Physical Journal C}\ }\textbf {\bibinfo {volume} {81}},\ \href {https://doi.org/10.1140/epjc/s10052-021-09454-5} {10.1140/epjc/s10052-021-09454-5} (\bibinfo {year} {2021}),\ \Eprint {https://arxiv.org/abs/2104.06850}
  {arXiv:2104.06850 [physics.ins-det]} \BibitemShut {NoStop}%
\bibitem [{\citenamefont {Cruciani}\ \emph {et~al.}(2022)\citenamefont {Cruciani}, \citenamefont {Bandiera}, \citenamefont {Calvo}, \citenamefont {Casali}, \citenamefont {Colantoni}, \citenamefont {Castello}, \citenamefont {del Gallo~Roccagiovine}, \citenamefont {Delicato}, \citenamefont {Giammei}, \citenamefont {Guidi}, \citenamefont {Goupy}, \citenamefont {Pettinacci}, \citenamefont {Pettinari}, \citenamefont {Romagnoni}, \citenamefont {Tamisari}, \citenamefont {Mazzolari}, \citenamefont {Monfardini},\ and\ \citenamefont {Vignati}}]{Cruciani2022}%
  \BibitemOpen
  \bibfield  {author} {\bibinfo {author} {\bibfnamefont {A.}~\bibnamefont {Cruciani}}, \bibinfo {author} {\bibfnamefont {L.}~\bibnamefont {Bandiera}}, \bibinfo {author} {\bibfnamefont {M.}~\bibnamefont {Calvo}}, \bibinfo {author} {\bibfnamefont {N.}~\bibnamefont {Casali}}, \bibinfo {author} {\bibfnamefont {I.}~\bibnamefont {Colantoni}}, \bibinfo {author} {\bibfnamefont {G.~D.}\ \bibnamefont {Castello}}, \bibinfo {author} {\bibfnamefont {M.}~\bibnamefont {del Gallo~Roccagiovine}}, \bibinfo {author} {\bibfnamefont {D.}~\bibnamefont {Delicato}}, \bibinfo {author} {\bibfnamefont {M.}~\bibnamefont {Giammei}}, \bibinfo {author} {\bibfnamefont {V.}~\bibnamefont {Guidi}}, \bibinfo {author} {\bibfnamefont {J.}~\bibnamefont {Goupy}}, \bibinfo {author} {\bibfnamefont {V.}~\bibnamefont {Pettinacci}}, \bibinfo {author} {\bibfnamefont {G.}~\bibnamefont {Pettinari}}, \bibinfo {author} {\bibfnamefont {M.}~\bibnamefont {Romagnoni}}, \bibinfo {author} {\bibfnamefont {M.}~\bibnamefont {Tamisari}}, \bibinfo {author}
  {\bibfnamefont {A.}~\bibnamefont {Mazzolari}}, \bibinfo {author} {\bibfnamefont {A.}~\bibnamefont {Monfardini}},\ and\ \bibinfo {author} {\bibfnamefont {M.}~\bibnamefont {Vignati}},\ }\bibfield  {title} {\bibinfo {title} {{BULLKID}: Monolithic array of particle absorbers sensed by kinetic inductance detectors},\ }\href {https://doi.org/https://doi.org/10.1063/5.0128723} {\bibfield  {journal} {\bibinfo  {journal} {Applied Physics Letters}\ }\textbf {\bibinfo {volume} {121}},\ \bibinfo {pages} {213504} (\bibinfo {year} {2022})},\ \Eprint {https://arxiv.org/abs/2209.14806} {arXiv:2209.14806 [physics.ins-det]} \BibitemShut {NoStop}%
\bibitem [{\citenamefont {Ren}\ \emph {et~al.}(2020)\citenamefont {Ren}, \citenamefont {Bathurst}, \citenamefont {Chang}, \citenamefont {Chen}, \citenamefont {Fink}, \citenamefont {Hong}, \citenamefont {Kurinsky}, \citenamefont {Mast}, \citenamefont {Mishra}, \citenamefont {Novati}, \citenamefont {Spahn}, \citenamefont {Theenhausen}, \citenamefont {Watkins}, \citenamefont {Williams}, \citenamefont {Wilson}, \citenamefont {Zaytsev}, \citenamefont {Bauer}, \citenamefont {Bunker}, \citenamefont {Figueroa-Feliciano}, \citenamefont {Hollister}, \citenamefont {Hsu}, \citenamefont {Lukens}, \citenamefont {Mahapatra}, \citenamefont {Mirabolfathi}, \citenamefont {Nebolsky}, \citenamefont {Platt}, \citenamefont {Ponce}, \citenamefont {Pyle}, \citenamefont {Reynolds},\ and\ \citenamefont {Saab}}]{Ren2020}%
  \BibitemOpen
  \bibfield  {author} {\bibinfo {author} {\bibfnamefont {R.}~\bibnamefont {Ren}}, \bibinfo {author} {\bibfnamefont {C.}~\bibnamefont {Bathurst}}, \bibinfo {author} {\bibfnamefont {Y.~Y.}\ \bibnamefont {Chang}}, \bibinfo {author} {\bibfnamefont {R.}~\bibnamefont {Chen}}, \bibinfo {author} {\bibfnamefont {C.~W.}\ \bibnamefont {Fink}}, \bibinfo {author} {\bibfnamefont {Z.}~\bibnamefont {Hong}}, \bibinfo {author} {\bibfnamefont {N.~A.}\ \bibnamefont {Kurinsky}}, \bibinfo {author} {\bibfnamefont {N.}~\bibnamefont {Mast}}, \bibinfo {author} {\bibfnamefont {N.}~\bibnamefont {Mishra}}, \bibinfo {author} {\bibfnamefont {V.}~\bibnamefont {Novati}}, \bibinfo {author} {\bibfnamefont {G.}~\bibnamefont {Spahn}}, \bibinfo {author} {\bibfnamefont {H.~M.~z.}\ \bibnamefont {Theenhausen}}, \bibinfo {author} {\bibfnamefont {S.~L.}\ \bibnamefont {Watkins}}, \bibinfo {author} {\bibfnamefont {Z.}~\bibnamefont {Williams}}, \bibinfo {author} {\bibfnamefont {M.~J.}\ \bibnamefont {Wilson}}, \bibinfo {author} {\bibfnamefont
  {A.}~\bibnamefont {Zaytsev}}, \bibinfo {author} {\bibfnamefont {D.}~\bibnamefont {Bauer}}, \bibinfo {author} {\bibfnamefont {R.}~\bibnamefont {Bunker}}, \bibinfo {author} {\bibfnamefont {E.}~\bibnamefont {Figueroa-Feliciano}}, \bibinfo {author} {\bibfnamefont {M.}~\bibnamefont {Hollister}}, \bibinfo {author} {\bibfnamefont {L.}~\bibnamefont {Hsu}}, \bibinfo {author} {\bibfnamefont {P.}~\bibnamefont {Lukens}}, \bibinfo {author} {\bibfnamefont {R.}~\bibnamefont {Mahapatra}}, \bibinfo {author} {\bibfnamefont {N.}~\bibnamefont {Mirabolfathi}}, \bibinfo {author} {\bibfnamefont {B.}~\bibnamefont {Nebolsky}}, \bibinfo {author} {\bibfnamefont {M.}~\bibnamefont {Platt}}, \bibinfo {author} {\bibfnamefont {F.}~\bibnamefont {Ponce}}, \bibinfo {author} {\bibfnamefont {M.}~\bibnamefont {Pyle}}, \bibinfo {author} {\bibfnamefont {T.}~\bibnamefont {Reynolds}},\ and\ \bibinfo {author} {\bibfnamefont {T.}~\bibnamefont {Saab}},\ }\bibfield  {title} {\bibinfo {title} {Design and characterization of a phonon-mediated cryogenic
  particle detector with an ev-scale threshold and 100 kev-scale dynamic range},\ }\bibfield  {journal} {\bibinfo  {journal} {Phys. Rev. D 104, 032010 (2021)}\ }\href {https://doi.org/10.1103/PhysRevD.104.032010} {10.1103/PhysRevD.104.032010} (\bibinfo {year} {2020}),\ \Eprint {https://arxiv.org/abs/2012.12430} {arXiv:2012.12430 [physics.ins-det]} \BibitemShut {NoStop}%
\bibitem [{\citenamefont {Romani}(2023)}]{Romani2023}%
  \BibitemOpen
  \bibfield  {author} {\bibinfo {author} {\bibfnamefont {R.}~\bibnamefont {Romani}},\ }\bibfield  {title} {\bibinfo {title} {Correlated and uncorrelated backgrounds and noise sources in athermal phonon detectors and other low temperature detectors},\ }in\ \href@noop {} {\emph {\bibinfo {booktitle} {Coordinating Panel for Advanced Detectors Workshop}}}\ (\bibinfo {year} {2023})\BibitemShut {NoStop}%
\bibitem [{\citenamefont {{D.G. Michael, P. Adamson, T. Alexopoulos et al. }}(2008)}]{Michael2008}%
  \BibitemOpen
  \bibfield  {author} {\bibinfo {author} {\bibnamefont {{D.G. Michael, P. Adamson, T. Alexopoulos et al. }}},\ }\bibfield  {title} {\bibinfo {title} {The magnetized steel and scintillator calorimeters of the minos experiment},\ }\href {https://doi.org/10.1016/j.nima.2008.08.003} {\bibfield  {journal} {\bibinfo  {journal} {Nuclear Instruments and Methods in Physics Research Section A: Accelerators, Spectrometers, Detectors and Associated Equipment}\ }\textbf {\bibinfo {volume} {596}},\ \bibinfo {pages} {190} (\bibinfo {year} {2008})}\BibitemShut {NoStop}%
\bibitem [{\citenamefont {Mei}\ and\ \citenamefont {Hime}(2006)}]{Mei2006}%
  \BibitemOpen
  \bibfield  {author} {\bibinfo {author} {\bibfnamefont {D.-M.}\ \bibnamefont {Mei}}\ and\ \bibinfo {author} {\bibfnamefont {A.}~\bibnamefont {Hime}},\ }\bibfield  {title} {\bibinfo {title} {Muon-induced background study for underground laboratories},\ }\href {https://doi.org/10.1103/physrevd.73.053004} {\bibfield  {journal} {\bibinfo  {journal} {Physical Review D}\ }\textbf {\bibinfo {volume} {73}},\ \bibinfo {pages} {053004} (\bibinfo {year} {2006})}\BibitemShut {NoStop}%
\bibitem [{hex()}]{hexadry}%
  \BibitemOpen
  \href {https://cryoconcept.com/product/hexa-dry-m/} {\bibinfo {title} {{HEXA-DRY M}}},\ \bibinfo {howpublished} {\url{https://cryoconcept.com/product/hexa-dry-m/}},\ \bibinfo {note} {accessed: 2023-11-09}\BibitemShut {NoStop}%
\bibitem [{Note01()}]{Note01}%
  \BibitemOpen
  Note01,\ \href@noop {} {}\bibinfo {note} {The volume of the inductor is estimated in the following way. The footprint area is known precisely by the design mask used in fabrication, while the film thickness is determined by the film deposition rate and exposure time. It was also measured via a Tencor$^\mathrm {TM}$ profilometer at the time of fabrication. These measurements confirm the thickness is as expected, with an uncertainty of 5\%. This uncertainty is subdominant to other sources of uncertainty, discussed below, but is included in the results presented.}\BibitemShut {Stop}%
\bibitem [{\citenamefont {Matthias}\ \emph {et~al.}(1963)\citenamefont {Matthias}, \citenamefont {Geballe},\ and\ \citenamefont {Compton}}]{Matthias1963}%
  \BibitemOpen
  \bibfield  {author} {\bibinfo {author} {\bibfnamefont {B.~T.}\ \bibnamefont {Matthias}}, \bibinfo {author} {\bibfnamefont {T.~H.}\ \bibnamefont {Geballe}},\ and\ \bibinfo {author} {\bibfnamefont {V.~B.}\ \bibnamefont {Compton}},\ }\bibfield  {title} {\bibinfo {title} {Superconductivity},\ }\href {https://doi.org/10.1103/revmodphys.35.1} {\bibfield  {journal} {\bibinfo  {journal} {Reviews of Modern Physics}\ }\textbf {\bibinfo {volume} {35}},\ \bibinfo {pages} {1} (\bibinfo {year} {1963})}\BibitemShut {NoStop}%
\bibitem [{\citenamefont {Martinez}\ \emph {et~al.}(2019)\citenamefont {Martinez}, \citenamefont {Cardani}, \citenamefont {Casali}, \citenamefont {Cruciani}, \citenamefont {Pettinari},\ and\ \citenamefont {Vignati}}]{Martinez2019}%
  \BibitemOpen
  \bibfield  {author} {\bibinfo {author} {\bibfnamefont {M.}~\bibnamefont {Martinez}}, \bibinfo {author} {\bibfnamefont {L.}~\bibnamefont {Cardani}}, \bibinfo {author} {\bibfnamefont {N.}~\bibnamefont {Casali}}, \bibinfo {author} {\bibfnamefont {A.}~\bibnamefont {Cruciani}}, \bibinfo {author} {\bibfnamefont {G.}~\bibnamefont {Pettinari}},\ and\ \bibinfo {author} {\bibfnamefont {M.}~\bibnamefont {Vignati}},\ }\bibfield  {title} {\bibinfo {title} {Measurements and simulations of athermal phonon transmission from silicon absorbers to aluminum sensors},\ }\href {https://doi.org/10.1103/physrevapplied.11.064025} {\bibfield  {journal} {\bibinfo  {journal} {Physical Review Applied}\ }\textbf {\bibinfo {volume} {11}},\ \bibinfo {pages} {064025} (\bibinfo {year} {2019})},\ \Eprint {https://arxiv.org/abs/1805.02495} {arXiv:1805.02495 [physics.ins-det]} \BibitemShut {NoStop}%
\bibitem [{\citenamefont {Fyhrie}\ \emph {et~al.}(2016)\citenamefont {Fyhrie}, \citenamefont {McKenney}, \citenamefont {Glenn}, \citenamefont {LeDuc}, \citenamefont {Gao}, \citenamefont {Day},\ and\ \citenamefont {Zmuidzinas}}]{Fyhrie2016}%
  \BibitemOpen
  \bibfield  {author} {\bibinfo {author} {\bibfnamefont {A.}~\bibnamefont {Fyhrie}}, \bibinfo {author} {\bibfnamefont {C.}~\bibnamefont {McKenney}}, \bibinfo {author} {\bibfnamefont {J.}~\bibnamefont {Glenn}}, \bibinfo {author} {\bibfnamefont {H.~G.}\ \bibnamefont {LeDuc}}, \bibinfo {author} {\bibfnamefont {J.}~\bibnamefont {Gao}}, \bibinfo {author} {\bibfnamefont {P.}~\bibnamefont {Day}},\ and\ \bibinfo {author} {\bibfnamefont {J.}~\bibnamefont {Zmuidzinas}},\ }\bibfield  {title} {\bibinfo {title} {Responsivity boosting in fir tin lekids using phonon recycling: simulations and array design},\ }in\ \href {https://doi.org/10.1117/12.2231476} {\emph {\bibinfo {booktitle} {Millimeter, Submillimeter, and Far-Infrared Detectors and Instrumentation for Astronomy VIII}}},\ \bibinfo {editor} {edited by\ \bibinfo {editor} {\bibfnamefont {W.~S.}\ \bibnamefont {Holland}}\ and\ \bibinfo {editor} {\bibfnamefont {J.}~\bibnamefont {Zmuidzinas}}}\ (\bibinfo  {publisher} {SPIE},\ \bibinfo {year} {2016})\BibitemShut {NoStop}%
\bibitem [{\citenamefont {Ulbricht}\ \emph {et~al.}(2015)\citenamefont {Ulbricht}, \citenamefont {Mazin}, \citenamefont {Szypryt}, \citenamefont {Walter}, \citenamefont {Bockstiegel},\ and\ \citenamefont {Bumble}}]{Ulbricht2015}%
  \BibitemOpen
  \bibfield  {author} {\bibinfo {author} {\bibfnamefont {G.}~\bibnamefont {Ulbricht}}, \bibinfo {author} {\bibfnamefont {B.~A.}\ \bibnamefont {Mazin}}, \bibinfo {author} {\bibfnamefont {P.}~\bibnamefont {Szypryt}}, \bibinfo {author} {\bibfnamefont {A.~B.}\ \bibnamefont {Walter}}, \bibinfo {author} {\bibfnamefont {C.}~\bibnamefont {Bockstiegel}},\ and\ \bibinfo {author} {\bibfnamefont {B.}~\bibnamefont {Bumble}},\ }\bibfield  {title} {\bibinfo {title} {Highly multiplexible thermal kinetic inductance detectors for x-ray imaging spectroscopy},\ }\bibfield  {journal} {\bibinfo  {journal} {Applied Physics Letters}\ }\textbf {\bibinfo {volume} {106}},\ \href {https://doi.org/10.1063/1.4923096} {10.1063/1.4923096} (\bibinfo {year} {2015})\BibitemShut {NoStop}%
\bibitem [{\citenamefont {Tinkham}(2004)}]{Tinkham2004}%
  \BibitemOpen
  \bibfield  {author} {\bibinfo {author} {\bibfnamefont {M.}~\bibnamefont {Tinkham}},\ }\href {https://doi.org/0486134725, 9780486134727} {\emph {\bibinfo {title} {Introduction to Superconductivity}}}\ (\bibinfo  {publisher} {Courier Corporation},\ \bibinfo {year} {2004})\BibitemShut {NoStop}%
\bibitem [{\citenamefont {Spahn}(2021)}]{Spahn2021}%
  \BibitemOpen
  \bibfield  {author} {\bibinfo {author} {\bibfnamefont {G.~C.}\ \bibnamefont {Spahn}},\ }\emph {\bibinfo {title} {Improving Performance of a Multiplexed Dark Matter Detector via Infrared Isolation}},\ \href {https://conservancy.umn.edu/handle/11299/225047} {Ph.D. thesis},\ \bibinfo  {school} {University of Minnesota} (\bibinfo {year} {2021})\BibitemShut {NoStop}%
\bibitem [{Note02()}]{Note02}%
  \BibitemOpen
  Note02,\ \href@noop {} {}\bibinfo {note} {This value does not affect our analysis in any way. We have estimated the noise temperature of our system at the HEMT output, however there are relatively large uncertainties on the in situ gain values. Using the maximum specified attenuation of the components in our post-HEMT signal chain, we find that the PSD of our noise is compatible with 3.6 K Johnson noise to within a factor of 2.}\BibitemShut {Stop}%
\bibitem [{\citenamefont {{Low Noise Factory}}(2022)}]{HEMT}%
  \BibitemOpen
  \bibfield  {author} {\bibinfo {author} {\bibnamefont {{Low Noise Factory}}},\ }\href {https://lownoisefactory.com/wp-content/uploads/2023/04/lnf-lnc0-3_14a.pdf} {\bibinfo {title} {{Datasheet LNF-LNC0.3\_14A}}} (\bibinfo {year} {2022})\BibitemShut {NoStop}%
\bibitem [{\citenamefont {Minutolo}\ \emph {et~al.}(2019)\citenamefont {Minutolo}, \citenamefont {Steinbach}, \citenamefont {Wandui},\ and\ \citenamefont {O'Brient}}]{Minutolo2019}%
  \BibitemOpen
  \bibfield  {author} {\bibinfo {author} {\bibfnamefont {L.}~\bibnamefont {Minutolo}}, \bibinfo {author} {\bibfnamefont {B.}~\bibnamefont {Steinbach}}, \bibinfo {author} {\bibfnamefont {A.}~\bibnamefont {Wandui}},\ and\ \bibinfo {author} {\bibfnamefont {R.}~\bibnamefont {O'Brient}},\ }\bibfield  {title} {\bibinfo {title} {A flexible gpu-accelerated radio-frequency readout for superconducting detectors},\ }\href {https://doi.org/10.1109/tasc.2019.2912027} {\bibfield  {journal} {\bibinfo  {journal} {IEEE Transactions on Applied Superconductivity}\ }\textbf {\bibinfo {volume} {29}},\ \bibinfo {pages} {1} (\bibinfo {year} {2019})}\BibitemShut {NoStop}%
\bibitem [{\citenamefont {Mazin}\ \emph {et~al.}(2002)\citenamefont {Mazin}, \citenamefont {Day}, \citenamefont {Zmuidzinas},\ and\ \citenamefont {Leduc}}]{Mazin2002}%
  \BibitemOpen
  \bibfield  {author} {\bibinfo {author} {\bibfnamefont {B.~A.}\ \bibnamefont {Mazin}}, \bibinfo {author} {\bibfnamefont {P.~K.}\ \bibnamefont {Day}}, \bibinfo {author} {\bibfnamefont {J.}~\bibnamefont {Zmuidzinas}},\ and\ \bibinfo {author} {\bibfnamefont {H.~G.}\ \bibnamefont {Leduc}},\ }\bibfield  {title} {\bibinfo {title} {Multiplexable kinetic inductance detectors},\ }in\ \href {https://doi.org/10.1063/1.1457652} {\emph {\bibinfo {booktitle} {AIP Conference Proceedings}}}\ (\bibinfo  {publisher} {American Institute of Physics},\ \bibinfo {year} {2002})\BibitemShut {NoStop}%
\bibitem [{\citenamefont {Zmuidzinas}(2012)}]{Zmuidzinas2012}%
  \BibitemOpen
  \bibfield  {author} {\bibinfo {author} {\bibfnamefont {J.}~\bibnamefont {Zmuidzinas}},\ }\bibfield  {title} {\bibinfo {title} {Superconducting microresonators: Physics and applications},\ }\href {https://doi.org/10.1146/annurev-conmatphys-020911-125022} {\bibfield  {journal} {\bibinfo  {journal} {Annual Review of Condensed Matter Physics}\ }\textbf {\bibinfo {volume} {3}},\ \bibinfo {pages} {169} (\bibinfo {year} {2012})}\BibitemShut {NoStop}%
\bibitem [{\citenamefont {Siegel}(2016)}]{Siegel2016}%
  \BibitemOpen
  \bibfield  {author} {\bibinfo {author} {\bibfnamefont {S.~R.}\ \bibnamefont {Siegel}},\ }\emph {\bibinfo {title} {A Multiwavelength Study of the Intracluster Medium and the Characterization of the Multiwavelength Sub/millimeter Inductance Camera}},\ \href@noop {} {Ph.D. thesis},\ \bibinfo  {school} {California Institute of Technology} (\bibinfo {year} {2016})\BibitemShut {NoStop}%
\bibitem [{\citenamefont {Gao}(2008)}]{Gao2008c}%
  \BibitemOpen
  \bibfield  {author} {\bibinfo {author} {\bibfnamefont {J.}~\bibnamefont {Gao}},\ }\emph {\bibinfo {title} {The Physics of Superconducting Microwave Resonators}},\ \href@noop {} {Ph.D. thesis},\ \bibinfo  {school} {California Institute of Technology} (\bibinfo {year} {2008})\BibitemShut {NoStop}%
\bibitem [{\citenamefont {Isaila}\ \emph {et~al.}(2012)\citenamefont {Isaila}, \citenamefont {Ciemniak}, \citenamefont {Feilitzsch}, \citenamefont {G\:{u}tlein}, \citenamefont {Kemmer}, \citenamefont {Lachenmaier}, \citenamefont {Lanfranchi}, \citenamefont {Pfister}, \citenamefont {Potzel}, \citenamefont {Roth}, \citenamefont {Sivers}, \citenamefont {Strauss}, \citenamefont {Westphal},\ and\ \citenamefont {Wiest}}]{Isaila2012}%
  \BibitemOpen
  \bibfield  {author} {\bibinfo {author} {\bibfnamefont {C.}~\bibnamefont {Isaila}}, \bibinfo {author} {\bibfnamefont {C.}~\bibnamefont {Ciemniak}}, \bibinfo {author} {\bibfnamefont {F.}~\bibnamefont {Feilitzsch}}, \bibinfo {author} {\bibfnamefont {A.}~\bibnamefont {G\:{u}tlein}}, \bibinfo {author} {\bibfnamefont {J.}~\bibnamefont {Kemmer}}, \bibinfo {author} {\bibfnamefont {T.}~\bibnamefont {Lachenmaier}}, \bibinfo {author} {\bibfnamefont {J.-C.}\ \bibnamefont {Lanfranchi}}, \bibinfo {author} {\bibfnamefont {S.}~\bibnamefont {Pfister}}, \bibinfo {author} {\bibfnamefont {W.}~\bibnamefont {Potzel}}, \bibinfo {author} {\bibfnamefont {S.}~\bibnamefont {Roth}}, \bibinfo {author} {\bibfnamefont {M.}~\bibnamefont {Sivers}}, \bibinfo {author} {\bibfnamefont {R.}~\bibnamefont {Strauss}}, \bibinfo {author} {\bibfnamefont {W.}~\bibnamefont {Westphal}},\ and\ \bibinfo {author} {\bibfnamefont {F.}~\bibnamefont {Wiest}},\ }\bibfield  {title} {\bibinfo {title} {Low-temperature light detectors: Neganov–luke amplification
  and calibration},\ }\href {https://doi.org/10.1016/j.physletb.2012.08.003} {\bibfield  {journal} {\bibinfo  {journal} {Physics Letters B}\ }\textbf {\bibinfo {volume} {716}},\ \bibinfo {pages} {160} (\bibinfo {year} {2012})}\BibitemShut {NoStop}%
\bibitem [{\citenamefont {Noroozian}\ \emph {et~al.}(2009)\citenamefont {Noroozian}, \citenamefont {Gao}, \citenamefont {Zmuidzinas}, \citenamefont {LeDuc},\ and\ \citenamefont {Mazin}}]{Noroozian2009}%
  \BibitemOpen
  \bibfield  {author} {\bibinfo {author} {\bibfnamefont {O.}~\bibnamefont {Noroozian}}, \bibinfo {author} {\bibfnamefont {J.}~\bibnamefont {Gao}}, \bibinfo {author} {\bibfnamefont {J.}~\bibnamefont {Zmuidzinas}}, \bibinfo {author} {\bibfnamefont {H.~G.}\ \bibnamefont {LeDuc}},\ and\ \bibinfo {author} {\bibfnamefont {B.~A.}\ \bibnamefont {Mazin}},\ }\bibfield  {title} {\bibinfo {title} {Two-level system noise reduction for microwave kinetic inductance detectors},\ }\bibfield  {journal} {\bibinfo  {journal} {AIP Conference Proceedings (LTD13), vol. 1185, pp. 148-151, 2009}\ }\href {https://doi.org/10.1063/1.3292302} {10.1063/1.3292302} (\bibinfo {year} {2009}),\ \Eprint {https://arxiv.org/abs/0909.2060} {arXiv:0909.2060 [physics.ins-det]} \BibitemShut {NoStop}%
\bibitem [{\citenamefont {Sueno}\ \emph {et~al.}(2021)\citenamefont {Sueno}, \citenamefont {Honda}, \citenamefont {Kutsuma}, \citenamefont {Mima}, \citenamefont {Otani}, \citenamefont {Oguri}, \citenamefont {Suzuki},\ and\ \citenamefont {Tajima}}]{Sueno2021}%
  \BibitemOpen
  \bibfield  {author} {\bibinfo {author} {\bibfnamefont {Y.}~\bibnamefont {Sueno}}, \bibinfo {author} {\bibfnamefont {S.}~\bibnamefont {Honda}}, \bibinfo {author} {\bibfnamefont {H.}~\bibnamefont {Kutsuma}}, \bibinfo {author} {\bibfnamefont {S.}~\bibnamefont {Mima}}, \bibinfo {author} {\bibfnamefont {C.}~\bibnamefont {Otani}}, \bibinfo {author} {\bibfnamefont {S.}~\bibnamefont {Oguri}}, \bibinfo {author} {\bibfnamefont {J.}~\bibnamefont {Suzuki}},\ and\ \bibinfo {author} {\bibfnamefont {O.}~\bibnamefont {Tajima}},\ }\bibfield  {title} {\bibinfo {title} {Characterization of two-level system noise for microwave kinetic inductance detector comprising niobium film on silicon substrate},\ }\bibfield  {journal} {\bibinfo  {journal} {Prog Theor Exp Phys (2022)}\ }\href {https://doi.org/10.1093/ptep/ptac023} {10.1093/ptep/ptac023} (\bibinfo {year} {2021}),\ \Eprint {https://arxiv.org/abs/2110.00127} {arXiv:2110.00127 [physics.ins-det]} \BibitemShut {NoStop}%
\bibitem [{\citenamefont {de~Visser}\ \emph {et~al.}(2011)\citenamefont {de~Visser}, \citenamefont {Baselmans}, \citenamefont {Diener}, \citenamefont {Yates}, \citenamefont {Endo},\ and\ \citenamefont {Klapwijk}}]{Visser2011}%
  \BibitemOpen
  \bibfield  {author} {\bibinfo {author} {\bibfnamefont {P.~J.}\ \bibnamefont {de~Visser}}, \bibinfo {author} {\bibfnamefont {J.~J.~A.}\ \bibnamefont {Baselmans}}, \bibinfo {author} {\bibfnamefont {P.}~\bibnamefont {Diener}}, \bibinfo {author} {\bibfnamefont {S.~J.~C.}\ \bibnamefont {Yates}}, \bibinfo {author} {\bibfnamefont {A.}~\bibnamefont {Endo}},\ and\ \bibinfo {author} {\bibfnamefont {T.~M.}\ \bibnamefont {Klapwijk}},\ }\bibfield  {title} {\bibinfo {title} {Number fluctuations of sparse quasiparticles in a superconductor},\ }\href {https://doi.org/10.1103/PhysRevLett.106.167004} {\bibfield  {journal} {\bibinfo  {journal} {Physical Review Letters}\ }\textbf {\bibinfo {volume} {106}},\ \bibinfo {pages} {167004} (\bibinfo {year} {2011})},\ \Eprint {https://arxiv.org/abs/1103.0758} {arXiv:1103.0758 [cond-mat.supr-con]} \BibitemShut {NoStop}%
\bibitem [{\citenamefont {de~Visser}\ \emph {et~al.}(2012)\citenamefont {de~Visser}, \citenamefont {Baselmans}, \citenamefont {Diener}, \citenamefont {Yates}, \citenamefont {Endo},\ and\ \citenamefont {Klapwijk}}]{Visser2012}%
  \BibitemOpen
  \bibfield  {author} {\bibinfo {author} {\bibfnamefont {P.~J.}\ \bibnamefont {de~Visser}}, \bibinfo {author} {\bibfnamefont {J.~J.~A.}\ \bibnamefont {Baselmans}}, \bibinfo {author} {\bibfnamefont {P.}~\bibnamefont {Diener}}, \bibinfo {author} {\bibfnamefont {S.~J.~C.}\ \bibnamefont {Yates}}, \bibinfo {author} {\bibfnamefont {A.}~\bibnamefont {Endo}},\ and\ \bibinfo {author} {\bibfnamefont {T.~M.}\ \bibnamefont {Klapwijk}},\ }\bibfield  {title} {\bibinfo {title} {Generation-recombination noise: The fundamental sensitivity limit for kinetic inductance detectors},\ }\href {https://doi.org/10.1007/s10909-012-0519-5} {\bibfield  {journal} {\bibinfo  {journal} {Journal of Low Temperature Physics}\ }\textbf {\bibinfo {volume} {167}},\ \bibinfo {pages} {335} (\bibinfo {year} {2012})}\BibitemShut {NoStop}%
\bibitem [{\citenamefont {Zadeh}\ and\ \citenamefont {Ragazzini}(1952)}]{Zadeh1952}%
  \BibitemOpen
  \bibfield  {author} {\bibinfo {author} {\bibfnamefont {L.}~\bibnamefont {Zadeh}}\ and\ \bibinfo {author} {\bibfnamefont {J.}~\bibnamefont {Ragazzini}},\ }\bibfield  {title} {\bibinfo {title} {Optimum filters for the detection of signals in noise},\ }\href {https://doi.org/10.1109/jrproc.1952.274117} {\bibfield  {journal} {\bibinfo  {journal} {Proceedings of the IRE}\ }\textbf {\bibinfo {volume} {40}},\ \bibinfo {pages} {1223} (\bibinfo {year} {1952})}\BibitemShut {NoStop}%
\bibitem [{\citenamefont {Golwala}(2000)}]{Golwala2000}%
  \BibitemOpen
  \bibfield  {author} {\bibinfo {author} {\bibfnamefont {S.~R.}\ \bibnamefont {Golwala}},\ }\emph {\bibinfo {title} {Exclusion Limits on the WIMP-Nucleon Elastic-Scattering Cross Section fromthe Cryogenic Dark Matter Search}},\ \href@noop {} {Ph.D. thesis},\ \bibinfo  {school} {University of California, Berkeley} (\bibinfo {year} {2000})\BibitemShut {NoStop}%
\bibitem [{\citenamefont {Pyle}(2012)}]{Pyle2012}%
  \BibitemOpen
  \bibfield  {author} {\bibinfo {author} {\bibfnamefont {M.~C.}\ \bibnamefont {Pyle}},\ }\emph {\bibinfo {title} {Optimizing the design and analysis of cryogenic semiconductor dark matter detectors for maximum sensitivity}},\ \href {https://doi.org/https://doi.org/10.2172/1127926} {Ph.D. thesis},\ \bibinfo  {school} {Stanford University} (\bibinfo {year} {2012})\BibitemShut {NoStop}%
\bibitem [{Note03()}]{Note03}%
  \BibitemOpen
  Note03,\ \href@noop {} {}\bibinfo {note} {The optimal filter estimator for resolution is constructed from a $\chi ^2$ fit, in frequency space, of the observed pulse to a signal template. The expression for $\chi ^2$ is then minimized with respect to the pulse amplitude estimator. The expression above is simply the expected variance of the distribution of this amplitude estimator. Taken another way, this is the discrete case of the integration of a normalized signal-to-noise ratio over the full pulse shape.}\BibitemShut {Stop}%
\bibitem [{Note04()}]{Note04}%
  \BibitemOpen
  Note04,\ \href@noop {} {}\bibinfo {note} {At the input of the room-temperature optical feed-through of the dilution refrigerator.}\BibitemShut {Stop}%
\bibitem [{\citenamefont {Bludau}\ \emph {et~al.}(1974)\citenamefont {Bludau}, \citenamefont {Onton},\ and\ \citenamefont {Heinke}}]{Bludau1974}%
  \BibitemOpen
  \bibfield  {author} {\bibinfo {author} {\bibfnamefont {W.}~\bibnamefont {Bludau}}, \bibinfo {author} {\bibfnamefont {A.}~\bibnamefont {Onton}},\ and\ \bibinfo {author} {\bibfnamefont {W.}~\bibnamefont {Heinke}},\ }\bibfield  {title} {\bibinfo {title} {Temperature dependence of the band gap of silicon},\ }\href {https://doi.org/10.1063/1.1663501} {\bibfield  {journal} {\bibinfo  {journal} {Journal of Applied Physics}\ }\textbf {\bibinfo {volume} {45}},\ \bibinfo {pages} {1846} (\bibinfo {year} {1974})}\BibitemShut {NoStop}%
\bibitem [{\citenamefont {Ramanathan}\ and\ \citenamefont {Kurinsky}(2020)}]{Ramanathan2020ab}%
  \BibitemOpen
  \bibfield  {author} {\bibinfo {author} {\bibfnamefont {K.}~\bibnamefont {Ramanathan}}\ and\ \bibinfo {author} {\bibfnamefont {N.}~\bibnamefont {Kurinsky}},\ }\bibfield  {title} {\bibinfo {title} {Ionization yield in silicon for ev-scale electron-recoil processes},\ }\href {https://doi.org/10.1103/PhysRevD.102.063026} {\bibfield  {journal} {\bibinfo  {journal} {Physical Review D}\ }\textbf {\bibinfo {volume} {102}},\ \bibinfo {pages} {063026} (\bibinfo {year} {2020})},\ \Eprint {https://arxiv.org/abs/2004.10709v2} {arXiv:2004.10709v2 [astro-ph.IM]} \BibitemShut {NoStop}%
\bibitem [{\citenamefont {Aralis}(2024)}]{Aralis2024}%
  \BibitemOpen
  \bibfield  {author} {\bibinfo {author} {\bibfnamefont {T.~B.}\ \bibnamefont {Aralis}},\ }\emph {\bibinfo {title} {SuperCDMS SNOLAB, HVeV Run 3, and Development of KIPM Detectors}},\ \href {https://doi.org/10.7907/GXJP-0863} {Ph.D. thesis},\ \bibinfo  {school} {California Institute of Technology} (\bibinfo {year} {2024})\BibitemShut {NoStop}%
\bibitem [{Note05()}]{Note05}%
  \BibitemOpen
  Note05,\ \href@noop {} {}\bibinfo {note} {Note that changing the post-pulse definition to start 5 ms (or longer) after the pulse rising edge (vs the 2.5 ms used in the analysis) does not significantly change the number of events passing this cut ($<1$\% difference in number of events passing this cut).}\BibitemShut {Stop}%
\bibitem [{Note06()}]{Note06}%
  \BibitemOpen
  Note06,\ \href@noop {} {}\bibinfo {note} {The larger RMS factor is used due to the long length of the post-pulse region relative to the pre-trigger region.}\BibitemShut {Stop}%
\bibitem [{Note07()}]{Note07}%
  \BibitemOpen
  Note07,\ \href@noop {} {}\bibinfo {note} {This cut retains nearly 100\% of all noise-only traces where the baseline is well-behaved and no spurious pulses are present.}\BibitemShut {Stop}%
\bibitem [{Note08()}]{Note08}%
  \BibitemOpen
  Note08,\ \href@noop {} {}\bibinfo {note} {This is due to noise from the first-stage amplifier at a power larger than that of the generation-recombination (GR) noise. See Appendix~\ref {supp-noise} for a brief discussion of GR noise.}\BibitemShut {Stop}%
\bibitem [{Note09()}]{Note09}%
  \BibitemOpen
  Note09,\ \href@noop {} {}\bibinfo {note} {In fact, one can see in Fig~\ref {fig:average_pulse} that there is a small contribution of a third, very fast, time constant which gives rise to the ``kink" shortly after the maximal pulse amplitude. The time span for this kink is of order 10 $\mu $s which corresponds to a frequency of $10^5$ Hz. The upper limit for the frequency PSDs that enter the optimal filter formalism is $10^4$ Hz, and as such any effects on timescales smaller than 100 $\mu $s do not have an effect on our result.}\BibitemShut {Stop}%
\bibitem [{\citenamefont {Moore}(2012)}]{Moore2012a}%
  \BibitemOpen
  \bibfield  {author} {\bibinfo {author} {\bibfnamefont {D.~C.}\ \bibnamefont {Moore}},\ }\emph {\bibinfo {title} {A Search for Low-Mass Dark Matter with the Cryogenic Dark Matter Search and the Development of Highly Multiplexed Phonon-Mediated Particle Detectors}},\ \href@noop {} {Ph.D. thesis},\ \bibinfo  {school} {California Institute of Technology} (\bibinfo {year} {2012})\BibitemShut {NoStop}%
\bibitem [{\citenamefont {Kaplan}\ \emph {et~al.}(1976)\citenamefont {Kaplan}, \citenamefont {Chi}, \citenamefont {Langenberg}, \citenamefont {Chang}, \citenamefont {Jafarey},\ and\ \citenamefont {Scalapino}}]{Kaplan1976}%
  \BibitemOpen
  \bibfield  {author} {\bibinfo {author} {\bibfnamefont {S.~B.}\ \bibnamefont {Kaplan}}, \bibinfo {author} {\bibfnamefont {C.~C.}\ \bibnamefont {Chi}}, \bibinfo {author} {\bibfnamefont {D.~N.}\ \bibnamefont {Langenberg}}, \bibinfo {author} {\bibfnamefont {J.~J.}\ \bibnamefont {Chang}}, \bibinfo {author} {\bibfnamefont {S.}~\bibnamefont {Jafarey}},\ and\ \bibinfo {author} {\bibfnamefont {D.~J.}\ \bibnamefont {Scalapino}},\ }\bibfield  {title} {\bibinfo {title} {Quasiparticle and phonon lifetimes in superconductors},\ }\href {https://doi.org/10.1103/physrevb.14.4854} {\bibfield  {journal} {\bibinfo  {journal} {Physical Review B}\ }\textbf {\bibinfo {volume} {14}},\ \bibinfo {pages} {4854} (\bibinfo {year} {1976})}\BibitemShut {NoStop}%
\bibitem [{\citenamefont {Ramanathan}\ \emph {et~al.}(2023)\citenamefont {Ramanathan}, \citenamefont {Klimovich}, \citenamefont {BasuThakur}, \citenamefont {Eom}, \citenamefont {Leduc}, \citenamefont {Shu}, \citenamefont {Beyer},\ and\ \citenamefont {Day}}]{Ramanathan2023}%
  \BibitemOpen
  \bibfield  {author} {\bibinfo {author} {\bibfnamefont {K.}~\bibnamefont {Ramanathan}}, \bibinfo {author} {\bibfnamefont {N.}~\bibnamefont {Klimovich}}, \bibinfo {author} {\bibfnamefont {R.}~\bibnamefont {BasuThakur}}, \bibinfo {author} {\bibfnamefont {B.~H.}\ \bibnamefont {Eom}}, \bibinfo {author} {\bibfnamefont {H.~G.}\ \bibnamefont {Leduc}}, \bibinfo {author} {\bibfnamefont {S.}~\bibnamefont {Shu}}, \bibinfo {author} {\bibfnamefont {A.~D.}\ \bibnamefont {Beyer}},\ and\ \bibinfo {author} {\bibfnamefont {P.~K.}\ \bibnamefont {Day}},\ }\bibfield  {title} {\bibinfo {title} {Wideband direct detection constraints on hidden photon dark matter with the {QUALIPHIDE} experiment},\ }\href {https://doi.org/10.1103/physrevlett.130.231001} {\bibfield  {journal} {\bibinfo  {journal} {Physical Review Letters}\ }\textbf {\bibinfo {volume} {130}},\ \bibinfo {pages} {231001} (\bibinfo {year} {2023})}\BibitemShut {NoStop}%
\bibitem [{\citenamefont {Kelsey}\ \emph {et~al.}(2023)\citenamefont {Kelsey}, \citenamefont {Agnese}, \citenamefont {Alam}, \citenamefont {Langroudy}, \citenamefont {Azadbakht}, \citenamefont {Brandt}, \citenamefont {Bunker}, \citenamefont {Cabrera}, \citenamefont {Chang}, \citenamefont {Coombes}, \citenamefont {Cormier}, \citenamefont {Diamond}, \citenamefont {Edwards}, \citenamefont {Figueroa-Feliciano}, \citenamefont {Gao}, \citenamefont {Harrington}, \citenamefont {Hong}, \citenamefont {Hui}, \citenamefont {Kurinsky}, \citenamefont {Lawrence}, \citenamefont {Loer}, \citenamefont {Masten}, \citenamefont {Michaud}, \citenamefont {Michielin}, \citenamefont {Miller}, \citenamefont {Novati}, \citenamefont {Oblath}, \citenamefont {Orrell}, \citenamefont {Perry}, \citenamefont {Redl}, \citenamefont {Reynolds}, \citenamefont {Saab}, \citenamefont {Sadoulet}, \citenamefont {Serniak}, \citenamefont {Singh}, \citenamefont {Speaks}, \citenamefont {Stanford}, \citenamefont {Stevens}, \citenamefont {Strube},
  \citenamefont {Toback}, \citenamefont {Ullom}, \citenamefont {VanDevender}, \citenamefont {Vissers}, \citenamefont {Wilson}, \citenamefont {Wilson}, \citenamefont {Zatschler},\ and\ \citenamefont {Zatschler}}]{Kelsey2023}%
  \BibitemOpen
  \bibfield  {author} {\bibinfo {author} {\bibfnamefont {M.}~\bibnamefont {Kelsey}}, \bibinfo {author} {\bibfnamefont {R.}~\bibnamefont {Agnese}}, \bibinfo {author} {\bibfnamefont {Y.}~\bibnamefont {Alam}}, \bibinfo {author} {\bibfnamefont {I.~A.}\ \bibnamefont {Langroudy}}, \bibinfo {author} {\bibfnamefont {E.}~\bibnamefont {Azadbakht}}, \bibinfo {author} {\bibfnamefont {D.}~\bibnamefont {Brandt}}, \bibinfo {author} {\bibfnamefont {R.}~\bibnamefont {Bunker}}, \bibinfo {author} {\bibfnamefont {B.}~\bibnamefont {Cabrera}}, \bibinfo {author} {\bibfnamefont {Y.-Y.}\ \bibnamefont {Chang}}, \bibinfo {author} {\bibfnamefont {H.}~\bibnamefont {Coombes}}, \bibinfo {author} {\bibfnamefont {R.}~\bibnamefont {Cormier}}, \bibinfo {author} {\bibfnamefont {M.}~\bibnamefont {Diamond}}, \bibinfo {author} {\bibfnamefont {E.}~\bibnamefont {Edwards}}, \bibinfo {author} {\bibfnamefont {E.}~\bibnamefont {Figueroa-Feliciano}}, \bibinfo {author} {\bibfnamefont {J.}~\bibnamefont {Gao}}, \bibinfo {author} {\bibfnamefont
  {P.}~\bibnamefont {Harrington}}, \bibinfo {author} {\bibfnamefont {Z.}~\bibnamefont {Hong}}, \bibinfo {author} {\bibfnamefont {M.}~\bibnamefont {Hui}}, \bibinfo {author} {\bibfnamefont {N.}~\bibnamefont {Kurinsky}}, \bibinfo {author} {\bibfnamefont {R.}~\bibnamefont {Lawrence}}, \bibinfo {author} {\bibfnamefont {B.}~\bibnamefont {Loer}}, \bibinfo {author} {\bibfnamefont {M.}~\bibnamefont {Masten}}, \bibinfo {author} {\bibfnamefont {E.}~\bibnamefont {Michaud}}, \bibinfo {author} {\bibfnamefont {E.}~\bibnamefont {Michielin}}, \bibinfo {author} {\bibfnamefont {J.}~\bibnamefont {Miller}}, \bibinfo {author} {\bibfnamefont {V.}~\bibnamefont {Novati}}, \bibinfo {author} {\bibfnamefont {N.}~\bibnamefont {Oblath}}, \bibinfo {author} {\bibfnamefont {J.}~\bibnamefont {Orrell}}, \bibinfo {author} {\bibfnamefont {W.}~\bibnamefont {Perry}}, \bibinfo {author} {\bibfnamefont {P.}~\bibnamefont {Redl}}, \bibinfo {author} {\bibfnamefont {T.}~\bibnamefont {Reynolds}}, \bibinfo {author} {\bibfnamefont {T.}~\bibnamefont {Saab}},
  \bibinfo {author} {\bibfnamefont {B.}~\bibnamefont {Sadoulet}}, \bibinfo {author} {\bibfnamefont {K.}~\bibnamefont {Serniak}}, \bibinfo {author} {\bibfnamefont {J.}~\bibnamefont {Singh}}, \bibinfo {author} {\bibfnamefont {Z.}~\bibnamefont {Speaks}}, \bibinfo {author} {\bibfnamefont {C.}~\bibnamefont {Stanford}}, \bibinfo {author} {\bibfnamefont {J.}~\bibnamefont {Stevens}}, \bibinfo {author} {\bibfnamefont {J.}~\bibnamefont {Strube}}, \bibinfo {author} {\bibfnamefont {D.}~\bibnamefont {Toback}}, \bibinfo {author} {\bibfnamefont {J.}~\bibnamefont {Ullom}}, \bibinfo {author} {\bibfnamefont {B.}~\bibnamefont {VanDevender}}, \bibinfo {author} {\bibfnamefont {M.}~\bibnamefont {Vissers}}, \bibinfo {author} {\bibfnamefont {M.}~\bibnamefont {Wilson}}, \bibinfo {author} {\bibfnamefont {J.}~\bibnamefont {Wilson}}, \bibinfo {author} {\bibfnamefont {B.}~\bibnamefont {Zatschler}},\ and\ \bibinfo {author} {\bibfnamefont {S.}~\bibnamefont {Zatschler}},\ }\bibfield  {title} {\bibinfo {title} {G4cmp: Condensed matter
  physics simulation using the geant4 toolkit},\ }\href {https://doi.org/10.1016/j.nima.2023.168473} {\bibfield  {journal} {\bibinfo  {journal} {Nuclear Instruments and Methods in Physics Research Section A: Accelerators, Spectrometers, Detectors and Associated Equipment}\ }\textbf {\bibinfo {volume} {1055}},\ \bibinfo {pages} {168473} (\bibinfo {year} {2023})}\BibitemShut {NoStop}%
\bibitem [{\citenamefont {Shu}\ \emph {et~al.}(2022)\citenamefont {Shu}, \citenamefont {Beyer}, \citenamefont {Day}, \citenamefont {Defrance}, \citenamefont {Sayers},\ and\ \citenamefont {Golwala}}]{Shu2022}%
  \BibitemOpen
  \bibfield  {author} {\bibinfo {author} {\bibfnamefont {S.}~\bibnamefont {Shu}}, \bibinfo {author} {\bibfnamefont {A.}~\bibnamefont {Beyer}}, \bibinfo {author} {\bibfnamefont {P.}~\bibnamefont {Day}}, \bibinfo {author} {\bibfnamefont {F.}~\bibnamefont {Defrance}}, \bibinfo {author} {\bibfnamefont {J.}~\bibnamefont {Sayers}},\ and\ \bibinfo {author} {\bibfnamefont {S.}~\bibnamefont {Golwala}},\ }\bibfield  {title} {\bibinfo {title} {A multi-chroic kinetic inductance detectors array using hierarchical phased array antenna},\ }\href {https://doi.org/10.1007/s10909-022-02890-x} {\bibfield  {journal} {\bibinfo  {journal} {Journal of Low Temperature Physics}\ }\textbf {\bibinfo {volume} {209}},\ \bibinfo {pages} {330} (\bibinfo {year} {2022})},\ \Eprint {https://arxiv.org/abs/2112.05366} {arXiv:2112.05366} \BibitemShut {NoStop}%
\bibitem [{Note10()}]{Note10}%
  \BibitemOpen
  Note10,\ \href@noop {} {}\bibinfo {note} {From~\cite {Siegel2016} Eq. 2.59, we see $\delta f/f = -\frac {\alpha }{2}\kappa _2 n_\mathrm {qp}$. We are interested in determining how this changes with respect to energy absorbed by the superconductor. Since $\alpha ,\kappa _2$ are independent of deposited energy, we must find $\mathrm {d}/\mathrm {d}E_\mathrm {abs} (n_\mathrm {qp})$. The density of quasiparticles increases by $2/V$ for every $2\Delta $ deposited, thus $\mathrm {d}/\mathrm {d}E_\mathrm {abs} (n_\mathrm {qp}) = \frac {1}{V\Delta }$.}\BibitemShut {Stop}%
\bibitem [{\citenamefont {Green}(2008)}]{Green2008}%
  \BibitemOpen
  \bibfield  {author} {\bibinfo {author} {\bibfnamefont {M.~A.}\ \bibnamefont {Green}},\ }\bibfield  {title} {\bibinfo {title} {Self-consistent optical parameters of intrinsic silicon at 300k including temperature coefficients},\ }\href {https://doi.org/10.1016/j.solmat.2008.06.009} {\bibfield  {journal} {\bibinfo  {journal} {Solar Energy Materials and Solar Cells}\ }\textbf {\bibinfo {volume} {92}},\ \bibinfo {pages} {1305} (\bibinfo {year} {2008})}\BibitemShut {NoStop}%
\bibitem [{\citenamefont {Loebich}(1972)}]{Loebich1972}%
  \BibitemOpen
  \bibfield  {author} {\bibinfo {author} {\bibfnamefont {O.}~\bibnamefont {Loebich}},\ }\bibfield  {title} {\bibinfo {title} {The optical properties of gold: A review of their technical utilisation},\ }\href {https://doi.org/10.1007/bf03215148} {\bibfield  {journal} {\bibinfo  {journal} {Gold Bulletin}\ }\textbf {\bibinfo {volume} {5}},\ \bibinfo {pages} {2} (\bibinfo {year} {1972})}\BibitemShut {NoStop}%
\end{thebibliography}
%

\end{document}